\documentclass[twocolumn,prd,aps,longbibliography,superscriptaddress,preprintnumbers,tightenlines,showpacs,nofootinbib,eqsecnum,amsfonts,amsmath]{revtex4-1}
\pdfoutput=1

\usepackage{epsfig}
\usepackage{graphics}
\usepackage{graphicx}
\usepackage{amsmath,amssymb,mathrsfs}
\usepackage{amsfonts}
\usepackage[usenames,dvipsnames]{xcolor}
\usepackage{wasysym}
\usepackage{times}
\usepackage{mathptmx}
\usepackage{gensymb}
\usepackage{appendix}
\usepackage{listings}
\usepackage{url}
\usepackage[normalem]{ulem} 
\usepackage{alltt}
\usepackage[colorlinks]{hyperref}
\usepackage{cleveref}
\usepackage{longtable}
\usepackage{enumitem}
\setlist{nosep}
\usepackage{color}
\usepackage{calc}
\usepackage{tensor}
\usepackage{bm}
\usepackage{times}
\usepackage{multirow}
\usepackage[varg]{txfonts}
\usepackage{float}
\usepackage{dcolumn}
\usepackage[nolist,nohyperlinks]{acronym}
\usepackage{xspace}
\usepackage[english]{babel}
\usepackage[abs]{overpic}
\usepackage{pict2e}
\usepackage[caption=false]{subfig} 
\allowdisplaybreaks[1]
\usepackage[utf8]{inputenc}
\usepackage{gensymb}
\usepackage{bm}
\usepackage{stackengine}
\usepackage{boldline,multirow}
\usepackage{braket}
\usepackage{longtable}

\usepackage{tabularx}
\usepackage{rotating}

\DeclareMathAlphabet{\mathcalstd}{OMS}{cmsy}{m}{n}
\DeclareMathAlphabet{\mathpzc}{OT1}{pzc}{m}{it}

\newcommand{\UIB}{Departament de F\'isica, Universitat de les Illes Balears, IAC3 -- IEEC, Crta. Valldemossa km 7.5, E-07122 Palma, Spain}

\newcommand{\UoB}{School of Physics and Astronomy and Institute for Gravitational Wave Astronomy, University of Birmingham, Edgbaston, Birmingham, B15 9TT, United Kingdom}

\definecolor{dodgerblue}{HTML}{1E90FF}
\definecolor{viennared}{HTML}{DA0A14}

\hypersetup{citecolor=dodgerblue}

\newcommand{\SXSqfiveID}{{\ensuremath{36}}}
\newcommand{\SXSqonepointfiveID}{{\ensuremath{10}}}

\begin{document}



\title{ 
Validity of common modelling approximations for precessing binary black holes with higher-order modes}

\author{Antoni Ramos-Buades}
\affiliation{\UIB}
\author{Patricia Schmidt}
\affiliation{\UoB}
\author{Geraint Pratten}
\affiliation{\UoB}
\affiliation{\UIB}
\author{Sascha Husa}
\affiliation{\UIB}

\date{\today}

\begin{abstract}
The current paradigm for constructing waveforms from precessing compact binaries is to first construct a waveform in a non-inertial, co-precessing binary source frame followed by a time-dependent rotation to map back to the physical, inertial frame. A key insight in the construction of these models is that the co-precessing waveform can be effectively mapped to some equivalent aligned spin waveform. Secondly, the time-dependent rotation implicitly introduces $m$-mode mixing, necessitating an accurate description of higher-order modes in the co-precessing frame. We assess the efficacy of this modelling strategy in the strong field regime using Numerical Relativity simulations. We find that this framework allows for the highly accurate construction of $(2,\pm 2)$ modes in our data set, while for higher order modes, especially the $(2,|1|), (3,|2|)$ and $(4,|3|)$ modes, we find rather large mismatches. We also investigate a variant of the approximate map between co-precessing and aligned spin waveforms, where we only identify the slowly varying part of the time domain co-precessing waveforms with the aligned-spin one, but find no significant improvement. 
Our results indicate that the simple paradigm to construct precessing waveforms does not provide an accurate description of higher order modes in the strong-field regime, and demonstrate the necessity for modelling mode asymmetries and mode-mixing to significantly improve the description of precessing higher order modes.
\end{abstract}

\maketitle

\acrodef{PN}{post-Newtonian}
\acrodef{EOB}{effective-one-body}
\acrodef{NR}{numerical relativity}
\acrodef{GW}{gravitational wave}
\acrodef{BBH}{binary black hole}
\acrodef{BH}{black hole}
\acrodef{BNS}{binary neutron star}
\acrodef{NSBH}{neutron star-black hole}
\acrodef{SNR}{signal-to-noise ratio}
\acrodef{aLIGO}{Advanced LIGO}
\acrodef{AdV}{Advanced Virgo}

\newcommand{\PN}[0]{\ac{PN}\xspace}
\newcommand{\EOB}[0]{\ac{EOB}\xspace}
\newcommand{\NR}[0]{\ac{NR}\xspace}
\newcommand{\BBH}[0]{\ac{BBH}\xspace}
\newcommand{\BH}[0]{\ac{BH}\xspace}
\newcommand{\BNS}[0]{\ac{BNS}\xspace}
\newcommand{\NSBH}[0]{\ac{NSBH}\xspace}
\newcommand{\GW}[0]{\ac{GW}\xspace}
\newcommand{\SNR}[0]{\ac{SNR}\xspace}
\newcommand{\aLIGO}[0]{\ac{aLIGO}\xspace}
\newcommand{\AdV}[0]{\ac{AdV}\xspace}

\newcommand{\citeme}[0]{{\color{purple}{Citation!}}}

\newcommand*{\vecL}[0]{${{\mathbf{L}}}$ }

\newcommand{\n}{\newline}

\section{Introduction}\label{sec:introduction}
\label{sec:intro}
The first observation of \acp{GW} from colliding black holes by Advanced LIGO~\cite{TheLIGOScientific:2014jea,PhysRevLett.116.061102} marked the beginning of a new era in astronomy. Since then, GWs from twelve coalescing compact binaries such \acp{BBH} and \acp{BNS} have been detected confidently~\cite{LIGOScientific:2018mvr, Abbott:2020uma} by \ac{aLIGO} and Virgo~\cite{TheVirgo:2014hva}, and many more GW candidates have been recorded since the start of the third observing run~\cite{GraceDBO3}. For all confident BBH detections, the emitted signal was found to be consistent with predictions from General Relativity \cite{LIGOScientific:2019fpa, Abbott:2018lct} and consistent with compact binaries whose spins are aligned with the orbital angular momentum \vecL~\cite{LIGOScientific:2018mvr}. The \GW signal of such aligned-spin binaries is well described by the current generation of semi-analytic waveform models~\cite{PhysRevD.93.044006,Khan2015,Pan2014,Bohe:2016gbl, phenX} governing the inspiral, merger and ringdown. More recent work~\cite{phenXHM,london2018,PhysRevD.98.084028,PhysRevD.99.064045} has focused on extending these waveform models to incorporate subdominant harmonics beyond the dominant quadrupolar ($2,|2|$) modes. 

Generic BBHs, however, can have arbitrarily oriented spin configurations, i.e., the spins are not (anti-)parallel to the orbital angular momentum. Relativistic couplings between the orbital and spin angular momenta induce precession of the spins and the orbital plane \cite{PhysRevD.49.6274,PhysRevD.52.821}, resulting in complex amplitude and phase modulations of GW signal. This complicates waveform modelling efforts and impedes brute force Numerical Relativity (NR) studies as the parameter space grows from three intrinsic parameters to seven for quasi-spherical binaries~\cite{Hannam:2013pra}. 
Recent attempts, guided by reduced order modelling strategies \cite{Field:2013cfa, Blackman:2014maa}, have been successful in accurately modelling precessing waveforms in very restricted domains of the parameter space \cite{Blackman:2017dfb,Blackman:2017pcm,PhysRevResearch.1.033015}. 

In recent years, a number of key breakthroughs in waveform modelling enabled the development of the first inspiral-merger-ringdown (IMR) waveforms for precessing compact binaries \cite{PhysRevD.84.024046,OShaughnessy2011,Boyle2011,PhysRevD.86.104063, PhysRevD.91.024043,PhysRevD.67.104025}. A key insight was the observation that the waveform of precessing binaries can be greatly simplified when transformed to an effectively co-precessing, non-inertial frame that tracks the leading-order precession of the orbital plane \cite{PhysRevD.84.024046, OShaughnessy2011, Boyle:2011dy}. This general framework has since been used to produce several IMR waveform models of precessing binaries~\cite{phenomp, Pan:2013rra, Blackman:2015pia, Blackman:2017pcm, Blackman:2017dfb, PhysRevD.100.024059}.
A second crucial insight was the realisation that a co-precessing waveform can be approximately mapped to a some equivalent aligned-spin waveform~\cite{PhysRevD.84.024046,PhysRevD.91.024043,Pekowsky:2013ska}. This identification is predicated on an approximate decoupling between the spin components parallel to the orbital angular momentum \vecL and the spin components perpendicular to \vecL (in-plane spins)~\cite{PhysRevD.91.024043}.
Schematically, we can construct an approximate precessing waveform using a time-dependent rotation of the co-precessing waveform modes given a model of the precessional motion of the orbital plane~\cite{PhysRevD.84.024046,PhysRevD.86.104063}.
Within this general framework, several approximations are commonly made, though different waveform models use different approximations. Here, we focus on the \emph{phenomenological} waveform family, a key tool for LIGO data analysis due to its computational efficiency. 
Precessing phenomenological waveform models~\cite{phenomp,Bohe:PPv2,PhysRevD.100.024059,Khan:2019kot} are constructed using three independent pieces: 1) an aligned-spin waveform model, 2) a model for the Euler angles describing the time-dependent rotation of the orbital plane, and 3) a modification of the final state that captures spin-precession effects. The most commonly for \GW analysis used model IMRPhenomPv2 \cite{phenomp,Bohe:PPv2}, has recently been upgraded to include double-spin effects in the inspiral~\cite{PhysRevD.100.024059}, and to incorporate (uncalibrated) subdominant spherical harmonic modes in the co-precessing frame~\cite{Khan:2019kot}, while a forthcoming phenomenological waveform model will include the calibration of these modes~\cite{phenXP}.

Precessing phenomenological waveform models are constructed based on a set of simplifying approximations. Each of these approximations will introduce systematic modelling errors. While current observations are dominated by the statistical uncertainty in the measurement, advances in the detector sensitivity will reveal systematic errors. We thus need to understand the impact of various modelling approximations to guide the model development for the coming years. Due to the dimensionality of the precessing parameter space, systematic studies are sparse. Here, we make a first attempt at scrutinizing two main approximations made in the phenomenological modelling paradigm:
\begin{enumerate}
\item (APX1) The identification between co-precessing and aligned-spin waveform modes.
\item (APX2) The choice of subdominant harmonic modes used in constructing the co-precessing waveform modes, i.e., the number of aligned-spin modes used to generate the approximate precessing modes.
\end{enumerate} 
In particular, we focus on the limitations of these two approximations when extended to higher order mode for both individual modes as well as the strain. We note that in the analyses presented here, we neglect modifications to the final state and compute the Euler angles directly from the precessing NR simulations.

The paper is organised as follows: In Sec.~\ref{sec:precession} we briefly summarise the general framework used to model precessing binaries. In Sec. \ref{sec:NRDataSet} we present the data set of NR waveforms used in this study, afterwards we present our results on the validity of (APX1) and (APX2) in Sec.~\ref{sec:systematic}. In Sec. \ref{sec:caveats} we discuss caveats and possible improvements of (APX1). We conclude in Sec.~\ref{sec:summary}. In Appendices~\ref{sec:AppendixA}-\ref{sec:AppendixE} we present details of the NR data set, additional results and supporting analyses.

Throughout we use geometric units $G=c=1$. To simplify expressions we set the total mass of the system to $M=m_1+m_2=1$ unless stated otherwise. We define the mass ratio as $q=m_1/m_2 \geq 1$ with $m_1 \geq m_2$. We also introduce the symmetric mass ratio $\eta=q/(1+q)^2$, and we will denote the black holes' dimensionless spin vectors by $\bm{\chi}_i= \bm{S}_i/m^2_i$, for $i=1,2$. 

\section{Modelling Precessing Binaries}
\label{sec:precession}
The orbital precession dynamics of a binary system is encoded in three time-dependent Euler angles $\lbrace \beta (t) , \alpha (t), \varepsilon (t) \rbrace$ \cite{PhysRevD.84.024046},
where $\beta$ is the angle between the total angular momentum $\bm{J}$ 
and \vecL and $\alpha$ is the azimuthal orientation of \vecL. These two angles track the direction of the maximal radiation axis, which is approximately normal to the orbital plane \cite{Schmidt:2012rh}. The final angle, $\varepsilon$, corresponds to a rotation around the maximal radiation axis given by enforcing the minimal rotation condition \cite{Boyle:2011gg}, $\varepsilon = -\int \dot{\alpha}(t) \cos\beta(t) \, dt$, which is related to the precession rate of the binary.

A coordinate frame, which tracks the orbital precession is referred to as co-precessing. In any such co-precessing frame, the waveform modes $h^{\rm{co-prec}}_{\ell m}$ can be obtained by an active rotation applied to the modes $h^{\rm{prec}}_{\ell m}$ obtained in an inertial coordinate system~\cite{PhysRevD.84.024046,PhysRevD.86.104063}:
\begin{equation}
h^{\rm{co-prec}}_{\ell m}(t) = \displaystyle\sum^{\ell}_{k = - \ell} {\bf{R}}_{\ell m k} (\beta, \alpha, \varepsilon ) \; h^{\rm{prec}}_{\ell k} (t) ,
\label{eq00}
\end{equation}
where $\mathbf{R}_{k\ell m}(\beta,\alpha,\varepsilon)$ is the $k\ell m$-element of the rotation operator which describes the inertial motion, adopting the $(z,y,z)$-convention.
It follows that the inverse transformation permits the generation of precessing waveform modes, i.e., given the modes in the co-precessing frame, we find
\begin{equation}
h^{\rm{prec}}_{\ell m} = \displaystyle\sum^{\ell}_{k = - \ell} {\bf{R}}^{-1}_{\ell m k} (\beta, \alpha, \varepsilon ) \; h^{\rm{co-prec}}_{\ell k} (t).
\label{eq:inv}
\end{equation}

While all available precessing IMR waveform models use this general framework, they make different assumptions about the RHS of Eq.~\eqref{eq:inv}. 
In particular, phenomenological waveform models~\cite{Schmidt:2012rh,phenomp,PhysRevD.100.024059} identify the co-precessing waveform modes in Eq.~\eqref{eq:inv} with aligned-spin (AS) modes obtained from a binary with the same mass ratio and spin component parallel to \vecL, i.e.,
\begin{equation}
\bar{h}^{\rm{prec}}_{\ell m}(t; q, \bm{S}_1, \bm{S}_2) = \sum_{k=-l}^l \mathbf{R}^{-1}_{\ell mk}(\beta,\alpha,\varepsilon) \; \; h^{\rm AS}_{lk}(t; q, S_{1||}(t_0), S_{2||}(t_0)),
\label{eq01}
\end{equation}
where $\bar{h}^{\rm{prec}}_{lk}$ and $h^{\rm{AS}}_{lk}$ denote the approximate precessing and AS waveform modes, respectively. Given an appropriate description of the rotation operator, the identification between $h_{\ell m}^{\rm co-prec} \simeq h_{\ell m}^{\rm AS}$ (APX1) provides a straightforward procedure to construct approximate precessing waveforms.

One key aspect of precessing waveforms that is not captured by this identification are mode asymmetries~\cite{Boyle:2014ioa}. For time domain aligned-spin waveforms the negative-$m$ modes are given by complex conjugation, i.e.,
\begin{equation}
    h^{\rm AS}_{\ell, -m}  = (-1)^\ell \left(h^{\rm AS}_{\ell m}\right)^*,
    \label{eq:eq001}
\end{equation}
where the symbol $*$ denotes complex conjugation. This relation is no longer true for precessing waveforms, which is neglected in the identification $h_{\ell m}^{\rm co-prec} \simeq h_{\ell m}^{\rm AS}$. We investigate in detail the effect of neglecting these mode-asymmetries in Sec.~\ref{sec:QAvsSA}.

\section{Numerical Relativity Dataset} 
\label{sec:NRDataSet}
\begin{table*}[]
    \centering
    \def\arraystretch{1.2 }
    \begin{tabular}{c c c c c c c c c c c }
    \hline
    \hline
    ID & Simulation &  Code &  q  &$\bm{\chi}_1$ & $\bm{\chi}_2$ & $ \chi_{\text{eff}}$ &$D/M$ & $M \Omega_0$ & $e_0 \cdot 10^{-3}$   \\
    \hline
         10 & \texttt{SXS:BBH:0023} & \texttt{SpEC} & 1.5& $(0.5, 0.05, 0.)$& $(0.08, -0.49, 0.)$& 0.& 16.& 0.0145& 0.28 \\
         
         36 &\texttt{SXS:BBH:0058} & \texttt{SpEC} & 5.& $(0.5, 0.03, 0.)$& $(0., 0., 0.)$& 0.& 15.& 0.0158& 2.12 \\
         
         28 &\texttt{q3.\_\_0.56\_0.56\_0.\_\_0.6\_0.\_0.\_T\_80\_400} & \texttt{BAM}  & 3  & $(0.75, -0.27, 0.)$ & $(0.3, 0.52, 0.)$   & 0 & 8.83 &  0.0329  &  2.94 \\
         \hline
         \hline
    \end{tabular}
    \caption{Parameters of three precessing simulations highlighted in various analyses. The full list of NR simulations and further details for all simulations can be found in Tab.~\ref{tab:tabNR3}.}
    \label{tab:cases}
\end{table*}

The set of NR simulations used in this study includes publicly available waveforms from the SXS Collaboration~\cite{PhysRevLett.111.241104}, as well as non-public waveforms generated with BAM \cite{Bruegmann:2006at,Husa:2007hp} and the open-source Einstein Toolkit \cite{Loffler:2011ay,maria_babiuc_hamilton_2019_3522086}. The simulations employed here, including their properties are listed in Tab. \ref{tab:tabNR3} of App. \ref{sec:AppendixA}.
Throughout the main text we will highlight results for three precessing cases: IDs 10, 28 and 36. Their parameters are listed in Tab.~\ref{tab:cases}. We choose these three cases due to the presence of particular features which we discuss in detail in Sec. \ref{sec:QAvsSA}.

\begin{figure}[hbt!]
\centering
\includegraphics[scale=0.22]{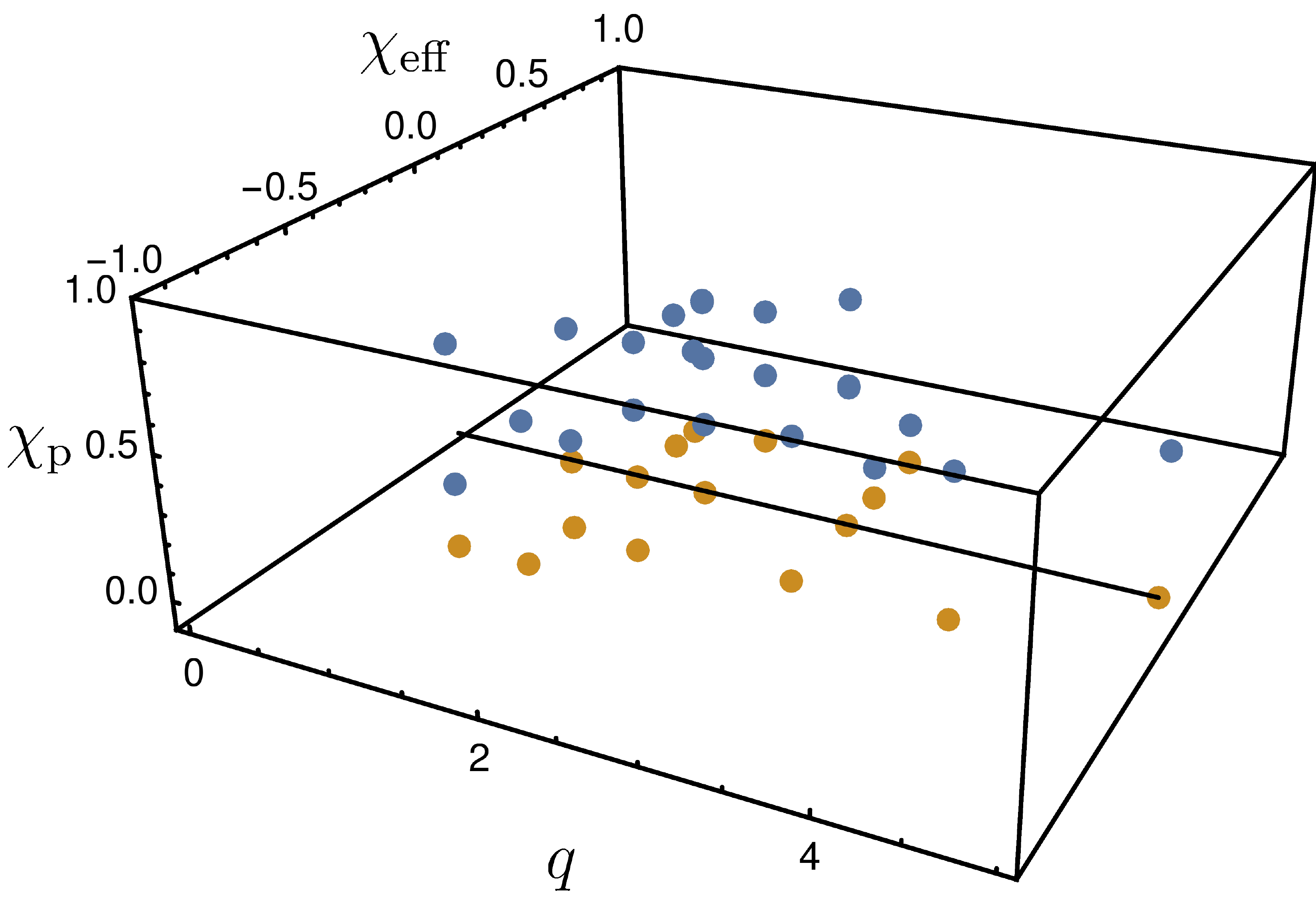}
\caption{Parameter space distribution in mass ratio, $q$, and effective spin parameters, $\chi_{\text{eff}}$ and $\chi_{p}$, of the NR simulations from Table \ref{tab:tabNR3}. The black thick line corresponds to $\chi_{\rm eff}=0$. The blue (yellow) dots represent precessing (aligned-spin) simulations. The simulations span the following parameter ranges: $q \in [1,5]$, $\chi_{\rm eff} \in [-0.5, 0.38]$ and $\chi_p \in [0,0.8]$.
}
\label{fig:NRcatalog}
\end{figure}

From all available NR simulations we pair the precessing and AS waveforms whose initial dimensionless spin vector projected onto the initial orbital angular momentum $\hat{\bm{L}}_0$ coincide, i.e.,
\begin{equation}
\hat{\bm{L}}^{\rm AS}_0 \cdot \bm{\chi}^{\rm AS}_{0,i} \equiv \hat{\bm{L}}^{\rm prec}_0 \cdot \bm{\chi}^{\rm prec}_{0,i},  \quad  i=1,2,
\label{eq:eq100}
\end{equation}
where $\hat{\bm{L}}_0= \bm{L}_0/ ||\bm{L}_0||$ is the unit orbital angular momentum vector after junk radiation. Note that satisfying Eq.~\eqref{eq:eq100} exactly is very difficult when working with NR simulations. We have thus chosen a tolerance of $10^{-3}$. Applying this criterion we obtain 36 unique precessing simulations with an AS counterpart. Figure~\ref{fig:NRcatalog} shows the distribution of the mass ratio $q$ as well as two spin parameters for the 36 NR simulations: (i) the effective inspiral spin parameter $\chi_\textrm{eff}$ ~\cite{Ajith2011,PhysRevD.78.044021} given by
\begin{equation}
\chi_\mathrm{eff}=\frac{m_1 \chi_{1L}+ m_2 \chi_{2L}}{m_1+m_2},
\label{eq101}
\end{equation}
where $\chi_{iL}= \bm{\chi}_i \cdot \hat{\bm{L}}$ with $i=1,2$, and (ii) the effective precession spin parameter $\chi_p$~\cite{PhysRevD.91.024043} defined as
\begin{equation}
\chi_{p}:=\frac{S_p}{A_2 m^2_2}, \quad S_p= \max(A_1 S_{1 \perp},A_2 S_{2 \perp}),
\label{eq102}
\end{equation}
where $A_1=2+3q/2$, $A_2=2+3/(2q)$, and $S_{i \perp}$, with $i=1,2$, is the norm of the spin components perpendicular to \vecL (in-plane spin components).
The effective spin parameter is a mass weighted combination of the spin components aligned with \vecL, which predominantly affects the inspiral rate~\cite{PhysRevLett.106.241101}. It is the best constrained spin parameter to date~\cite{LIGOScientific:2018mvr}. The in-plane spin components source the precession of the binary system. The average precession exhibited by the binary is approximated by $\chi_p$.

The NR simulations have been selected according to the following criteria:
\begin{itemize}
    \item[\textbf{1)}] \emph{Waveform accuracy}. When multiple resolutions of a simulation  are available, we use the highest resolution and the waveforms extracted at largest extraction radius. For SXS waveforms we choose the second order extrapolation to future-null infinity. 
    \item[\textbf{2)}] \emph{NR code}. We only compare simulations produced with the same NR code to avoid systematics coming from the different numerical methods and ambiguities due to the use of different gauges.
    \item[\textbf{3)}] \emph{Length requirements}. Due to the lack of a robust hybridization procedure between precessing NR and post-Newtonian inspiral waveforms as well as the introduction of additional systematics, we restrict this study to NR waveforms only. We select NR waveforms long enough to cover a total mass below 100 $M_\odot$ at 20 Hz for all the considered $(\ell m)$ modes, except for one BAM case, ID 28, whose length is shorter but it is interesting as it has a high value of $\chi_p=0.8$. 
    \item[\textbf{4)}] \emph{Residual eccentricity}. We only select NR simulations that have a residual initial eccentricity $e \leq 3 \times 10^{-3}$. The low-eccentricity initial parameters of the ET simulations have been computed using the method developed in \cite{PhysRevD.99.023003}.
\end{itemize}

\section{Methodology}
\label{sec:methods}

\subsection{Quadrupole Alignment}  
\label{sec:introQA }
Several ways have been put forward to compute the waveform modes in a co-precessing frame~\cite{PhysRevD.84.024046,OShaughnessy2011,Boyle2011}. We choose the method developed in~\cite{PhysRevD.84.024046} referred to as \emph{quadrupole-alignment}, henceforth abbreviated QA.
It is based on finding the coordinate frame that maximises the mean magnitude of the $(2,\pm 2)$ modes~\cite{PhysRevD.84.024046,OShaughnessy2011,Boyle2011,PhysRevD.86.104063, PhysRevD.91.024043}. 

Once the three time-dependent Euler angles that define this frame have been computed, each precessing waveform mode can be rotated to this QA frame through Eq. \eqref{eq00}. Conversely, given AS modes, these can be rotated through Eq. \eqref{eq01} into an inertial frame where they resemble precessing modes. 

Furthermore, in order to minimize the effect of the difference between the inertial frames of the rotated AS and the precessing simulations, we perform an additional global rotation of the $(\ell m)$ modes specified by the three Euler angles which rotate the z-axis onto the direction of the final spin of the black-hole. The direction of the final spin is computed from the data of the apparent horizon of the simulations, while its magnitude is computed using two different procedures, from the apparent horizon of the simulations and from fits to the ringdown orbital frequency as in \cite{PhysRevD.95.064024}. Note that another approximated fixed direction for a precessing system is the asymptotic total angular of the system \cite{Schmidt:2012rh}, which one could in principle compute from the initial total angular momentum of the system read from the NR simulations and evolve it backwards in time using PN equations of motion. However, this is a difficult procedure due to the gauge differences between PN and NR. We have also tested that the  differences between the directions of the initial and final angular momentum of the system are small $( \sim 1 \degree)$ for the cases discussed here, thus, the choice of one or the other does not modify the subsequent analysis.

\subsection{Match calculation}
\label{sec:introMatch}
The GW signal of a quasicircular  binary black hole system with arbitrary spins is described by 15 parameters \cite{PhysRevD.79.104023}. Some of these parameters are properties intrinsic to the \GW emitting source: the total mass and mass ratio of the binary as well as the six components of the two spin vectors. The remaining parameters are extrinsic and describe the relation between the binary source frame and the observer; they are: the luminosity distance $d_L$, the coalescence time $t_c$, the inclination $\iota$ , the azimuthal angle $\varphi$, the right ascension $\phi$, declination $\theta$ and polarization angle $\psi$.

The real-valued GW strain observed in a detector is given by~\cite{PhysRevD.47.2198}
\begin{align}
h_{resp}(t;\zeta,\Theta) = & F_+ (\theta,\phi,\psi)h_+(t-t_c, d_L, \iota,\varphi, \zeta)  \nonumber \\
&  +  F_\times (\theta,\phi,\psi)h_\times (t-t_c,d_L,\iota,\varphi, \zeta)  ,
\label{eq1}
\end{align}
where $\Theta=\{ t_c, d_L, \theta, \varphi,\alpha, \delta, \psi  \}$ and $\zeta=\{M, q, {\bm S}_1, {\bm S}_2\}$  are the set of extrinsic and intrinsic parameters, respectively. The two waveform polarisations $h_+,h_\times$ are defined as
\begin{equation}
h(t)=h_+ - i h_\times = \sum_{l=2}^{\infty}\sum_{m=-l}^{l} {}^{-2}Y_{\ell m}(\iota, \varphi) h_{\ell m} (t-t_c;\zeta),
\label{eq2}
\end{equation}
where ${}^{-2}Y_{\ell m}$ denotes the spin-weighted spherical harmonics of spin weight $-2$.

The comparison between two waveforms is commonly quantified by the match -- the noise-weighted inner product between the signals \cite{Jaranowski2012}. Given a real-valued detector response, the inner product between the signal $h_{resp}^S(t)$ and a model $h_{resp}^M(t)$, is defined as
\begin{equation}
	\braket{h^S_{resp}|h^M_{resp}}= 2 \int^\infty_{-\infty} \frac{\tilde{h}^S_{resp}(f) \tilde{h}^{M*}_{resp}(f)}{S_n(|f|)}df,
\label{eq3}
\end{equation}
where $\tilde{h}$ denotes the Fourier transform of $h$, $h^*$ the complex conjugate of $h$ and $S_n(|f|)$ is the one-sided noise power spectral density (PSD) of the detector.

In order to reduce the dimensionality of the parameter space we can combine declination, right ascension and polarization angle $(\theta,\phi,\psi)$ into an effective polarization angle $\kappa$ defined as \cite{PhysRevD.89.102003}
\begin{equation}
	\kappa (\theta,\phi,\psi) := \arctan \left( \frac{F_\times}{F_+} \right),  \quad  \mathcal{A}(\theta,\phi)= \sqrt{F^2_\times+ F^2_+}.
\label{eq3.1}
\end{equation}
The detector response can then be rewritten in terms of the effective polarization angle $\kappa$ as 
\begin{equation}
	h_{resp}(t)=\frac{\mathcal{A}}{d_L}\left[h_+(t) \cos\kappa  + h_\times (t) \sin\kappa\right].
\label{eq4}
\end{equation}
The normalized match is then defined as the inner product optimized over a relative time shift, the initial orbital phase and the polarization angle given by
\begin{equation}
\mathcal{M} = \max_{t^M_0, \varphi^M_0, \kappa^M} 
\left[ 	\frac{\braket{h^S_{resp}|h^M_{resp}}}{\sqrt{\braket{h^S_{resp}|h^S_{resp}}\braket{h^M_{resp}|h^M_{resp}}}} \right],
\label{eq5}
\end{equation}
where the values of the signal angles $(\iota^S,\varphi^S_0,\kappa^S)$ are fixed. The procedure to compute the match is described in detail in App. B of \cite{PhysRevD.91.024043}. A match $\mathcal{M} \simeq 1$ indicates good agreement between the signal and the model, while $\mathcal{M} \simeq 0$ indicates orthogonality between the two waveforms. 

We perform an analytical maximization over $\kappa^M$ and compute numerically the maximum for $t^M_0$ and $\varphi^M_0$ through an inverse Fourier transform and numerical maximization. To ease the comparisons we introduce the mismatch,	$1- \mathcal{M}$. 

\subsection{Radiated energy}
\label{sec:introEnergy}

In addition to the commonly used mismatch calculation to quantify the disagreement between two waveforms, we also compute the radiated energy per $(\ell m)$-mode,
\begin{align}
E_{\ell m}=\frac{1}{16 \pi} \int^{t_f}_{t_0}d\tau \left|\dot{h}_{\ell m}(\tau)\right|^2,
\label{eq999}
\end{align}
where $t_0$ is the relaxed time after the burst of junk of radiation, $t_f$ is the final time of the simulation, $\dot{h}_{\ell m}(\tau) \equiv dh_{\ell m}(\tau)/d\tau$. This quantity is more sensitive to discrepancies in the amplitude of the waveforms than the mismatch, which is more sensitive to phase differences. We will use this measure in particular as a diagnostic tool to quantify mode asymmetries. Note that given the fact that we have set set the scale of the total mass to $1$, the radiated energy scales consistently with that choice.

\section{Testing the accuracy of modelling approximations}
\label{sec:systematic}
We quantify and discuss the impact of the two approximations (APX1) and (APX2) used in the phenomenological framework to model precessing binaries including higher order modes. The higher order modes analyzed in this paper are $(\ell, m)=\{(2, |2|),(2, |1|), (3, |2|), (3, |3|), (4, |3|), (4, |4|)\}$. These modes can be grouped in three subsets: the $\ell=2$ modes, where at least for the $(2,|2|)$ we expect high accuracy of the approximations, the $(3,|2|),(4,|3|)$ modes for which poor accuracy is expected due to the significant mode-mixing \cite{PhysRevD.90.064012} which the approximations are not able to reproduce, and the $(4,|4|)$ and $(3,|3|)$ modes as the next dominant higher order modes.

The analyses are carried out in two different coordinate frames, the non-inertial co-precessing frame and the inertial precessing frame. We discuss the interpretation of the results in both frames and show the suitability of one or another to assess the accuracy of the approximations.

\subsection{Co-precessing waveforms: QA vs. AS} 
\label{sec:QAvsSA}
We first study the validity of the identification of AS and co-precessing waveforms (APX1), where the latter are constructed via the QA method described in Sec.~\ref{sec:introQA }. For this comparison we use all available waveform modes of each simulation in order to generate the QA modes, i.e., we take all terms in the sum of Eq. \eqref{eq00}. For instance, we generate the QA (2,2) and (3,3) modes as follows:
\begin{equation}
\begin{split}
h^{\rm{QA}}_{22}(t) & = \mathbf{R}_{222} h^{\rm{prec}}_{22}(t)+\mathbf{R}_{22,-2} h^{\rm{prec}}_{2,-2}(t)+\mathbf{R}_{22,0} h^{\rm{prec}}_{2,0}(t) \\
& + \mathbf{R}_{221} h^{\rm{prec}}_{21}(t)+\mathbf{R}_{22,-1}  h^{\rm{prec}}_{2,-1}(t).\\
\end{split}
\label{eq51}
\end{equation}
\begin{equation}
\begin{split}
h^{\rm{QA}}_{33}(t) & = \mathbf{R}_{333} h^{\rm{prec}}_{33}(t)+\mathbf{R}_{33,-3} h^{\rm{prec}}_{3,-3}(t)+\mathbf{R}_{33,0} h^{\rm{prec}}_{3,0}(t) \\
& + \mathbf{R}_{332} h^{\rm{prec}}_{32}(t)+\mathbf{R}_{32,-2}  h^{\rm{prec}}_{3,-2}(t)\\
& + \mathbf{R}_{331} h^{\rm{prec}}_{31}(t)+\mathbf{R}_{33,-1}  h^{\rm{prec}}_{3,-1}(t).
\end{split}
\label{eq52}
\end{equation}

\begin{figure}[hbt!]
\centering
\includegraphics[scale=0.42]{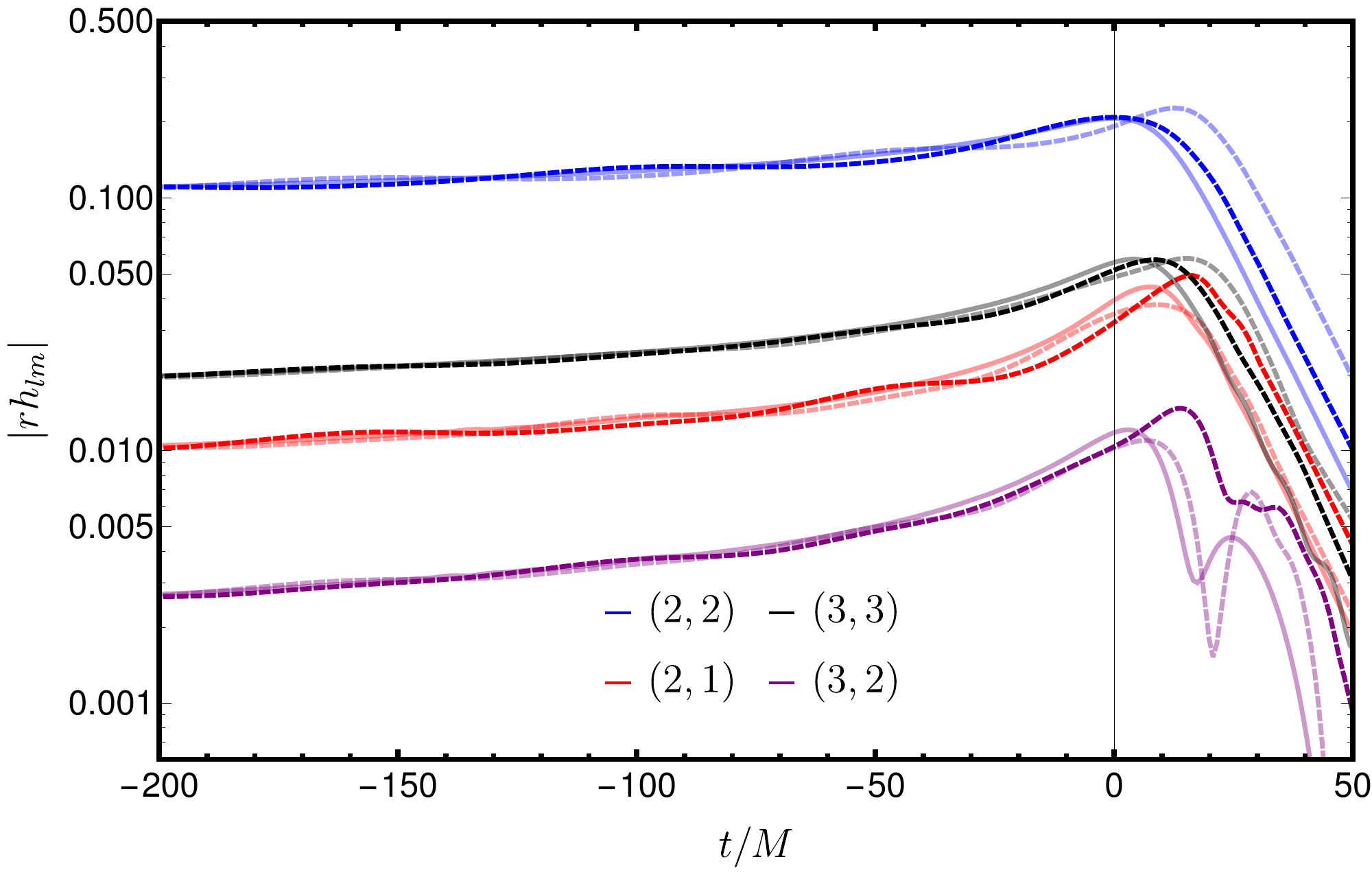}
\caption{Time-domain amplitude of the strain for the $\{(2,\pm 2),(2,\pm 1),(3,\pm 3),(3,\pm 2))\}$ modes. The solid lines with low opacity represent the amplitude of the AS $(\ell, m)$ mode, while the dashed lines with high (low) opacity represent the QA $(\ell, m)$ $((\ell,-m))$ modes for the configuration with ID 36 from Table \ref{tab:tabNR3} of App. \ref{sec:AppendixA}.
}
\label{fig:plotASvsQA}
\end{figure}

The qualitative behavior of (APX1) is illustrated in Fig.~\ref{fig:plotASvsQA}, where we show a selection of higher-order modes in the co-precessing frame with the corresponding AS modes for the configuration with ID \SXSqfiveID{}. We observe the well-known hierarchy between the amplitudes of the AS higher-order modes\cite{PhysRevD.88.024034}, which is also reproduced by the QA modes. Furthermore, we see clear asymmetries between positive and negative m QA modes at merger. 

Note that in Fig. \ref{fig:plotASvsQA} there is not only an amplitude discrepancy but also a time shift between positive and negative m QA modes. This is due to the fact that for the strain, which is two time derivatives of the Newman-Penrose scalar $\psi_4$, the amplitude discrepancies in the $\psi_{4,\ell m}$ modes translate also into time-shifts in the $h_{\ell m}$ modes.
However, only amplitude asymmetries are relevant for the subsequent analysis as the mismatch calculation maximizes over possible time-shifts between waveforms by performing an inverse Fourier transform. These amplitude asymmetries are not captured by (APX1), and reduce the accuracy of the QA-AS identification, especially for higher order modes, where these effects are exacerbated (see Fig. \ref{fig:plotASvsQA}).

We now quantify the agreement between the QA and AS modes by calculating the mismatch between individual modes optimized over a time shift and phase offset for all pairs of NR simulations in Tab.~\ref{tab:tabNR3}. 
The integral in Eq. \eqref{eq3} is evaluated between $20$ Hz and a maximum frequency below $2000$ Hz which varies depending on the total mass of the system and the length of the NR waveform. We use the Advanced LIGO design sensitivity PSD \cite{LIGOPSD,TheLIGOScientific:2014jea}.

\begin{figure}[!]
\centering
\includegraphics[scale=0.35]{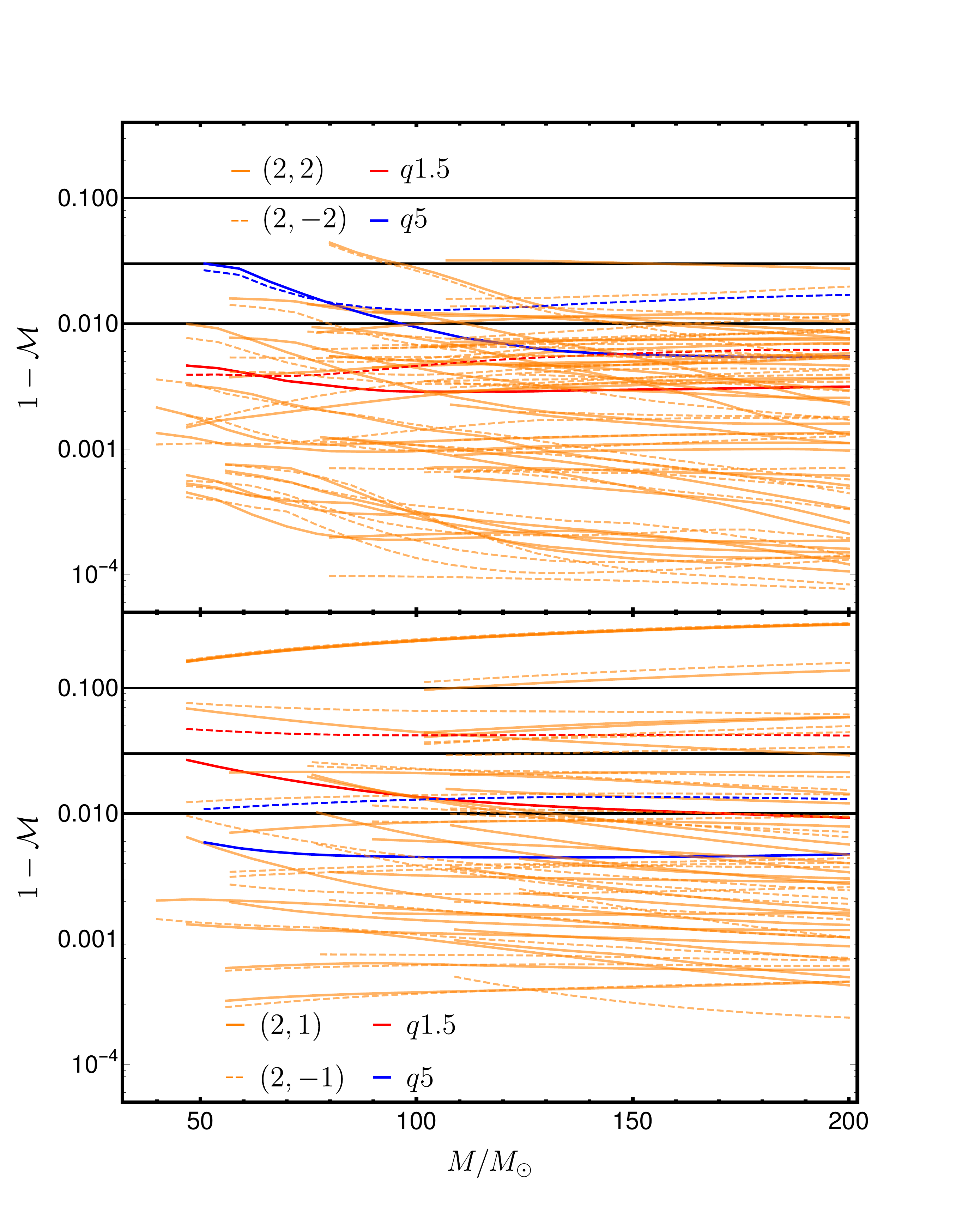}
\vspace*{-0.8cm}
\caption{Mode by mode mismatches between QA and AS modes for all NR configurations in Table \ref{tab:tabNR3} as a function of the total mass of the system. Top: Results for the $(2,\pm 2)$; Bottom: Result for the $(2, \pm 1)$ modes. The configurations with IDs \SXSqonepointfiveID{} (red) and \SXSqfiveID{} (blue) are highlighted. The solid and dashed curves correspond to positive and negative $m$-modes, respectively. The horizontal lines mark mismatches of $1\%$, $3 \%$ and $10\%$. Configurations with PI symmetry (IDs 1, 2, 3 and 4) have been removed from the bottom panel. 
}
\label{fig:mmQA}
\end{figure}

Figure~\ref{fig:mmQA} shows the mismatch between single QA modes and AS ones for the $\{\ell, m\}=\{(2,\pm 2)$ (top panel) and $\{\ell, m\}=(2, \pm 1)\}$ modes (bottom panel) as a function of the total mass compatible with the length of the NR waveforms. 
The results for the other modes can be found in Fig. \ref{fig:mmQAHMs} in App. \ref{sec:AppendixB}. In each panel of Figures \ref{fig:mmQA} and \ref{fig:mmQAHMs} we mark with horizontal lines the  $1\%$, $3 \%$ and $10 \%$ values of the mismatch.
Moreover, we highlight two cases with IDs \SXSqonepointfiveID{} (red) and \SXSqfiveID{} (blue): ID \SXSqonepointfiveID{} is selected as a representative of the bulk of available NR waveforms, with a small mass ratio, $q=1.5$, and moderate precession spin, $\chi_p=0.5$, while the case ID \SXSqfiveID{} has the highest mass ratio in our data set of NR waveforms, $q=5$.

For the $(2,\pm 2)$-modes, top panel in Fig. \ref{fig:mmQA}, we observe mismatches well below $3\%$, except for the $(2,2)$ mode of the case with ID 28. This configuration has a moderate mass ratio, $q=3$, and high values of the in-plane spin components $\chi_{1\perp}=0.8$, $\chi_{2\perp}=0.6$ on both BHs. A closer look (see Fig.~\ref{fig:plotq3} in App.~\ref{sec:AppendixB}) reveals that while the $(2,2)$-QA mode resembles the AS $(2,2)$-mode reasonably well during the inspiral, at merger the amplitude of the QA-mode is significantly higher than for the AS-mode. Additionally, we identify a clear asymmetry between the $(2,2)$ and $(2,-2)$ QA modes.
In order to quantify this asymmetry between positive and negative $m$-modes, as well as the difference between AS and QA modes, we also compute the radiated energy per $(\ell, m)$-mode for this case. The amount of energy radiated per $m$-mode is given in Tab.~\ref{tab:tabErad}. We also calculate the ratio of radiated energy between positive and negative $m$-modes. The large differences in radiated energy between positive and negative $m$-modes translate into great differences in the peak of the waveforms, which is the cause of the high mismatch for this particular case.

\begin{table}[!]
\begin{center}
 \def\arraystretch{1.6 }
\begin{tabular}{c r c c c  c }
\hline  
\hline  
 $(\ell,|m|)$& $ 10^{-3} \left[ E^{AS}_{\ell, m} \right.$ & $E^{QA}_{\ell, m} $& $E^{AS}_{\ell,-m} $ & $\left.E^{QA}_{\ell,-m} \right]$ &$E^{QA}_{\ell, m}/E^{QA}_{\ell,-m} $\\
  \hline
 $(2,2)$ &   8.912  & 13.619  &  8.912  & 8.083 &1.68  \\
 $(2,1)$ &   0.104 & 0.098  &  0.104 &  0.192&0.51 \\
$(3,2)$ &    0.012 & 0.026 & 0.012 & 0.049 &0.53\\
$(3,3)$ &    0.852 & 1.124 & 0.852  &  1.003& 1.21 \\
$(4,3)$ &    0.005 & 0.008  & 0.005 & 0.016 & 0.52\\
$(4,4)$ &   0.164 &  0.201 & 0.164 &  0.195 &1.03\\
 \hline
 \hline
 \end{tabular}
\end{center}
\caption{Radiated energy in the $(2,|2|)$, $(2,|1|)$, $(3,|2|)$, $(3,|3|)$, $(4,|3|)$, $(4,|4|)$ modes of the AS and QA configurations for the case with ID 28 in Table \ref{tab:tabNR3}.}
\label{tab:tabErad}
\end{table}

This picture changes quite significantly for the $(2,\pm 1)$ modes, bottom panel of Fig. \ref{fig:mmQA}: We identify five configurations with a mass averaged mismatch larger than $10\%$: ID 4, which is an equal mass, equal spin binary; IDs 9, 10 and 12, which correspond to a series of $q=1.5$ simulations with the same $\chi_{1L}$ but differently oriented in-plane spin components for the smaller black hole; and ID 20, a $q=2$ simulation. 
For those cases we find that the QA-mode is not represented well by the corresponding AS-mode (see App.~\ref{sec:AppendixC} for details). We note that odd-m modes, in particular the (2,1)-mode, are very sensitive to asymmetries in the binary, which may be reflected in the values of the final recoil of the system~\cite{PhysRevD.77.124047}. Thus, we have computed the recoil velocity for all available simulations in Tab. \ref{tab:tabNR3}. However, we do not find a direct correlation between the recoil velocity and large mismatches. We observe that some configurations with mass ratios 1.5 and 2, the same $\chi_{\text{eff}}$ but different in-plane spin components have mismatches $<3 \%$, while others have mismatches $\geq 10 \%$.

Furthermore, there are also four cases with mass averaged mismatches $\langle 1- \mathcal{M} \rangle$ between 5\% and 10\%, corresponding to the simulations with IDs 1, 15, 16 and 21. Simulation with ID 1 is an equal mass equal, spin configuration with PI symmetry, hence, with mathematically vanishing odd m modes, while IDs 15, 16 and 21 are $q=2$ simulations with $\chi_{\text{eff}}=0$ and $\chi_{\rm eff}=-0.33$ , respectively, and small AS (2,1) modes. Further discussion can be found in App.~\ref{sec:AppendixC}. For the remaining simulations, i.e., $75 \%$ of the NR data set, we find mass averaged mismatches $\langle 1- \mathcal{M} \rangle \leq 3\%$.

We have also investigated the QA-AS correspondence for other higher-order modes. Overall, we find that the number of cases with $\langle 1-\mathcal{M} \rangle \leq 3\%$ is significantly smaller than for the quadrupolar modes. This indicates a clear degradation of (APX1) for higher order modes. We identify the QA mode asymmetries as well as strongly pronounced residual oscillations due to nutation as the cause. 
See Fig. \ref{fig:mmQAHMs} in App. \ref{sec:AppendixB} for the details.
We further remark that the $(3, |2|)$ and $(4, |3|)$ modes are affected strongly by mode mixing, which requires a transformation to a spheroidal harmonic basis. In addition, all higher order modes suffer from more numerical noises in comparison to the quadrupolar mode, which necessarily impacts the mismatch. Possible ways to address such limitations are discussed in Sec. \ref{sec:WaveformDecomp}.

\subsection{Approximate precessing waveforms: Impact of higher-order modes}
\label{sec:PrecvsP}
We are now turning our attention to (APX2), analyzing the impact of the number of AS higher order modes used in the construction of approximate precessing waveforms in the inertial frame.
To do so, we use the inverse QA-transformation. In contrast to the previous section, where all available higher-order modes were taken into account (see Eqs. \eqref{eq51} and \eqref{eq52}), in this section we restrict the number of available AS modes in the sum of Eq. \eqref{eq01} to the same set of modes used in current Phenom/EOB waveform models~\cite{london2018,PhysRevD.98.084028}: $(\ell,m)=\{(2,\pm 2),(2,\pm 1),(3,\pm 2),(3,\pm 3),(4,\pm 3),(4,\pm 4) \}$. The impact of these higher order modes in the map between the co-precessing and the inertial frame is assessed via truncating different terms in the sum. For instance, in the case of the approximate precessing $(2,2)$ mode, we calculate it taking into account either only the AS $(2,\pm 2)$ modes or the AS $(2,\pm 2),(2,\pm 1)$ modes, i.e.,
\begin{equation}
h^{\rm{P,k:\{\pm 2\}}}_{22}(t) = \mathbf{R}^{-1}_{222}   h^{\rm{AS}}_{22}(t)+\mathbf{R}^{-1}_{22,-2}   h^{\rm{AS}}_{2,-2}(t),
\label{eq07}
\end{equation}
\begin{equation}
\begin{split}
h^{\rm{P,k:\{\pm 2, \pm 1\}}}_{22}(t) & = \mathbf{R}^{-1}_{222}   h^{\rm{AS}}_{22}(t)+\mathbf{R}^{-1}_{22,-2}  h^{\rm{AS}}_{2,-2}(t) \\
&+ \mathbf{R}^{-1}_{221}   h^{\rm{AS}}_{21}(t)+\mathbf{R}^{-1}_{22,-1}  h^{\rm{AS}}_{2,-1}(t).\\
\end{split}
\label{eq08}
\end{equation}
The notation $h^{\rm{P,k:\{\pm r,\pm s\}}}_{\ell m}$ refers to the approximate precessing $(\ell, m)$ waveform mode constructed with the AS $(\ell,\pm r)$, $(\ell,\pm s)$ modes.

\begin{figure}[!]
\centering
\vspace*{-1cm}
\hspace*{-0.3cm}
\includegraphics[scale=0.31]{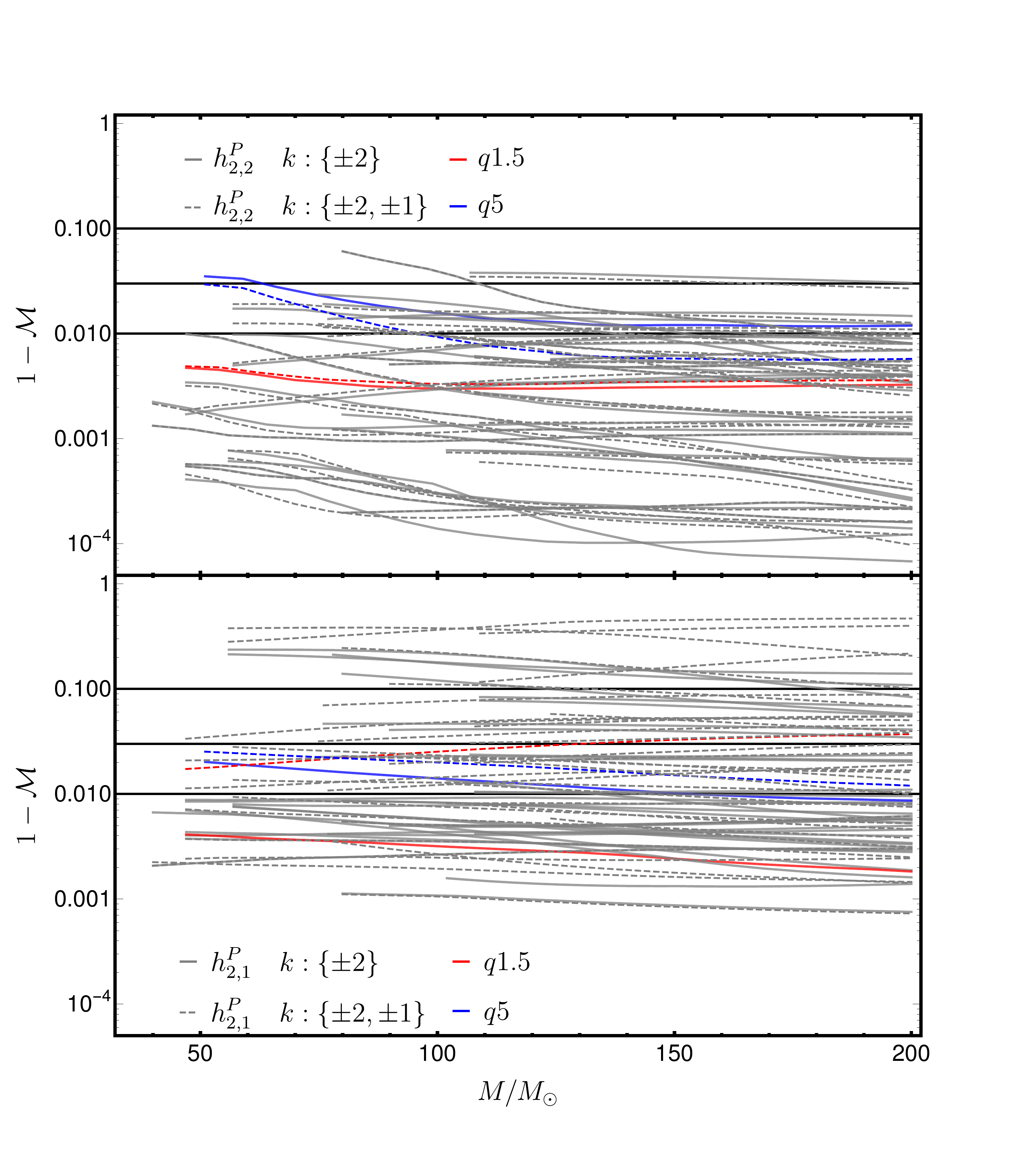}
\caption{Mode by mode mismatches between precessing and approximate precessing modes in the inertial frame for all NR configurations from Table \ref{tab:tabNR3} as a function of the total mass of the system. In the top and bottom panels we show the results for the $(2,2)$ and $(2,1)$ modes, respectively. The thick and dashed lines correspond to taking two, Eq. (\ref{eq07}), and four AS modes, Eq. (\ref{eq08}), to generate the approximate precessing waveforms, respectively. The letter $k$ represents the index of the rotation operator given in Eq. \eqref{eq01}. In addition, configurations with IDs \SXSqonepointfiveID{} and \SXSqfiveID{} are highlighted with red and blue colors, respectively. The horizontal lines mark a mismatch of $1\%$, $3 \%$ and $10 \%$ respectively. 
}
\label{fig:mmPrec}
\end{figure}

The agreement between fully precessing and approximate precessing modes is first quantified via single-mode mismatches following the same procedure as in Sec. \ref{sec:QAvsSA}. The results for the $(2,\pm 2)$ and $(2,\pm 1)$ modes are shown in Fig. \ref{fig:mmPrec}, the results for the other  modes in Fig. \ref{fig:mmPrecHMs}. Solid and dashed lines represent the mismatches calculated with two AS modes, as per Eq. \eqref{eq07}, or with four AS modes, as per Eq. \eqref{eq08}, respectively. The configurations with IDs \SXSqonepointfiveID{} (blue) and \SXSqfiveID{} (red) are again highlighted; horizontal lines indicate mismatches of $1\%$, $3\%$ and $10\%$.

The precessing $(2,2)$-mode mismatches (top panel of Fig. \ref{fig:mmPrec}) are below $3\%$ for all cases except for the case with ID 28, which shows mismatches $>3\%$ for all total masses. This outlier configuration is the same as in Sec. \ref{sec:QAvsSA} when testing (APX1) for the $(2,2)$-mode and it corresponds to a short BAM simulation with $q=3$ and $\chi_p=0.8$, the highest value in our NR data set. We identify an amplitude asymmetry as the underlying cause (see App. \ref{sec:AppendixC} for details).

In Tab. \ref{tab:tabmmPrec} the percentages of cases with a mass average mismatch within different threshold values are shown. For the $(2,2)$ mode $97.2 \%$ of the cases in our data set have an average mismatch below the $3 \%$. The addition of the AS $(2,\pm 1)$ modes does not change the percentage of simulations with an average mismatch below $3 \%$. This indicates that the inclusion of the AS $(2,\pm 1)$ modes in the construction of the approximate precessing $(2,2)$ mode has little impact, although we generally observe improved mismatches (see top panel of Fig. \ref{fig:mmPrecHMs}).

\begin{table}[!]
\begin{center}
 \def\arraystretch{1.2}
\begin{tabular}{c l c c c }
\hline  
\hline  
P Mode & AS Modes & $N_{3 \%  \leq \langle 1- \mathcal{M}  \rangle } $  & $N_{3 \%  \leq \langle 1- \mathcal{M} \rangle \leq 10 \%}$  &   $N_{\langle 1- \mathcal{M} \rangle \geq 10 \%}$  \\
 \hline
\multirow{2}{*}{$(2,2$)} &   $(2,\pm 2)$       &      $97.2  \%$   &    $2.8  \%$    & $0 \%$  \\
                         &$(2,\pm 2),(2,\pm 1)$  &      $97.2  \%$   &    $2.8  \%$    & $0 \%$  \\
 \hline 
\multirow{2}{*}{$(2,1$)} &   $(2,\pm 2)$         &      $77.8\%$   &    $13.9 \%$    & $ 8.3 \%$    \\
                         &$(2,\pm 2),(2,\pm 1)$  &     $63.9  \%$   &    $22.2 \%$    & $13.9\%$     \\
  \hline 
\multirow{2}{*}{$(3,2$)} &   $(3,\pm 3)$         &   $8.3  \%$   &    $13.9  \%$    & $77.8 \%$   \\
                         &$(3,\pm 3),(3,\pm 2)$  &     $25. \%$   &    $33.3 \%$    & $41.7 \%$   \\
 \hline 
\multirow{2}{*}{$(3,3$)} &   $(3,\pm 3)$         &     $86.1  \%$   &    $5.6  \%$    & $8.3 \%$    \\
\cline{2-5}
                         &$(3,\pm 3),(3,\pm 2)$  &    $88.9 \%$   &    $8.3 \%$    & $2.8 \%$  \\
 \hline 
\multirow{2}{*}{$(4,3$)} &   $(4,\pm 4)$         &    $27.8  \%$   &    $33.3  \%$    & $38.9 \%$    \\
\cline{2-5}
                         &$(4,\pm 4),(4,\pm 3)$  &       $27.8  \%$   &    $30.6 \%$    & $41.7 \%$   \\
\hline 
\multirow{2}{*}{$(4,4$)} &   $(4,\pm 4)$         &      $83.3  \%$   &    $16.7 \%$    & $0 \%$   \\
                         &$(4,\pm 4),(4,\pm 3)$  &       $83.3  \%$   &    $16.7  \%$    & $0 \%$   \\
 \hline
 \hline
 \end{tabular}
\end{center}
\caption{Distribution of single mode mismatches shown in Figs. \ref{fig:mmPrec} and \ref{fig:mmPrecHMs}. The notation  $N_{X \%  \leq \langle 1- \mathcal{M}  \rangle \leq Y} $ refers to the percentage of cases in the NR data set, with a mismatch averaged over the mass range between the $X \%$ and $Y \%$. The first column indicates the precessing $(\ell,m)-$mode for which the mismatches are calculated; the second column the AS modes used to constructed the approximate precessing mode; the remaining columns give the percentage of cases in our NR data set with an average mismatch $\leq 3 \%$, between $3 \%$ and $10 \%$, and $\geq 10 \%$ , respectively.
}
\label{tab:tabmmPrec}
\end{table}

The bottom panel of Fig. \ref{fig:mmPrec} shows the results for the precessing $(2,1)$ mode; we see a higher number of cases with mismatches above $3 \%$ than for the $(2,2)$ mode.   
In particular, we find that the inclusion of the AS $(2,\pm 1)$ decreases the total percentage of simulations with mismatched below $3 \%$ as shown in Table \ref{tab:tabmmPrec}, see e.g. the red and blue curves in the bottom panel of Fig. \ref{fig:mmPrec}. We attribute this decrease to the less accurate identification between the QA and AS $(2,1)$ mode. For the configuration ID~\SXSqonepointfiveID{} we see in the right panel of Fig. \ref{fig:plotq3} that the approximate precessing $(2,1)$-mode constructed with four AS modes, although it reproduces more accurately the shape of the precessing mode during the inspiral, it has larger errors at merger than the one built with only two AS modes. This error at merger dominates the value of the mismatch and it also indicates the inability of the approximation to accurately reproduce the merger part of the precessing $(2,1)$-mode for this case. Further, high mismatches for the $(2, \pm 1)$ modes are also obtained for configurations for which the $(2,|1|)$-modes have a particularly small amplitude. This poses a challenge for NR codes to resolve such small signals. We discuss possible systematics for the AS $(2,\pm 1)$ mode in Sec. \ref{sec:NRerrors}.

The mismatches for the remaining higher order modes  $\{(3, 3),(3, 2),(4, 4),(4, 3)\}$ are shown in Fig. \ref{fig:mmPrecHMs} of App. \ref{sec:AppendixB} and summarized in Table \ref{tab:tabmmPrec}. We observe a clear difference between the higher order modes affected by mode-mixing, $(3,2)$ and $(4,3)$ modes, which show poor mismatches with less than $30 \%$ of cases below the $3\%$ mismatch; the next dominant higher order modes, $(3,3)$ and $(4,4)$ modes, which are not affected significantly by mode-mixing and have  $80 \%$ of configurations with mass-averaged mismatches below the $3 \%$ mismatch. Note that the mismatches of the $(3,2)$ and $(4,3)$ modes are higher in the inertial frame than in the co-precessing, indicating that the effects of mode-mixing become more relevant in the former due to the more complicated structure caused by the precession of the orbital plane of the binary. Generally, the addition of the AS $(3,\pm 2)$ or the $(4, \pm 3)$  modes tends to improve the mismatches. However, for a non-negligible subset of configurations their inclusion increases the mismatch, see e.g. the blue and red curves in the left panels of Fig. \ref{fig:mmPrecHMs}. This indicates the necessity to disentangle the effects of the two sources of mode-mixing in the approximate precessing waveforms, the one coming from using different AS modes in the map between the co-precessing and inertial frame, and the one from the contribution of approximate precessing higher order modes with the same m index. One possible approach to that problem would be to study the map between inertial and co-precessing frames with the spheroidal harmonic basis for the ringdown part of the waveform for these modes, which we leave for future work. 

We also compute mismatches for negative $m$ modes in Fig. \ref{fig:mmNegPrec}. Computing the average mismatch for each configuration, we find similar results to the positive $m$-modes. 


The analysis of the single mode mismatches indicates that the inclusion of more AS higher order modes can lower the mismatch between the precessing and approximate precessing waveforms quite significantly. Therefore, it is not necessarily optimal to include an arbitrary number of AS modes when constructing approximate precessing waveform modes. 

\begin{figure*}[!]

\noindent\begin{minipage}{\textwidth}

\noindent\begin{minipage}[h]{.45\textwidth}
\includegraphics[scale=0.35]{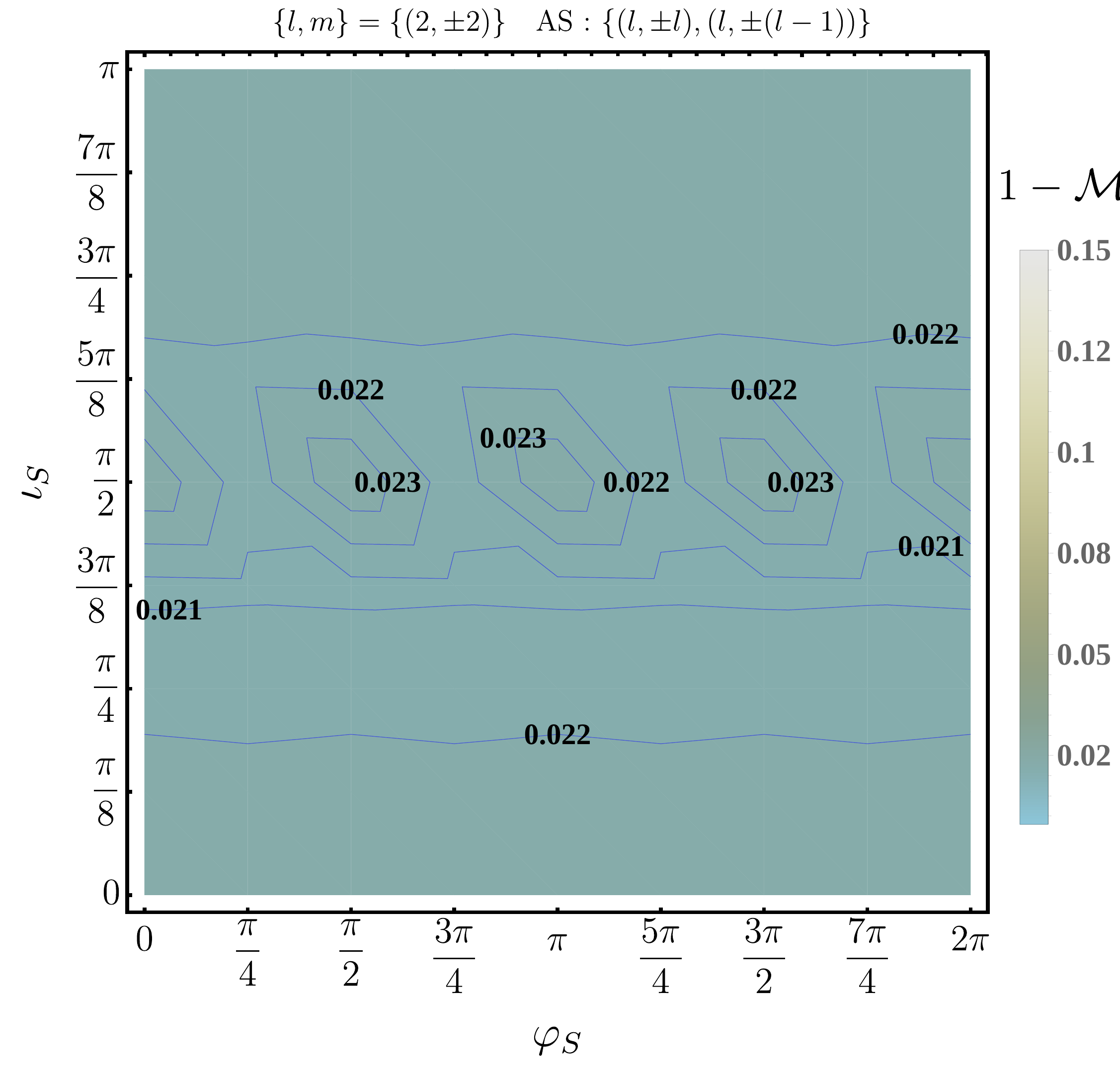}
\end{minipage} 
\hspace{1cm}
\begin{minipage}[h]{.45\textwidth}
\includegraphics[scale=0.35]{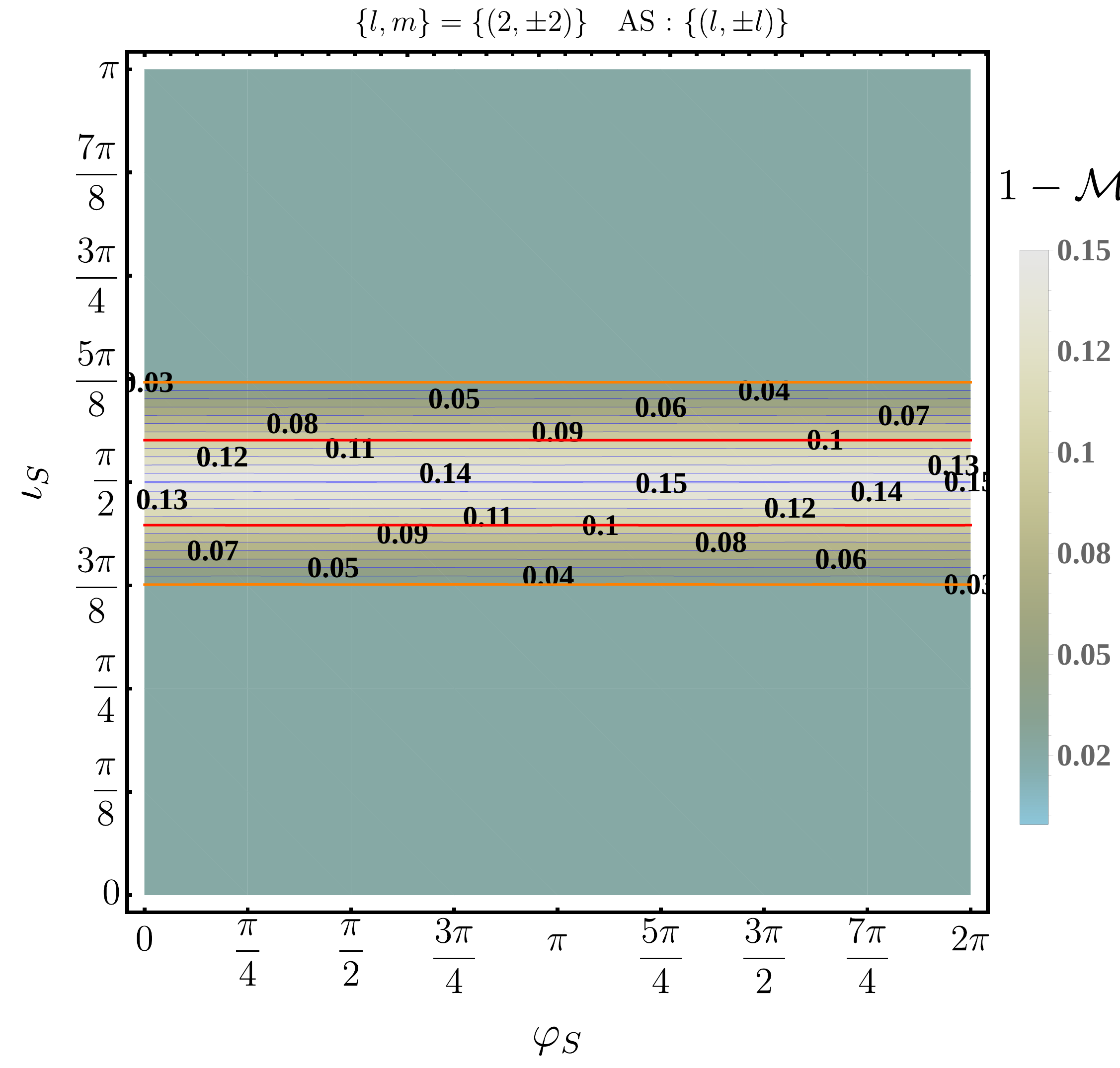}
\end{minipage}

\end{minipage}


\noindent\begin{minipage}{\textwidth}

\noindent\begin{minipage}[h]{.45\textwidth}
\includegraphics[scale=0.35]{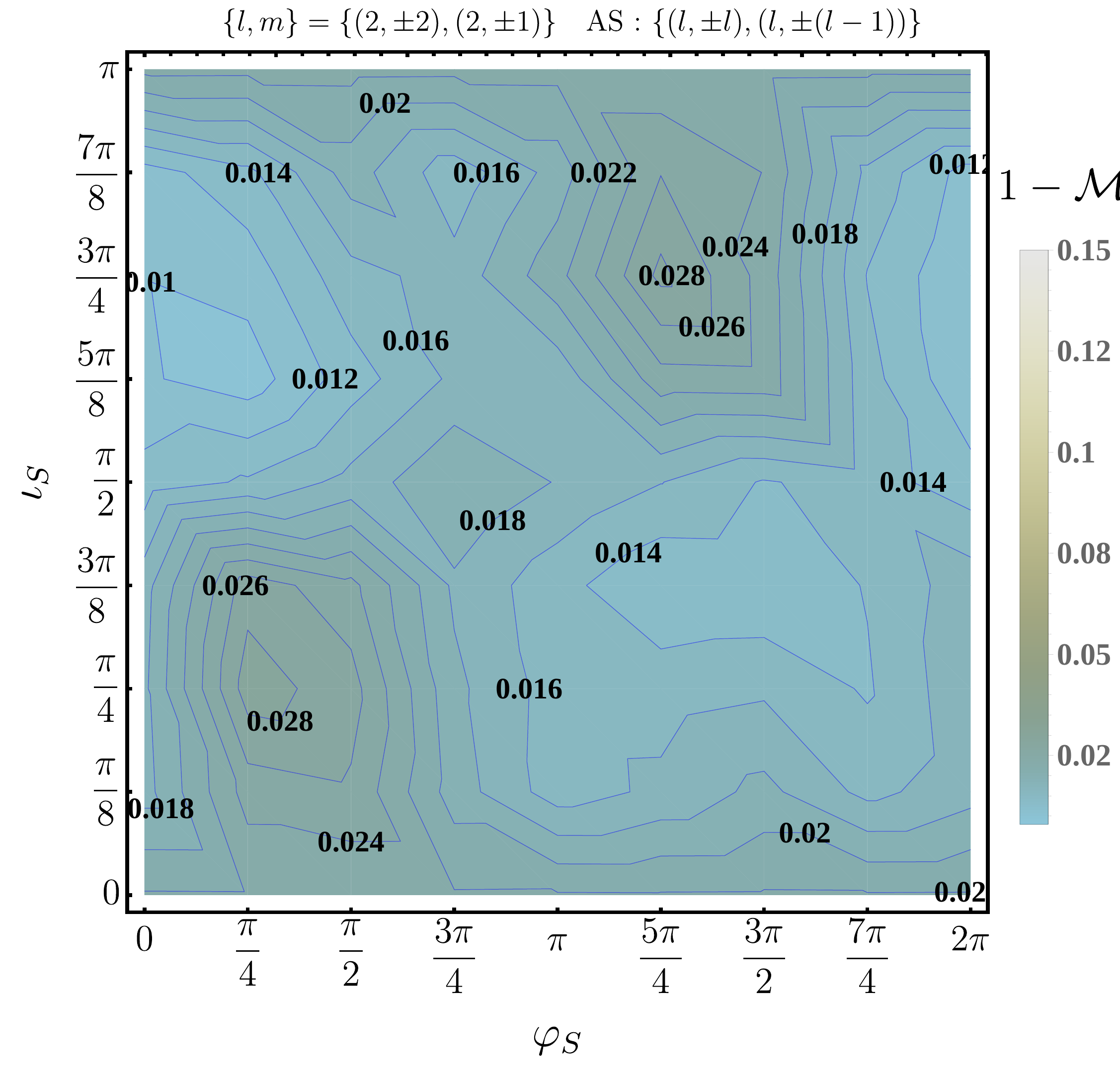}
\end{minipage} 
\hspace{1cm}
\begin{minipage}[h]{.45\textwidth}
\includegraphics[scale=0.35]{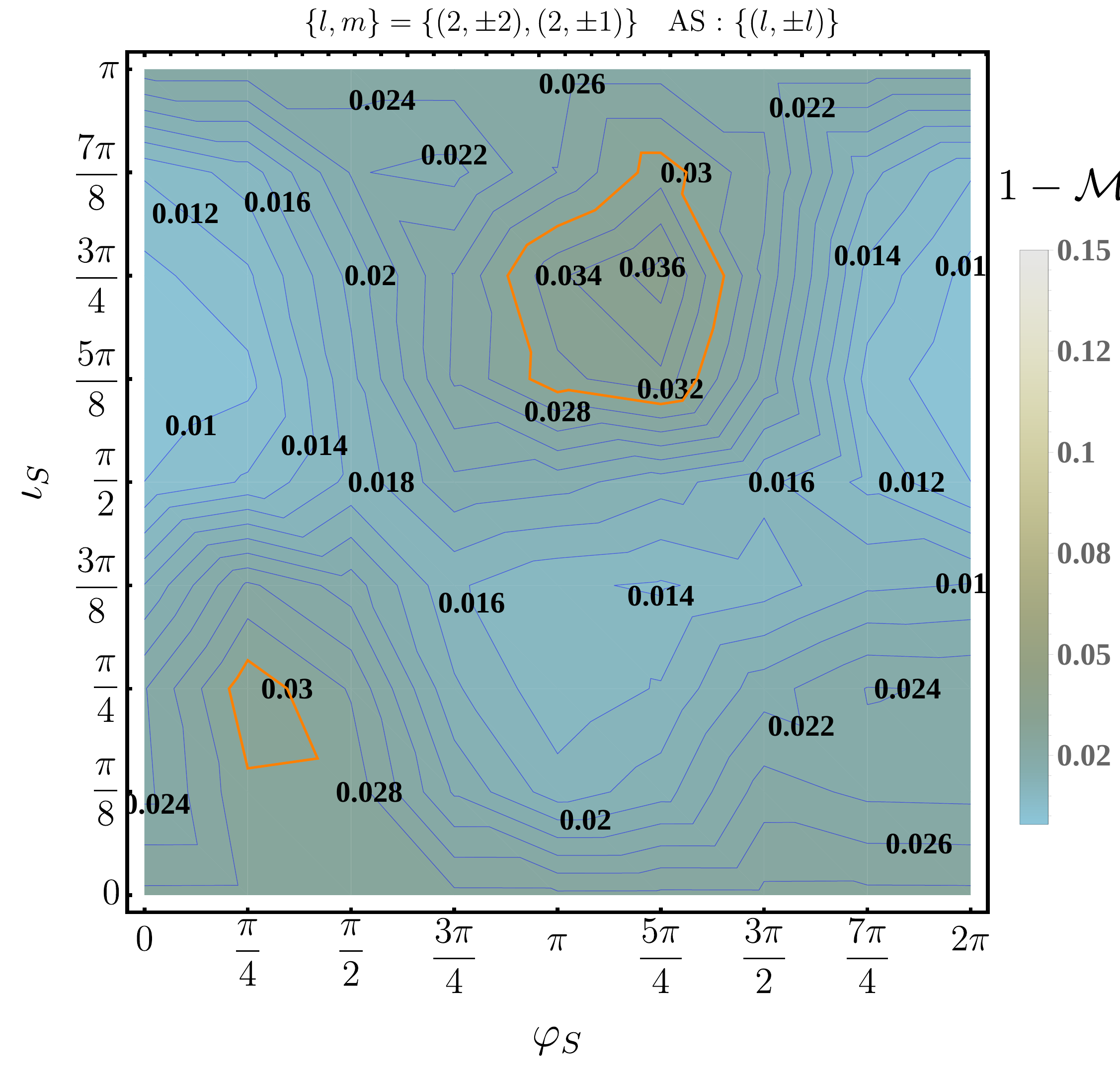}
\end{minipage}

\end{minipage}

\noindent\begin{minipage}{\textwidth}

\noindent\begin{minipage}[h]{.45\textwidth}
\includegraphics[scale=0.35]{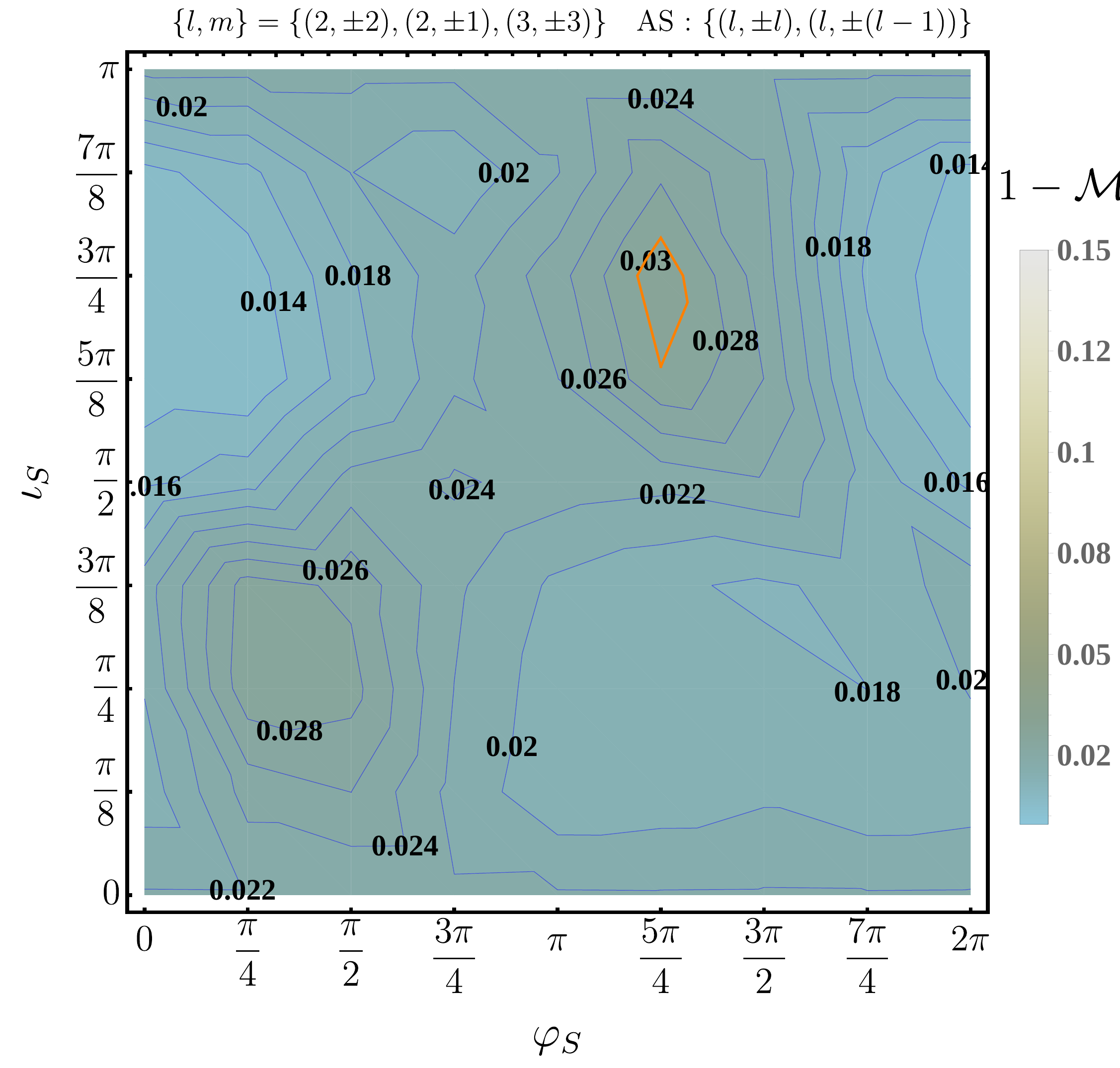}
\end{minipage} 
\hspace{1cm}
\begin{minipage}[h]{.45\textwidth}
\includegraphics[scale=0.35]{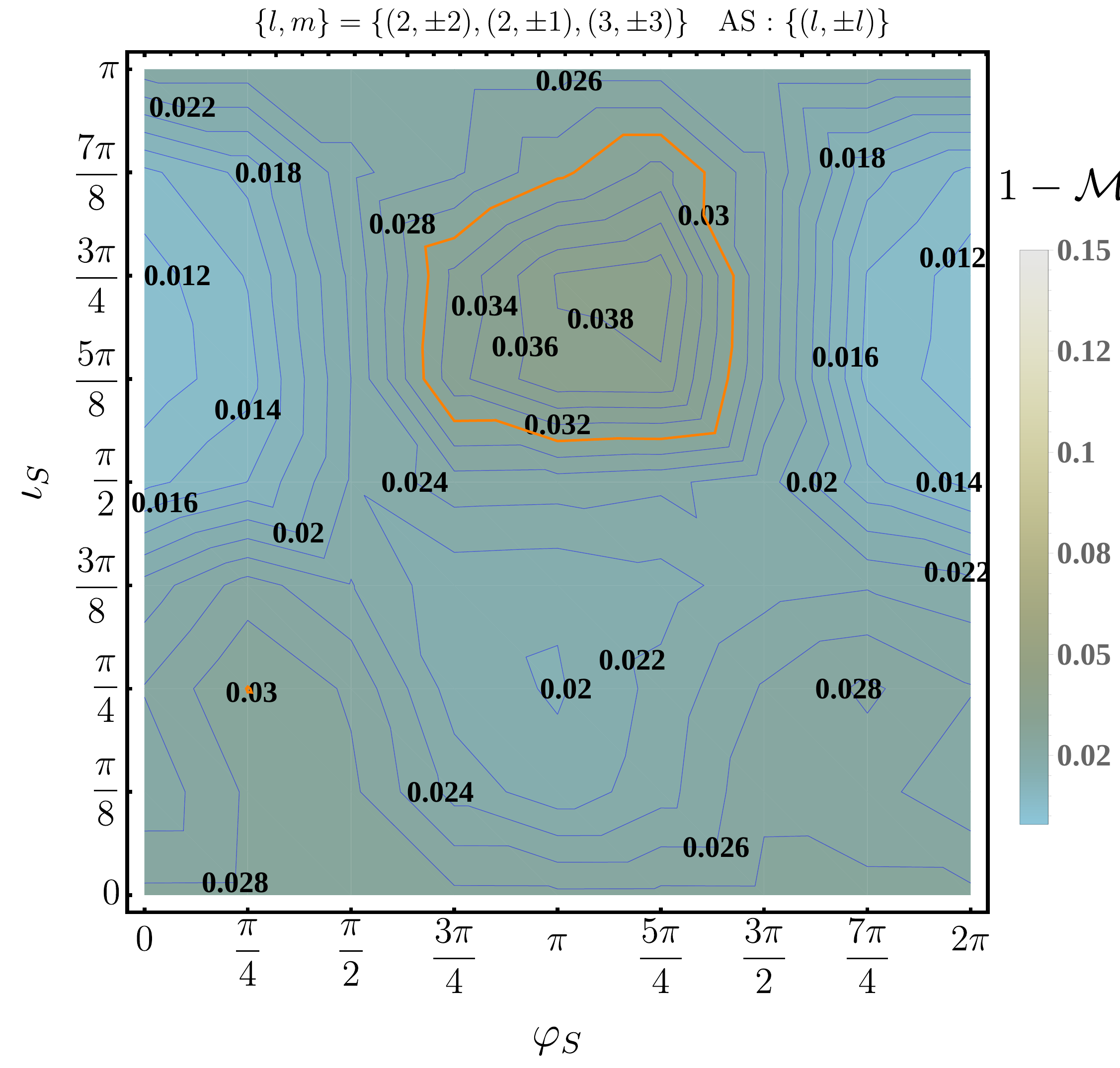}
\end{minipage}
\end{minipage}
\caption{Strain mismatch between precessing and approximate precessing waveforms  in the inertial frame averaged over the angle $\kappa_S$ for a total mass of 65 $M_\odot$ for the configuration  with  ID \SXSqfiveID{}  as a function of the inclination and the azimuthal angle of the signal (precessing waveform). In the plot labels $\{\ell,m\}$ denotes the modes used in the sum of the complex strain given in Eq. \eqref{eq2}, while AS represent the aligned-spin modes taken into account in Eq. \eqref{eq01}. In addition, the  $3 \%$ and $10 \%$  mismatch values are highlighted with orange and red curvess, respectively. }
\label{fig:strainPrec}
\end{figure*}

However, this analysis concerns only the individual modes, thus neglects the geometric coefficients which reweight the modes depending on the orientation of the source. Therefore, we now take this into account and compute mismatches between the detector response (Eq. \eqref{eq4}) constructed from the precessing NR modes and the approximate precessing modes calculated with either two or four AS modes as per Eqs. \eqref{eq07} and \eqref{eq08}. When computing the mismatches for the detector response we optimize over time shifts and phase offsets as in the case of the single mode mismatches, but we also optimize analytically over the effective polarization angle of the template, $\kappa_t$, following the procedure described in \cite{PhysRevD.91.024043}. The mismatches are calculated using the same number of $(\ell,m)$ modes in the signal and the template. For instance, when using only the approximate precessing $(2,\pm 2)$ modes in the complex strain,
\begin{equation}
h^{P, k:\{\pm 2\}}(t)=Y^{-2}_{22}(\iota, \varphi ) h^{P, k:\{\pm 2\}}_{22} +Y^{-2}_{2,-2}(\iota, \varphi ) h^{P, k:\{\pm 2\}}_{2,-2} ,
\label{eq09}
\end{equation}
and the AS $(2,\pm 2)$ modes in the rotation operator as in Eq. \eqref{eq07}, we use the label $(\ell,m)=(2,\pm 2) /  \text{AS}:\{(\ell,\pm \ell)\}$. Figures \ref{fig:strainPrec} and \ref{fig:strainPrecHMs} show contour plots of the mismatches between precessing and approximate precessing waveforms averaged over $\kappa_S$ for the configuration with ID \SXSqfiveID{} as a function of inclination and azimuthal angle for a total mass of $M= 65 M_\odot$.

In the top right panel of Fig. \ref{fig:strainPrec} the mismatches for the strain computed with the $(\ell,m)=(2,\pm 2)/  \text{AS}:\{(\ell,\pm \ell)\}$ modes are displayed. The mismatches increase above $3\%$ in a range of inclinations $67.5 \degree <\iota_S <112.5 \degree$. In addition close to $\iota_S = 90 \degree$ (edge-on configuration) the values reach a maximum of $\sim 15 \%$ value. On the left panel, where the AS $(2,\pm 1)$ modes have been included in the calculation of the approximate precessing modes, the maximum value at $\iota_S = 90 \degree$ has decreased to $ \sim 2 \%$. For small inclinations the benefit of adding the AS $(2,\pm 1)$ modes is more moderate. Hence, the inclusion of the AS $(2,\pm 1)$ modes significantly improves the description of the strain constructed with the $(2,\pm 2)$ modes, especially for edge-on configurations.

In the mid panels  the $(2, \pm 1)$ modes are added to the complex strain. The right mid panel, where only the AS $(2, \pm 2)$ modes are taken into account, displays mismatches above the $3 \%$  in small regions around $\iota_S=\{45 \degree, 135 \degree\}$ , while the left mid panel, where the AS $(2, \pm 2), (2, \pm 1)$ modes are taken into account, shows mismatch values below $3 \%$ for all orientations. Therefore, the inclusion of the $(2,\pm 1)$ AS modes reduces the mismatch with respect to the case where only the $(2,\pm 2)$ AS modes are available. This result also indicates that the improvement in the $(2,\pm 2)$ approximate precessing modes, due to the inclusion of the $(2,\pm 1)$ AS modes, is higher than the degradation of the single $(2,\pm 1)$ approximate precessing modes as observed in the bottom panel of Fig. \ref{fig:mmPrec}. The choice of an inertial frame where the $(2,\pm 2)$ modes have more power than the $(2,\pm 1)$ modes alleviates the inaccuracy in the description of the precessing $(2,\pm 1)$ modes. 

In the bottom panels of Fig. \ref{fig:strainPrec} the strain is constructed from the $(2, \pm 2),(2, \pm 1),(3, \pm 3)$ modes. The bottom right panel, which uses the AS $(2,\pm2),(3,\pm 3)$ modes to generate the approximate precessing waveforms, shows higher mismatches than the left panel, which employs the AS $(2, \pm 2),(2, \pm 1),(3, \pm 3),(3, \pm 2)$ modes. The results show an overall increase in the mismatch due to the addition of the $(3,\pm 3)$, $(3,\pm 2)$ modes whose inaccuracy, as shown in the single mode mismatches of Fig. \ref{fig:mmPrecHMs} of App. \ref{sec:AppendixB}, is higher than for the $(2, \pm 2)$, $(2,\pm 1)$ modes.

In Fig. \ref{fig:strainPrecHMs} of App. \ref{sec:AppendixB} strain mismatch contour plots between precessing and approximate precessing waveforms in the inertial frame for the same configuration as in Fig. \ref{fig:strainPrec} with more higher order modes in the sum of the complex strain are shown. In the top, middle and bottom panels the $(2, \pm 2),(2, \pm 1),(3, \pm 3),(4, \pm 4)$; $(2, \pm 2),(2, \pm 1),(3, \pm 3),(3, \pm 2),(4, \pm 4)$ and $(2, \pm 2),(2, \pm 1)$, $(3, \pm 3)$, $(3, \pm 2)$, $(4, \pm 4)$, $(4, \pm 3)$  modes are taken into account in the sum of the complex strain, respectively. In the left and right panels the $(\ell,|\ell|),(\ell,|\ell-1|)$ AS modes are taken into account, respectively. The mismatches tend to increase slightly when adding more higher order modes in the sum of the complex strain, consistent with the single mode mismatches of Figs. \ref{fig:mmPrec} and \ref{fig:mmPrecHMs}, while the inclusion of more AS higher order tends to lower the mismatch, although its effect is restricted due to the small power of the AS higher order modes. Note that the results of Figs. \ref{fig:strainPrec} and \ref{fig:strainPrecHMs}, although different quantities, are consistent with the signal-to-noise ratio weighted mismatches of references \cite{PhysRevD.100.024059,Khan:2019kot}.

We conclude that the inclusion of AS higher order modes in the construction of approximate precessing waveforms tends to decrease the mismatches when the full strain is considered. However, individual modes are not always better described when adding more AS modes due to the inaccuracy of (APX1) for higher order modes, especially those significantly affected by mode-mixing like the $(3,|2|)$ and $(4,|3|)$ modes.  
Furthermore, the analysis showed that the inclusion of AS higher order modes, like the $(2,\pm 1)$, in the precessing strain can reduce the mismatches by up to an order of magnitude. We stress, however, that the addition of even more higher order modes can also have a negative impact, especially when modes, where (APX1) is clearly not applicable, are included.


\section{Caveats and possible improvements}
\label{sec:caveats}

\subsection{Systematic errors and $(2,1)$-amplitude minima} 
\label{sec:NRerrors}
Let us now discuss possible sources of systematic errors in the NR waveforms which can affect our results.

The first systematic error source we consider is the quantity used to calculate the Euler angles that encode the precession of the orbital plane. For the SXS simulations we compute the angles from the gravitational radiation extrapolated to infinite radius \cite{PhysRevD.80.124045},  while for the ET and BAM simulations we use the Newman-Penrose scalar \cite{Bruegmann2008}, $\psi_4$, at the outermost extraction radius available in the simulation. Alternatively, we could also integrate $\psi_4$ twice in time to obtain the strain and calculate the angles from it. However, integrating twice in time can add extra oscillations in the waveforms which can be as large as those coming from the difference between using $\psi_4$ or the strain. Therefore, we restrict to compute the angles from the $\psi_4$ in the case of the BAM and ET simulations. 

Aligned-spin configurations with PI symmetry, i.e. the two black holes can be exchanged under a reflection in the orbital plane, have vanishing odd $m$-modes, which reduce to noise in NR simulations. Naturally, 
this poses a clear limitation to the identification between AS and QA modes. In our data set simulations with IDs 1-4 show this particular feature. From those four we note that the non-spinning configuration ID 1 has a small AS spin amplitude with respect to the precessing counterpart leading to higher mismatches than the spinning configurations IDs 2, 3 and 4. We discuss this point in more detail in App. \ref{sec:AppendixC}.

Another known issue concerns the occurrence of minima in the amplitude of the AS $(2,\pm 1)$, which are not observed in the corresponding QA modes. In order to obtain some insight into these minima, we have reproduced the AS configuration $\texttt{q2.\_0.6\_-0.6\_\_pcD12}$ simulation from Tab. I of Ref.~\cite{PhysRevD.98.084028} with the Einstein Toolkit (ID 37 in our data set).
In addition, we also produced two precessing simulations to test not only the existence of the minimum with a finite difference code, but also to check its relevance for the QA approximation. We summarize the properties of these three simulations in Tab.~\ref{tab:tabNR21}. 

\begin{figure}[!]
\centering
\includegraphics[scale=0.4]{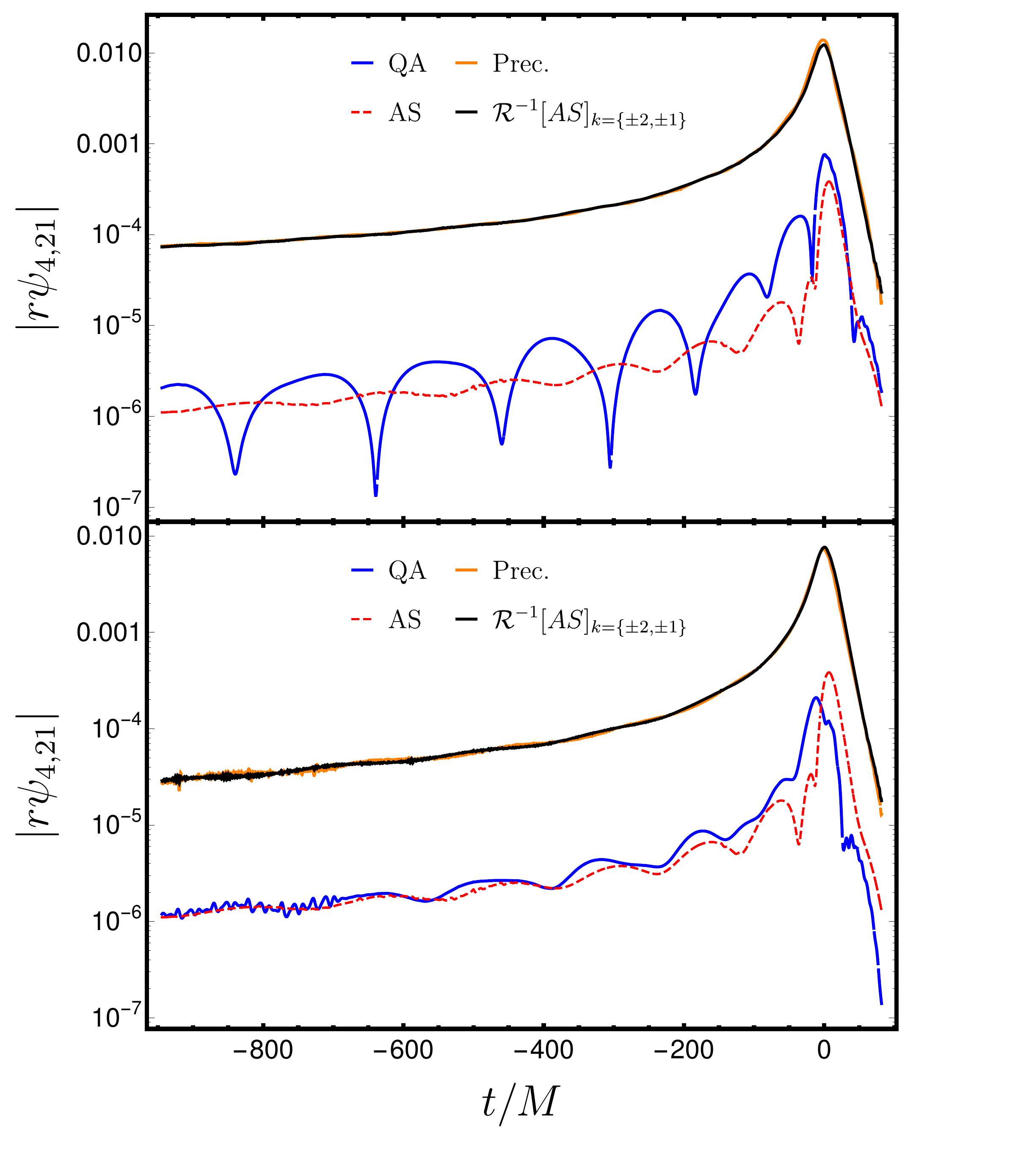}
\vspace*{-0.75cm}
\caption{Time domain amplitude of the $r\psi_4$ for the $(2,1)$ mode of the aligned-spin (AS), quadrupole-aligned (QA), precessing (Prec.) and approximate precessing ($\mathcal{R}^{-1}[AS]_{k=\{\pm 2, \pm 1\}}$) configurations. In the upper plot we compare simulations with ID 37 and 38, and in the bottom panel we compare simulations with ID 37 and 39 of Table \ref{tab:tabNR21}.}
\label{fig:plotNR21mode}
\end{figure}

Figure \ref{fig:plotNR21mode} shows the time domain amplitudes of the $(2,1)$ mode of $\psi_4$ for the three simulations of Tab. \ref{tab:tabNR21}. We clearly identify a minimum in the AS $(2,1)$ mode shortly before $t=0$. The minimum occurs at an orbital frequency of $\Omega^{\text{ET}}_{\text{min}}=0.19$, which is slightly different from the one obtained from the original SpEC simulation, $\Omega^{\text{SpEC}}_{\text{min}}=0.17$. This small difference could be due to differences in the initial data and numerical errors, such as the inaccuracies in the wave extraction or in the double time integration of $\psi_4$ to obtain the strain. Additionally, we also display the approximate precessing $(2,1)$ modes constructed from all available AS modes, and the corresponding QA modes. We see that the QA $(2,1)$ modes do not accurately resemble the AS $(2,1)$ modes. The mismatches between the $(2,1)$-modes are of the order of $15(20)\%$ for  configuration 38 (39), while the mismatches between the $(2,2)$-AS and $(2,2)$-QA modes are $0.2(0.04)\%$, respectively. Furthermore, the precessing $(2,2)[(2,1)]$ modes are faithfully reproduced by the approximate precessing ones with mismatches of $0.2(0.04)[0.2(0.2)]\%$ for simulation with ID 38 (39). The agreement of the precessing modes is due to the negligible contribution of the rather small AS $(2,1)$-mode in comparison to the large AS $(2,2)$ mode, while the poor recovery of the AS mode by the QA mode confirms the inability of the AS-QA identification to reproduce the amplitude minima observed in the AS $(2,1)$-mode.

\begin{table*}[!]
\begin{center}
 \def\arraystretch{1.2 }
\begin{tabular}{ c c c c c c c c c c }
\hline  
\hline  
ID & Simulation &  Code &  q  &$\bm{\chi}_1$ & $\bm{\chi}_2$ & $D/M$ & $M \Omega_0$ & $e_0 \cdot 10^{-3}$ \\
\hline
37 & \texttt{q2.\_0.6\_-0.6\_\_pcD12} & \texttt{ET} & 2& $(0., 0., 0.6)$& $(0., 0., -0.6)$&    11.72 & 0.022 & 1.17  \\
38 & \texttt{q2.\_\_0.4\_0.\_0.6\_\_0.\_0.\_-0.6\_\_pcD12} & \texttt{ET} & 2& $(0.4, 0., 0.6)$& $(0., 0., -0.6)$&    11.68 & 0.022  & 1.47 \\
39 & \texttt{q2.\_\_0.\_0.\_0.6\_\_0.4\_0.\_-0.6\_\_pcD12} & \texttt{ET} & 2& $(0., 0., 0.6)$& $(0.4, 0., -0.6)$&    11.71 & 0.022 & 1.08 \\
\hline
\hline
 \end{tabular}
\end{center}
\caption{Summary of the properties of the simulations used for the analysis of the impact of the $(2,1)$ minimum. Each simulation is specified by its mass ratio $q=m_1/m_2\geq 1 $, the initial dimensionless spin vectors, $\bm{\chi}_1$, $\bm{\chi}_2$, the orbital separation $D/M$, the orbital frequency $\Omega_0$ and the orbital eccentricity, $e_0$, at the relaxation time.}
\label{tab:tabNR21}
\end{table*}

\subsection{Waveform decomposition in the co-precessing frame}
\label{sec:WaveformDecomp}
We have seen previously that the identification between QA and AS modes does not capture mode asymmetries as well as residual modulations due to nutation. This can ultimately lead to a poor reconstructions of the fully precessing GW strain. We now study an extension to (APX1), following the strategy adopted by the precessing surrogate models NRSur4d2s and NRSur7dq2~\cite{PhysRevD.95.104023,PhysRevD.96.024058}, where the time domain co-precessing waveforms are decomposed into slowly-varying functions and small oscillatory ones such that
\begin{align}
A_{\ell,m}^{\pm}(t) & =\frac{1}{2}\left[ A_{\ell,m}(t) \pm A_{\ell,-m}(t) \right], \quad m>0,  \label{eq:eq149} \\
\phi_{\ell,m}^{\pm}(t) & =\frac{1}{2}\left[ \phi_{\ell,m}(t) \pm \phi_{\ell,-m}(t) \right], \quad m>0,  \label{eq:eq150}
\end{align}
where $A_{\ell,m}=|h^{\rm co-prec}_{\ell,m}(t)|$ and $\phi_{\ell,m}=\arg\left(h^{\rm co-prec}_{\ell,m}(t)\right)$. The symmetric amplitude $A^+_{\ell,m}$ and the antisymmetric phase $\phi_{\ell,m}^{-}$ are monotonic functions similar to aligned-spin waveforms, while the antisymmetric amplitude,  $A^-_{\ell,m}$ and the symmetric phase $\phi_{\ell,m}^{+}$ are small real-valued oscillatory functions whose modelling is challenging. In Ref.~\cite{Blackman:2017dfb} apply a Hilbert transform is applied to $A^-_{\ell,m}$ and $\phi_{\ell,m}^{+}$ to convert them into slowly-varying functions easier to model.

\begin{figure}[!]
\centering
\hspace*{-0.5cm}
\includegraphics[scale=0.36]{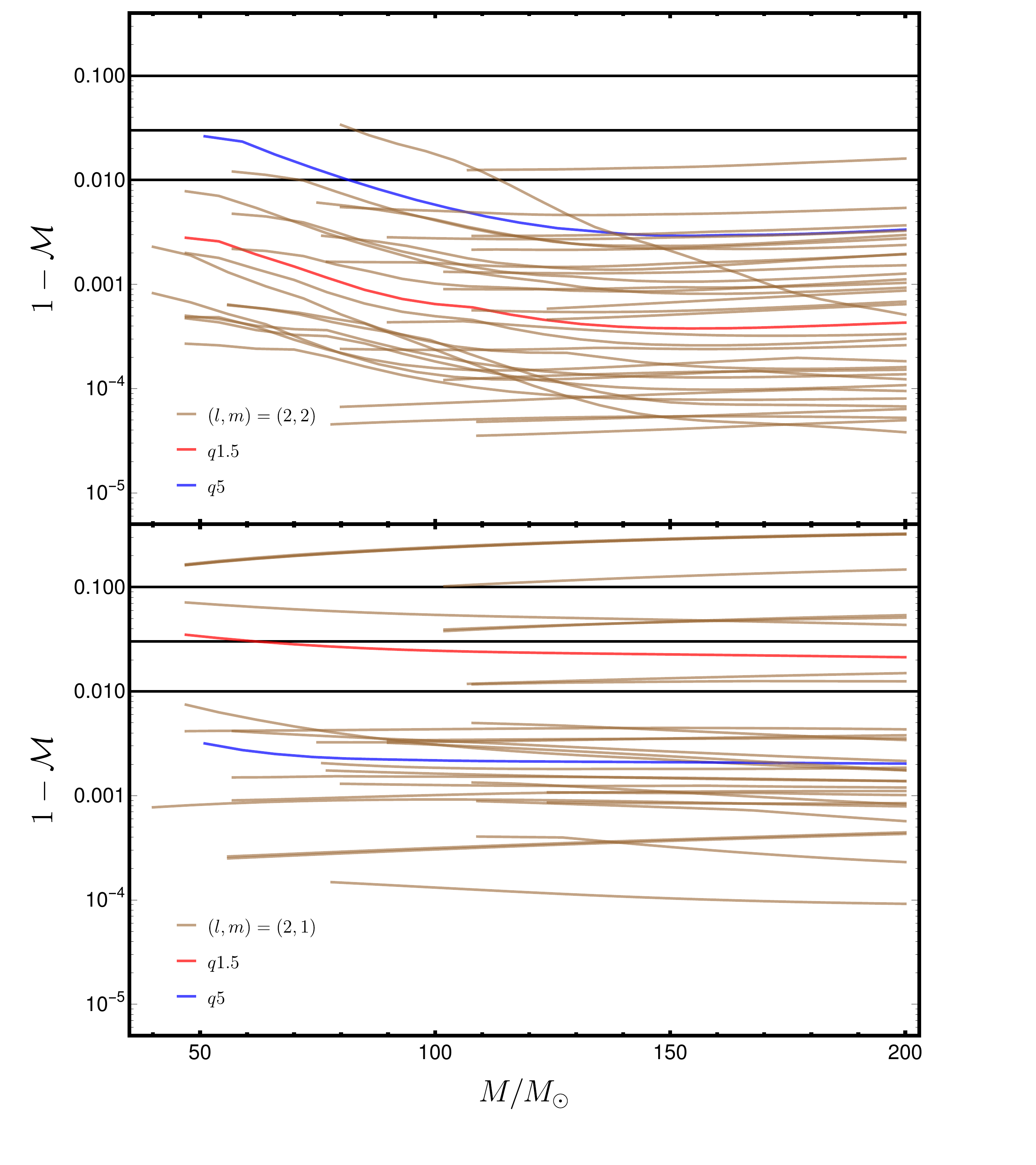}
\vspace*{-1cm}
\caption{Single mode mismatches between the AS modes, $h^{AS}_{\ell,m}$, and the symmetric QA modes, $h^+_{\ell m}$, for all the configurations in Table \ref{tab:tabNR3} as a function of the total mass of the system. In the top and bottom panels we show the results for the $(2, 2)$ and $(2, 1)$ modes, respectively. The configurations  with  IDs \SXSqonepointfiveID{} and \SXSqfiveID{} are highlighted with red and blue colors, respectively. The horizontal lines mark the $1\%$, $3 \%$ and $10 \%$ value of the mismatch.}
\label{fig:mmQASymvsAS2221}
\end{figure}

We pursue to assess the identification between what we call the symmetric waveform, constructed with the symmetric amplitude and the antisymmetric phase, i.e., $h^{+}_{\ell m}=A^+_{\ell,m} e^{i \phi_{\ell,m}^{-}}$, and the aligned-spin modes. In order to quantify that comparison we calculate single mode mismatches following the procedure of Sec. \ref{sec:QAvsSA}. In Fig. \ref{fig:mmQASymvsAS2221} we show the single mode mismatches between the $h^+_{\ell m}$ and $h^{AS}_{\ell m}$ for the $(2,2)$ and $(2,1)$ modes.  The mismatches for higher order modes are displayed in Fig. \ref{fig:mmQASymvsASHMs} of App. \ref{sec:AppendixB}. For odd-m modes we have removed the cases with PI symmetry. The results of Fig.  \ref{fig:mmQASymvsAS2221} suggest that for the $(2, \pm 2)$ modes the identification between $h^+_{2,\pm 2}$ and $h^{AS}_{2, \pm 2}$ is an outstanding approximation as all the mismatches are  below $3 \%$. For higher order modes the mismatches increase one or two orders of magnitude for some particular cases as shown in Fig. \ref{fig:mmQASymvsASHMs} of App. \ref{sec:AppendixB}, although the bulk of cases are below $3 \%$.

\begin{figure*}[!]
\centering
\hspace*{-0.5cm} 
\includegraphics[scale=0.36]{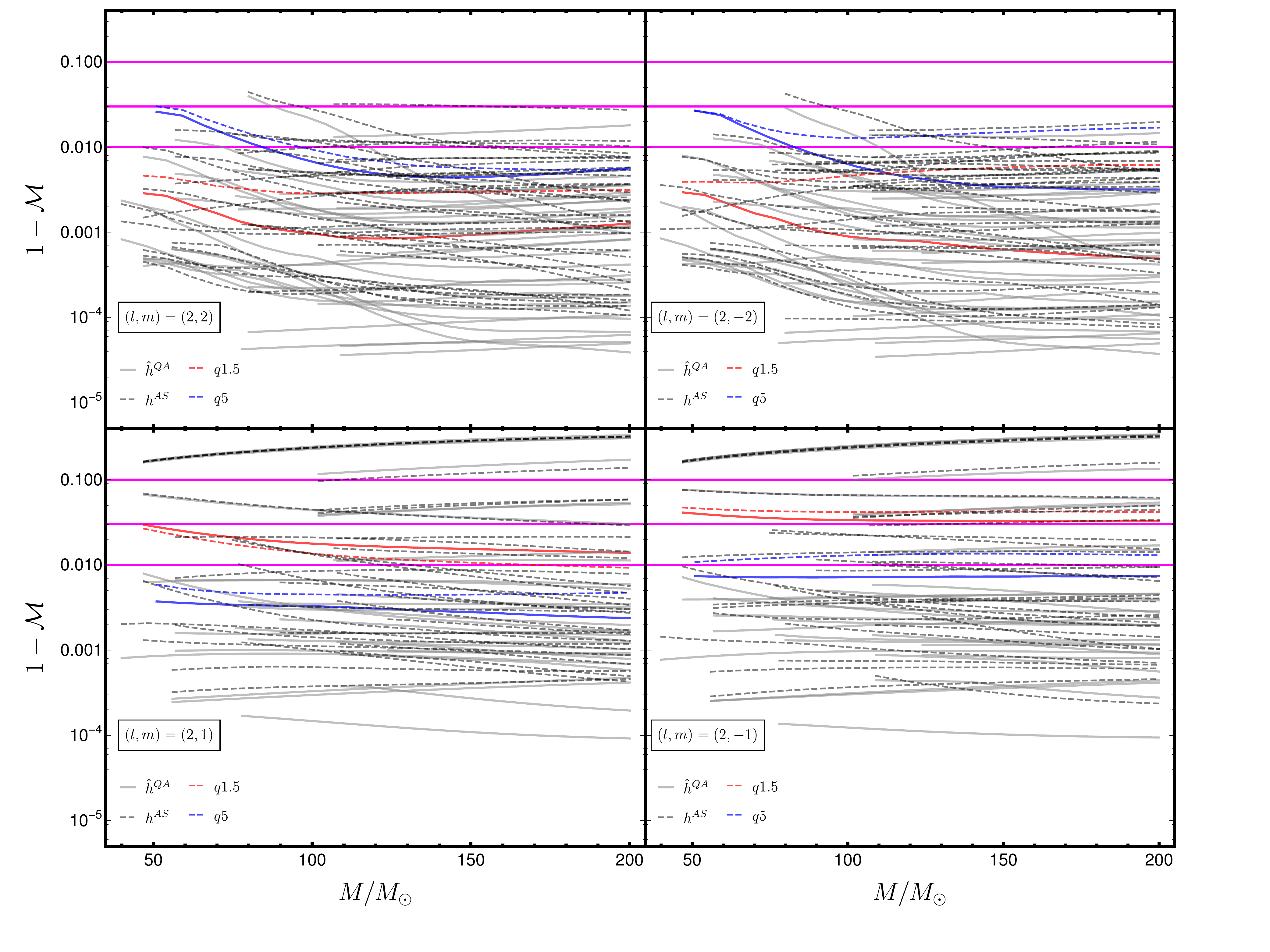}
\vspace*{-0.75cm}
\caption{Single mode mismatches between the QA modes, $h^{QA}_{\ell m}$, and the approximate QA, $\hat{h}^{QA}_{\ell m}$, and the AS, $h^{AS}_{\ell m}$, modes; for all the configurations in Table \ref{tab:tabNR3} of App. \ref{sec:AppendixA} as a function of the total mass of the system. In the top [bottom] left and right panels we show the results for the $(2,\pm 2)$ [$(2,\pm 1)$], respectively.  The thick gray (dashed black) lines correspond to mismatches between $h^{QA}_{\ell m}$ and $\hat{h}^{QA}_{\ell m}$ ($h^{AS}_{\ell m}$). In addition, configurations  with  IDs \SXSqonepointfiveID{} and \SXSqfiveID{} are highlighted with red and blue colors, respectively. The horizontal lines mark the $1\%$, $3 \%$ and $10 \%$ value of the mismatch.}
\label{fig:mmQAvsQAAS2221}
\end{figure*}

Given this, which suggests a good approximation between the symmetric QA and the AS waveforms, we can also study the impact of constructing the QA waveform modes replacing $A_+^{\ell,m}$ and $\phi^{\ell,m}_{-}$ with the AS the amplitude and phases, $A^{AS}_{\ell,m}$, $\phi^{AS}_{\ell,m}$, 
\begin{align}
A^{QA}_{l, \pm m} & =A^+_{\ell,m} \pm A^-_{\ell,-m}   \quad \rightarrow \quad  \hat{A}^{QA}_{l, \pm m} \approx A^{AS}_{\ell,m} \pm A^-_{\ell,-m},\label{eq:eq1510}\\
\phi^{QA}_{l, \pm m} & =\phi^+_{\ell,m} \pm \phi^-_{\ell,-m} \quad \rightarrow \quad  \hat{\phi}^{QA}_{l, \pm m} \approx \phi^{+}_{\ell,m} \pm \phi^{AS}_{\ell,-m}. 
\label{eq:eq151}
\end{align}
Therefore, one can compute an approximate QA waveform as $\hat{h}^{QA}_{\ell m}= \hat{A}^{QA}_{l,  m} e^{i  \hat{\phi}^{QA}_{l,  m} }$. Once, this waveform is constructed we quantify the difference between the QA modes, $h^{QA}_{\ell m}$, and the approximate ones, $\hat{h}^{QA}_{\ell m}$ via single mode mismatches. In Fig. \ref{fig:mmQAvsQAAS2221} one observes that $\hat{h}^{QA}_{\ell m}$ produces lower mismatches than  $h^{AS}_{\ell m}$. This indicates that the approximation of the symmetric amplitude and asymmetric phase by the AS amplitude and phase can be used with high accuracy for the $(2,\pm 2)$ modes, while for higher order modes, especially the weak $(2,\pm 1),(3,\pm 2)$ and $(4,\pm 3)$ modes this approximation degrades as shown in Fig. \ref{fig:mmQAvsQAASHMs}. This degradation is mainly due to the fact that the small difference between $A^{AS}_{\ell,m}$ and $A^{+}_{\ell,m}$ is a significant fraction of the power of the modes.

These results suggest a modification to the modelling strategy of precessing waveforms as follows: Instead of directly identifying the QA with corresponding AS modes, one should use the symmetric amplitude and antisymmetric phases constructed from the AS modes as per Eqs.~\eqref{eq:eq1510} and \eqref{eq:eq151}.

\section{Summary and Conclusions} 
\label{sec:summary}
We have assessed and quantified the accuracy of two main approximations commonly used to construct phenomenological inspiral-merger-ringdown waveforms from precessing \acp{BBH}. The first approximation (APX1) is the identification between aligned-spin and co-precessing waveforms~ \cite{PhysRevD.84.024046,OShaughnessy2011,Boyle2011,PhysRevD.86.104063}. The second approximation (APX2) concerns the inclusion of higher-order aligned-spin modes in the construction of approxmiate precessing modes.

Focusing exclusively on the late inspiral and merger regime, we use NR waveforms from the first SXS catalog \cite{PhysRevLett.111.241104} and additional waveforms produced with the private BAM code and the open-source Einstein Toolkit. Our analysed  NR data set consists of a total of 36 pairs of AS and precessing configurations, and we restrict our analyses to comparisons of waveforms generated with the same NR code to avoid the introduction of systematics due to numerical errors. We note that during the preparation of this manuscript a much larger SXS catalog \cite{Boyle:2019kee} was released. This allows for the extension of the presented analyses to a larger parameter space, which we leave for future work.

We first quantified the efficacy of the QA-AS mapping (APX1) via single-mode mismatches and the radiated energy per mode. We find that this approximation yields mismatches below 3\% for the $(2,\pm 2)$ modes for the majority of configurations in our sample. However, the picture changes dramatically for higher-order modes. For modes that are prone to mode-mixing such as the $(3,|2|)$ and $(4,|3|)$ mode, the approximation is particularly poor, but the matches drop significantly also for the $(2,|1|)$-modes. 

Furthermore, we find that the QA-AS map breaks down for configurations with highly asymmetric energy content between the $+m$ and $-m$ modes as quantified by the radiated energy per mode. The mode asymmetries are one of the clear limitations of this approximate mapping due to the tight symmetry condition of the AS waveforms which is not fulfilled by precessing and therefore the QA waveforms. We conclude that it will become increasingly important to correctly model these mode asymmetries in order to improve the accuracy of waveform models, which will particularly important in the coming years as ground-based GW detectors are set to improve their sensitivity~\cite{PhysRevD.91.062005,AplusDesignCurve,Punturo_2010,Hild_2011,Abbott_2017}.

To alleviate some of the shortcomings of (APX1), we have investigated a modification used in surrogate models  \cite{Blackman:2017dfb,PhysRevResearch.1.033015}, where rather than identifying the co-precessing modes with AS modes, a combination of slowly varying amplitude and phase functions is used to model the co-precessing modes. We find that the symmetric amplitude and antisymmetric phase of the co-precessing modes can be identified with the amplitude and phase of the AS modes to high accuracy for the $(2,\pm 2)$ modes. For certain higher order modes such as the $(3,\pm 3)$ and $(4,\pm 4)$ we find comparable results, but the weakest modes such as $(2,\pm 1)$, $(3,\pm 2)$ and $(4, \pm 3)$ still have significantly larger mismatches. 

Our study of (APX2) shows that the addition of the AS $(2, \pm 1)$ modes to construct the approximate precessing $(2,\pm 2)$ modes, does not significantly impact the mode accuracy. Again, we find that the opposite is true for higher order modes, where the inclusion of higher order AS modes improves the accuracy of the approximate precessing modes. And similarly to (APX1), we find that the $(2,|1|), (3,|2|)$ and $(4,|3|)$ modes are most strongly affected.

Beyond the individual modes, we have also analyzed the GW strain, which takes into account the different contributions from higher order modes depending on the orientation of the binary. Similarly, we find that the inclusion of AS higher order modes to construct approximate precessing waveforms improves the mismatches by an order of magnitude for edge-on configurations. However, care needs to be taken as the inclusion of even more higher order modes in the strain can increase the mismatch due to the accumulation of approximation errors when summing up the individual modes to construct the strain. 

To highlight some additional error sources, we have studied a particular configuration which shows a minimum in the aligned-spin $(2,1)$ mode. We find that while the QA $(2,1)$ mode is not able to resemble the AS mode accurately, the precessing $(2,1)$ is hardly affected since the main contribution in its construction stems from the AS $(2,2)$ mode. Nevertheless, we have also seen that the inclusion of higher order modes in the construction of approximate precessing waveforms does matter for the majority of cases and therefore their accuracy is crucial. 

Overall, our results show larger mismatches than what has previously been found for precessing phenomenological and EOB models \cite{PhysRevD.100.024059,PhysRevD.95.024010}. We attribute this difference  to the impact of NR errors in our waveforms, which are much higher than those described in  \cite{Bohe:PPv2,PhysRevD.100.024059,PhysRevD.95.024010} due to the inclusion of higher order modes, although the strain mismatches in Sec. \ref{sec:PrecvsP} are consistent with those obtained in \cite{Khan:2019kot} for the same configurations. We also note that we neglect modifications to the final state that capture spin precession effects. However, we have verified using other phenomenological waveform models~\cite{PhysRevD.93.044006,Khan2015, phenT} that such modifications are a subdominant effect in the whole framework. Hence, the intrinsic limitations of the two modelling approximation (APX1) and (APX2), combined with the impact of NR errors for higher order modes are responsible for the reduced accuracy pf precessing higher order modes produced in this paradigm.

Our studies show that in order to produce accurate phenomenological precessing waveform models necessary to facilitate the maximal science return from future GW observations, modifications to the simple paradigm that take into account mode asymmetries and subdominant effects will be crucial.

\section{Acknowledgements} 
We would like to thank Vijay Varma for useful comments on the manuscript. A. Ramos-Buades was supported by the Spanish Ministry of Education and Professional Formation grants EST17/00421 and FPU15/03344, also Sascha Husa and Geraint Pratten by European Union FEDER funds, the Ministry of Science, Innovation and Universities and the Spanish Agencia Estatal de Investigacion grants FPA2016-76821-P, RED2018-102661-T, RED2018-102573-E, FPA2017-90687-REDC, FPA2017-90566-REDC, Vicepresidència i Conselleria d’Innovació, Recerca i Turisme del Govern de les Illes Balears i Fons Social Europeu, Generalitat Valenciana (PROMETEO/2019/071), EU COST Actions CA18108,CA17137, CA16214, and CA16104. The authors thankfully acknowledge the computer resources at MareNostrum and the technical support provided by Barcelona Supercomputing Center (BSC) through Grants No. AECT-2019-2-0010, AECT-2019-1-0022, AECT-2018-3-0017, AECT-2018-2-0022, AECT-2018-1-0009, AECT-2017-3-0013, AECT-2017-2-0017, AECT2017-1-0017, AECT-2016-3-0014, AECT2016-2-0009, from the Red Española de Supercomputación (RES) and PRACE (Grant No. 2015133131). {\tt BAM} and {\tt ET} simulations were carried out on the BSC MareNostrum computer center under PRACE and RES allocations and on the FONER cluster at the University of the Balearic Islands.  
P. Schmidt acknowledges support from the Netherlands Organisation for Scientific Research (NWO) Veni grant no. 680-47-460.
A. Ramos-Buades is grateful to Radboud University, Nijmegen, The Netherlands for hospitality during stages of this work. 
This paper has LIGO document number P1900388.


\appendix

\section{Numerical Relativity Simulations} 
\label{sec:AppendixA}
The \texttt{Einstein Toolkit} (ET)~\cite{Loffler:2011ay,maria_babiuc_hamilton_2019_3522086} is an open source NR code built around the Cactus framework~\cite{cactuscode, Goodale}. The numerical setup of our simulations is similar to that used in \cite{Pollney:2009yz}, though we present the details here for completeness. 

We use standard Bowen-York initial data \cite{Bowen1980,Brandt1997} computed using the \texttt{TwoPunctures} thorn \cite{Ansorg2004}, in which the punctures are initially placed on the x-axis at positions $x_1=D/(1+q)$ and $x_2=-qD/(1+q)$, where $D$ is the coordinate separation and we assume $m_1 \geq m_2$. The initial momenta are chosen such that $\textbf{p}=\left( \mp p_r, \pm p_t,0 \right)$. We use low-eccentricity initial data following the prescription detailed in \cite{PhysRevD.99.023003}. 

The time evolution is performed using the $W$-variant \cite{Marronetti:2007wz} of the BSSN formulation \citep{PhysRevD.52.5428,Baumgarte:1998te} of the Einstein field equations as implemented by \texttt{McLachlan} \cite{Brown:2008sb}. The black holes are evolved using the standard moving punctures gauge conditions \cite{Baker:2005vv,Campanelli:2005dd} with the lapse being evolved according to the "$1+\log$" condition \cite{Bona:1994dr} and the shift being evolved using a hyperbolic $\tilde{\Gamma}$-driver equation \cite{Alcubierre:2002kk}.

The simulations were performed using 8th order accurate finite differencing and Kreiss-Oliger dissipation~\cite{Kreiss1973}. Adaptive mesh refinement is provided by \texttt{Carpet}~\cite{carpet, Schnetter:2003rb, Schnetter:2006pg}, with the near zone being computed with high resolution Cartesian grids that track the motion of the BHs, while the wave extraction zone uses spherical grids provided the \texttt{Llama} multipatch infrastructure \cite{Pollney:2009yz}. By using grids adapted to the spherical topology of the wave extraction zone, we are able to efficiently compute high-accuracy waveforms at large extraction radii relative to standard Cartesian grids. The apparent horizons are computed using \texttt{AHFinderDirect} \cite{Thornburg:2003sf} and a calculation of the spins is performed in the dynamical horizon formalism using the \texttt{QuasiLocalMeasures} thorn \cite{Dreyer:2002mx}.  

The gravitational waves are computed using the \texttt{WeylScal4} thorn and the GW strain $h$ is calculated from $\Psi_4$ via fixed-frequency integration \cite{Reisswig:2010di}. The thorns \texttt{McLachlan} and \texttt{WeylScal4} are generated by the automated-code-generation package \texttt{Kranc} \cite{Husa:2004ip, kranc}. The ET simulations are managed using \texttt{Simulation Factory} \cite{SimulationFactory} and the analysis and post-processing of ET waveforms was performed using the open source \texttt{Mathematica} package \texttt{Simulation Tools} \cite{SimulationTools}.

The SXS waveform data used here are described in detail in~\cite{Mroue:2013xna, Boyle:2019kee} and can be obtained from~\cite{SXS}.

The BAM simulations use the same numerical setup as described in App. C 1 of \cite{PhysRevD.99.023003}. In brief, the BAM code~\cite{Bruegmann:2006at} evolves black-hole binary initial data \cite{PhysRevLett.78.3606,PhysRevD.21.2047} using the $\chi$-variant version of the moving puncture \cite{Baker:2005vv,Campanelli:2005dd} version of the BSSN  formulation \cite{PhysRevD.52.5428,Baumgarte:1998te} of the Einstein equations. The black-hole punctures are initially located on the y-axis at positions $y_1=-qD/(1+q)$ and $y_2=D/(1+q)$, where $D$ is the coordinate distance between the two punctures and the mass ratio is $q=m_2/m_1 > 1$.
The code uses sixth-order spatial finite-difference derivatives, fourth-order Runge-Kutta algorithm and Kreiss-Oliger dissipation terms~\cite{Kreiss1973} which converge at fifth order. Furthermore, the code uses sixteen mesh-refinement buffer points and the base configuration consists of $n_1$ nested mesh-refinement boxes with $N^3$ points surrounding each black hole, and $n_2$ nested boxes with $(2N)^3$ points surrounding the entire system. On the levels where the extraction of gravitational radiation is performed, $(4N)^3$ points are used in order to extract more accurately the gravitational waves emitted by the binary. These waves are computed from the Newman-Penrose scalar $\Psi_4$ \cite{Bruegmann:2006at} and converted into strain via fixed-frequency integration~\cite{Reisswig:2010di}.

Table \ref{tab:tabNR3} summarizes some key properties of the main set of NR simulations used for this work. 
We arrange the simulations in pairs, each pair consisting of a different precessing simulation and its corresponding aligned-spin counterpart following Eq.~\eqref{eq:eq100}.

\renewcommand{\thetable}{\Alph{table}}
\setcounter{table}{0}

\begin{turnpage}
\begin{table*}[!]
\caption{Summary of NR simulations used in this work. Each simulation is specified by its mass ratio $q=m_1/m_2\geq 1 $, the initial dimensionless spin vectors, $\vec{\chi}_1$, $\vec{\chi}_2$, the orbital separation $D/M$, the orbital frequency $\Omega_0$ and the orbital eccentricity, $e_0$, at the relaxation time, final dimensionless spin vector, $\bm{\chi}_f$, its magnitude $\chi_f$, final recoil vector $\bm{v}_f$ ($km/s$) and its magnitude  $v_f$ ($km/s$).}\label{tab:tabNR3} 
\resizebox{23.5cm}{!}{
\def\arraystretch{1.4}
\begin{tabular}{ c c c c c c c c c c  c  c  c c c }
\hline
\hline
ID & Simulation &  Code &  q  &$\bm{\chi}_1$ & $\bm{\chi}_2$ & $ \chi_{\text{eff}}$ &$D/M$ & $M \Omega_0$ & $e_0 \cdot 10^{-3}$ & $M_f$ & $\bm{\chi}_f$& $\chi_f$& $\bm{v}_f (km/s)$ & $v_f (km/s)$  \\
\hline
\multirow{2}{*}{1} & \texttt{q1.\_\_0.2\_0.1\_-0.5\_\_0.1\_0.2\_-0.5\_\_pcD12}& \texttt{ET}  &  1.& $(0.2, 0.1, -0.5)$& $(0.1, 0.2, -0.5)$& -0.5 & 11.38 & 0.0232 & 2.14 & 0.9618 & $(0.047,0.036,0.533)$ & 0.536 &$(3.88,-34.97,-157.09)$  & 160.98  \\
 & \texttt{q1.\_-0.5\_-0.5\_\_pcD12} & \texttt{ET} & 1.& $(0., 0., -0.5)$& $(0., 0., -0.5)$& -0.5 & 11.38 & 0.0232 &  2.31  & 0.9621 & $(0.,0.,0.527)$ &0.527 & $(-0.94,0.81,0.00)$ & 1.24 \\
 \hline
\multirow{2}{*}{2} & \texttt{q1.\_\_-0.5\_0.5\_-0.5\_\_0.5\_-0.5\_-0.5\_\_pcD12}& \texttt{ET}  &  1.& $(-0.5, 0.5, -0.5)$& $(0.5, -0.5, -0.5)$& -0.5  & 11.38 & 0.0232 & 1.59  & 0.9595 & $(0.,0.,0.517)$  & 0.517 &$(-3.15,2.78,1693.88)$  & 1693.89\\
 & \texttt{q1.\_-0.5\_-0.5\_\_pcD12} & \texttt{ET} & 1.& $(0., 0., -0.5)$& $(0., 0., -0.5)$& -0.5 & 11.38 & 0.0232 &  2.31  &  0.9621 & $(0.,0.,0.527)$ & 0.527  & $(-0.94,0.81,0.)$  &  1.24 \\
  \hline
\multirow{2}{*}{3} & \texttt{q1.\_\_0.5\_0.5\_-0.5\_\_0.5\_0.5\_-0.5\_\_pcD12}& \texttt{ET}  &  1.& $(0.5, 0.5, -0.5)$& $(0.5, 0.5, -0.5)$& -0.5 & 11.25 & 0.0233 & 1.21  & 0.9576 &$(0.194,0.144,0.567)$ & 0.618 &  $(-1.71,-1.31,3.79)$  & 4.36\\
 & \texttt{q1.\_-0.5\_-0.5\_\_pcD12} & \texttt{ET} & 1.& $(0., 0., -0.5)$& $(0., 0., -0.5)$& -0.5 & 11.38 & 0.0232 &  2.31  & 0.9621 & $(0.,0.,0.527)$ &0.527  &  $(-0.94,0.81,0.)$  & 1.24   \\
 \hline
\multirow{2}{*}{4} & \texttt{SXS:BBH:0003}& \texttt{SpEC}  &  1.& $(0.5, 0.05, 0.)$& $(0., 0., 0.)$& 0.& 19.& 0.0113& 0.26  & 0.9511 & $(0.066,0.009,0.691)$ & 0.6947 &$(122.91,6.75,697.04)$ & 707.82 \\
 & \texttt{SXS:BBH:0001} & \texttt{SpEC} & 1.& $(0., 0., 0.)$& $(0., 0., 0.)$& 0.& 18.& 0.0113& 0.17  &  0.9516 & $(0.,0.,0.686)$ & 0.686 & $(0.,0.,0.)$ &  0.00 \\
 \hline
\multirow{2}{*}{5} & \texttt{q1.5\_\_0.\_0.\_0.5\_\_0.\_-0.5\_0.\_\_pcD11}& \texttt{ET}  &  1.5& $(0., -0.5, 0.)$& $(0., 0., 0.5)$& 0.2 & 10.44.& 0.0260 & 1.50  & 0.9487 &$(-0.007,-0.129,0.710)$ & 0.722 &$(-73.58,-266.64,1539.6)$  & 1564.25 \\
 & \texttt{q1.5\_0.5\_0.\_\_pcD11} & \texttt{ET} & 1.5& $(0., 0., 0.5)$& $(0., 0., 0.)$& 0.2  & 10.46   & 0.0260  &  1.10   & 0.9506 & $(0.,0.,0.707)$ & 0.707 & $(-122.89,-89.37,0.)$  & 151.95 \\
 \hline
\multirow{2}{*}{6} & \texttt{q1.5\_\_0.\_0.\_-0.5\_\_0.5\_0.\_0.\_\_pcD12}& \texttt{ET}  &  1.5& $(0., 0., -0.5)$& $(0.5, 0., 0.)$& -0.3 & 11.47& 0.0229 & 1.30  & 0.9621 & $(0.052,0.0,0.54)$ & 0.542 &$(-62.27,208.92,-644.27)$  & 680.15 \\
 & \texttt{q1.5\_-0.5\_0.\_\_pcD12} & \texttt{ET} & 1.5& $(0., 0., -0.5)$& $(0., 0., 0.)$&  -0.3& 11.49& 0.0229 & 1.51  &0.9624 & $(0.,0.,0.540)$ & 0.540 & $(-6.32,221.25,0)$ & 221.34 \\
 \hline
\multirow{2}{*}{7} & \texttt{q1.5\_\_0.\_0.5\_-0.5\_\_0.\_0.\_-0.5\_\_pcD12}& \texttt{ET}  &  1.5& $(0., 0.5, -0.5)$& $(0., 0., -0.5)$& -0.5 & 11.42 & 0.0230  & 2.60  &0.9643 &$(0.001,0.090,0.514)$  & 0.522 &$(-113.37,-72.82,-321.97)$  & 349.03 \\
 & \texttt{q1.5\_-0.5\_-0.5\_\_pcD12} & \texttt{ET} & 1.5& $(0., 0., -0.5)$& $(0., 0., -0.5)$& -0.5  & 11.43 & 0.0230 & 2.03   & 0.9651 & $(0.,0.,0.491)$ & 0.491 &  $(-108.15,112.94,0.)$  & 156.37 \\
 \hline
\multirow{2}{*}{8} &\texttt{SXS:BBH:0015}& \texttt{SpEC}  & 1.5& $(0.5, 0.06, 0.)$& $(0., 0., 0.)$& 0.& 18.& 0.0123& 0.34   & 0.9548 & $(0.115,-0.014,0.675)$ &  $0.685$ & $(119.5,65.27,324.21)$ &  351.64 \\
& \texttt{SXS:BBH:0007} & \texttt SpEC & 1.5& $(0., 0., 0.)$& $(0., 0., 0.)$& 0.& 18.& 0.0123& 0.43  & 0.9552 & (0.,0.,0.664) & 0.664 & $(103.38,-1.,0.)$ &  103.38\\
 \hline
\multirow{2}{*}{9} &\texttt{SXS:BBH:0020} & \texttt{SpEC} & 1.5& $(0., 0., 0.5)$& $(0.49, 0.07, 0.)$& 0.3& 16.& 0.0144& 0.12   &  0.9450 & $(0.045,0.009,0.788)$ &  0.79 &  $(-14.38,-25.56,20.23)$ & 35.63 \\
& \texttt{SXS:BBH:0013} & \texttt{SpEC} & 1.5& $(0., 0., 0.5)$& $(0., 0., 0.)$& 0.3& 16.& 0.0144& 0.14  &  0.9446 &$(0.,0.,0.781)$ &  0.781 & $(-58.40,27.41,0.01)$ & 64.51\\
\hline
\multirow{2}{*}{10} &\texttt{SXS:BBH:0023} & \texttt{SpEC} & 1.5& $(0.5, 0.05, 0.)$& $(0.08, -0.49, 0.)$& 0.& 16.& 0.0145& 0.28  &  0.9535 & $(0.1,-0.046,0.675)$ & 0.684  & $(156.67,101.15,968.07)$ & 985.87\\
 & \texttt{SXS:BBH:0007} & \texttt SpEC & 1.5& $(0., 0., 0.)$& $(0., 0., 0.)$& 0.& 18.& 0.0123& 0.43  & 0.9552 & (0.,0.,0.664) & 0.664 & $(103.38,-1.,0.)$ &  103.38 \\
 \hline
\multirow{2}{*}{11} &\texttt{SXS:BBH:0024} & \texttt{SpEC} & 1.5& $(0.5, 0.05, 0.)$& $(-0.08, 0.49, 0.)$& 0.& 16.& 0.0145& 0.21   & 0.9550 & $(0.092,0.053,0.68)$ &  0.688 & $(-161.54,73.31,-129.14)$ & 219.42 \\
& \texttt{SXS:BBH:0007} & \texttt SpEC & 1.5& $(0., 0., 0.)$& $(0., 0., 0.)$& 0.& 18.& 0.0123& 0.43  &  0.9552 & (0.,0.,0.664) & 0.664 & $(103.38,-1.,0.)$ &  103.38 \\
\hline
\multirow{2}{*}{12} &\texttt{SXS:BBH:0026} & \texttt{SpEC} & 1.5& $(0., 0., 0.5)$& $(-0.49, -0.07, 0.)$& 0.3& 16.& 0.0144& 0.12   & 0.9450 & $(-0.046,-0.009,0.788)$ &0.789 & $(-15.01,-25.75,-20.12)$ & 35.90 \\
&\texttt{SXS:BBH:0013} & \texttt{SpEC} & 1.5& $(0., 0., 0.5)$& $(0., 0., 0.)$& 0.3& 16.& 0.0144& 0.14  & 0.9446 &$(0.,0.,0.781)$ &  0.781 & $(-58.40,27.41,0.01)$ & 64.51\\
 \hline
\multirow{2}{*}{13} & \texttt{SXS:BBH:0027 }& \texttt{SpEC}  & 1.5& $(0.5, 0.05, 0.)$& $(-0.49, -0.07, 0.)$& 0.& 16.& 0.0145& 0.07  & 0.9545 & $(0.044,0.004,0.673)$ & 0.675  & $(105.94,251.68,1128.58)$ & 1161.14 \\
 & \texttt{SXS:BBH:0007} & \texttt SpEC & 1.5& $(0., 0., 0.)$& $(0., 0., 0.)$& 0.& 18.& 0.0123& 0.43  &  0.9552 & (0.,0.,0.664) & 0.664 & $(103.38,-1.,0.)$ &  103.38 \\
\hline
\multirow{2}{*}{14} & \texttt{SXS:BBH:0028} & \texttt{SpEC} & 1.5& $(0., 0., 0.5)$& $(0.08, -0.49, 0.)$& 0.3& 16.& 0.0144& 0.18  &  0.9435 & $(0.009,-0.041,0.785)$ &  0.786 & $(-42.34,-32.47,-41.01)$ & 67.30\\
 &\texttt{SXS:BBH:0013} & \texttt{SpEC} & 1.5& $(0., 0., 0.5)$& $(0., 0., 0.)$& 0.3& 16.& 0.0144& 0.14  & 0.9446 &$(0.,0.,0.781)$ &  0.781 & $(-58.40,27.41,0.01)$ & 64.51\\
 \hline
\multirow{2}{*}{15}&\texttt{SXS:BBH:0029} & \texttt{SpEC} & 1.5& $(0.5, 0.05, 0.)$& $(0.49, 0.07, 0.)$& 0.& 16.& 0.0145& 0.47  & 0.9536 & $(0.151,-0.001,0.678)$ & 0.694 &$(107.59,120.27,474.1)$ & 500.81 \\
& \texttt{SXS:BBH:0007} & \texttt SpEC & 1.5& $(0., 0., 0.)$& $(0., 0., 0.)$& 0.& 18.& 0.0123& 0.43  & 0.9552 & (0.,0.,0.664) & 0.664 & $(103.38,-1.,0.)$ &  103.38 \\
\hline
\multirow{2}{*}{16} & \texttt{q2.\_\_-0.35\_0.35\_0.5\_\_0.\_0.\_0.\_\_pcD10.8} & \texttt{ET} & 2.& $(-0.35, 0.35, 0.5)$& $(0., 0., 0.)$& 0.333& 10.28& 0.0265& 1.11  & 0.9476 & $(-0.126, 0.120, 0.785)$ & 0.804 &  $(-104.70,95.90, 551.66)$ & 569.64\\
&  \texttt{q2.\_0.\_0.5\_\_pcD11} & \texttt{ET} & 2.& $(0., 0., 0.5)$& $(0., 0., 0.)$& 0.333& 10.55& 0.0257& 0.96  & 0.9494 & $(0.,0.,0.778)$  & 0.778 & $(37.33, -43.68, 0.)$  & 57.46\\
\hline
 \multirow{2}{*}{17} & \texttt{q2.\_\_-0.35\_0.35\_0.\_\_0.\_0.\_0.\_\_pcD10.8} & \texttt{ET} & 2.& $(-0.35, 0.35, 0.)$& $(0., 0., 0.)$& 0.& 10.24& 0.0270& 0.78  & 0.9602  & $(-0.095,0.114,0.641)$  & 0.658 & $(96.36,-159.30,-93.83)$  & 208.48 \\
&  \texttt{q2.\_0.\_0.\_\_pcD11} & \texttt{ET} & 2.& $(0., 0.,0.)$& $(0., 0., 0.)$& 0.& 10.52& 0.0261& 1.34   & $0.9612$ & $(0.,0.,0.623)$  & 0.623 & $(111.94,89.01,0.)$  & 143.02\\
 \hline
\hline
\end{tabular}
}
\end{table*}
\end{turnpage}

\begin{turnpage}
\begin{table*}[!]

\raggedright \textit{(Continued)}
\label{tab:tabNR3}
\resizebox{22.5cm}{!}{
\def\arraystretch{1.4}
\begin{tabular}{ c c c c c c c c c c  c  c  c c c }
\hline
\hline
ID & Simulation &  Code &  q  &$\bm{\chi}_1$ & $\bm{\chi}_2$ & $ \chi_{\text{eff}}$ &$D/M$ & $M \Omega_0$ & $e_0 \cdot 10^{-3}$ & $M_f$ & $\bm{\chi}_f$& $\chi_f$& $\bm{v}_f (km/s)$ & $v_f (km/s)$  \\
\hline
 \hline
\multirow{2}{*}{18} &\texttt{q2.\_\_-0.35\_0.35\_-0.5\_\_0.\_0.\_0.\_\_pcD10.8} & \texttt{ET} & 2.& $(-0.35, 0.35, -0.5)$& $(0., 0., 0.)$& -0.333& 10.06& 0.0277& 2.65  & $0.9674$  & $(-0.115,0.071,0.489)$ & 0.507 &  $(308.13,95.87,0.04)$  & 322.70\\
& \texttt{q2.\_0.\_-0.5\_\_pcD11} & \texttt{ET} & 2.& $(0., 0.,-0.5)$& $(0., 0., 0.)$& -0.333& 10.37& 0.0267& 1.88  & 0.9683 & $(0.,0.,0.460)$ & 0.460  & $(-239.88,91.74,0.)$ & 256.82\\
 \hline
\multirow{2}{*}{19} & \texttt{q2.\_\_0.35\_0.35\_-0.5\_\_0.\_0.\_0.\_\_pcD10.8} & \texttt{ET} & 2.& $(0.35, 0.35, -0.5)$& $(0., 0., 0.)$& -0.333& 10.06& 0.0277& 1.51  & 0.9669 & $(0.071,0.116,0.486)$ &0.505 &  $(-5.13,-207.31,-758.34)$  & 786.18\\
& \texttt{q2.\_0.\_-0.5\_\_pcD11} & \texttt{ET} & 2.& $(0., 0.,-0.5)$& $(0., 0., 0.)$& -0.333& 10.37& 0.0267& 1.88  &  0.9683 &$(0.,0.,0.460)$ &   0.460  & $(-239.88,91.74,0.)$ & 256.82\\
 \hline
\multirow{2}{*}{20} &  \texttt{q2.\_\_0.35\_0.35\_0.5\_\_0.\_0.\_0.\_\_pcD10.8} & \texttt{ET} & 2.& $(0.35, 0.35, 0.5)$& $(0., 0., 0.)$& 0.333& 10.28& 0.0265& 0.24  & 0.9464  & $(0.119,0.125,0.783)$ &0.802 &  $(-356.8,-141.84,-1279.92)$  & 1336.27 \\
&  \texttt{q2.\_0.\_0.5\_\_pcD11} & \texttt{ET} & 2.& $(0., 0., 0.5)$& $(0., 0., 0.)$& 0.333& 10.55& 0.0257& 0.96  & 0.9494 & $(0.,0.,0.778)$  & 0.778 & $(37.33,-43.68,0.)$  & 57.46\\
 \hline
\multirow{2}{*}{21} & \texttt{q2.\_\_-0.18\_0.18\_0.5\_\_0.\_0.\_0.\_\_pcD10.8} & \texttt{ET} & 2.& $(-0.18, 0.18,0.5)$& $(0., 0., 0.)$& 0.333& 10.31& 0.0265& 0.27  &0.9490 & $(-0.062,0.060,0.780)$  & 0.785 &$(-79.86,34.08,308.69)$ &  320.67 \\
&  \texttt{q2.\_0.\_0.5\_\_pcD11} & \texttt{ET} & 2.& $(0., 0., 0.5)$& $(0., 0., 0.)$& 0.333& 10.55& 0.0257& 0.96  &  0.9494 & $(0.,0.,0.778)$  & 0.778 & $(37.33,-43.68,0.)$ & 57.46 \\
 \hline
\multirow{2}{*}{22} & \texttt{q2.\_\_-0.18\_0.18\_0.\_\_0.\_0.\_0.\_\_pcD10.8} & \texttt{ET} & 2.& $(-0.18, 0.18, 0.)$& $(0., 0., 0.)$& 0.& 10.25& 0.027& 1.47  & 0.9609 &$(-0.045,0.053,0.628)$ & 0.632 &  $(-28.97,-139.95,-294.76)$ & 327.58 \\
&   \texttt{q2.\_0.\_0.\_\_pcD11} & \texttt{ET} & 2.& $(0., 0.,0.)$& $(0., 0., 0.)$& 0.& 10.52& 0.0261& 1.34   &    $0.9612$ & $(0.,0.,0.623)$  & 0.623 & $(111.94,89.01,0.)$  &143.02 \\
 \hline
\multirow{2}{*}{23} & \texttt{q2.\_\_0.18\_0.18\_0.\_\_0.\_0.\_0.\_\_pcD10.8} & \texttt{ET} & 2.& $(0.18, 0.18, 0.)$& $(0., 0., 0.)$& 0.& 10.25& 0.0270& 0.50   & 0.9608 & $(0.053,0.046,0.627)$ & 0.631 & $(-48.56,-95.22,388.39)$  & 402.83 \\
&   \texttt{q2.\_0.\_0.\_\_pcD11} & \texttt{ET} & 2.& $(0., 0.,0.)$& $(0., 0., 0.)$& 0.& 10.52& 0.0261& 1.34  & $0.9612$ & $(0.,0.,0.623)$ & 0.623&  $(111.94,89.01,0.)$ & 143.02 \\
 \hline
\multirow{2}{*}{24} &  \texttt{q2.\_\_-0.49\_0.49\_0.\_\_0.\_0.\_0.\_\_pcD10} & \texttt{ET} & 2.& $(-0.49, 0.49, 0.)$& $(0., 0., 0.)$& 0.& 10.2& 0.0270& 1.62 & 0.9591 & $(-0.141,0.168,0.654)$ & 0.69 & $(171.54,-157.08,-108.77)$ & 256.77 \\
&   \texttt{q2.\_0.\_0.\_\_pcD11} & \texttt{ET} & 2.& $(0., 0.,0.)$& $(0., 0., 0.)$& 0.& 10.52& 0.0261& 1.34   &  $0.9612$ & $(0.,0.,0.623)$ & 0.623 & $(111.94,89.01,0.)$  & 143.02\\
 \hline
\multirow{2}{*}{25} &  \texttt{q2.\_\_0.49\_0.49\_0.\_\_0.\_0.\_0.\_\_pcD10.8} & \texttt{ET} & 2.& $(0.49, 0.49, 0.)$& $(0., 0., 0.)$& 0.& 10.19& 0.0270& 0.20  & 0.9575 & $(0.167,0.139,0.649)$ &0.684 &  $(179.99,243.12,1436.06)$  &  1467.57\\
&   \texttt{q2.\_0.\_0.\_\_pcD11} & \texttt{ET} & 2.& $(0., 0.,0.)$& $(0., 0., 0.)$& 0.& 10.52& 0.0261& 1.34   & $0.9612$ & $(0.,0.,0.623)$ & 0.623 & $(111.94,89.01,0.)$ & 143.02 \\
 \hline
 \multirow{2}{*}{26} &  \texttt{q2.\_\_0.\_0.\_0.\_\_0.7\_0.\_0.\_\_pcD11.5} & \texttt ET & 2.& $(0.7, 0., 0.)$& $(0., 0., 0.)$& 0.& 11.03& 0.0241& 1.87  & 0.9578 & $(0.227,-0.022,0.645)$   & 0.684 & $(-303.04,18.28,-1281.16)$  &  1316.64 \\
&   \texttt{q2.\_0.\_0.\_\_pcD11} & \texttt{ET} & 2.& $(0., 0.,0.)$& $(0., 0., 0.)$& 0.& 10.52& 0.0261& 1.34   &  $0.9612$ & $(0.,0.,0.623)$  & 0.623 & $(111.94,89.01,0.)$  & 143.02  \\
 \hline
\multirow{2}{*}{27} &  \texttt{q3.\_\_0.5\_0.\_-0.3\_\_0.\_0.\_-0.3\_\_pcD12} & \texttt ET & 3.& $(0.5, 0.,-0.3)$& $(0., 0., -0.3)$& -0.3 & 11.66 & 0.0226 & 1.47  & 0.9744 & $(0.169,0.019,0.450)$ & 0.481 &$(-135.64,-283.05,-426.4)$  & 529.46 \\
&   \texttt{q3.\_-0.3\_-0.3\_\_pcD12} & \texttt{ET} & 3.& $(0., 0.,-0.3)$& $(0., 0., -0.3)$& -0.3 & 11.73 & 0.0226 & 1.19  & $0.9757$ & $(0.,0.,0.396)$  & 0.396 & $(-6.63,-205.79,0.)$ & 205.90 \\
 \hline
\multirow{2}{*}{28} &\texttt{q3.\_\_0.56\_0.56\_0.\_\_0.6\_0.\_0.\_T\_80\_400} & \texttt{BAM}  & 3  & $(0.75, -0.27, 0.)$ & $(0.3, 0.52, 0.)$   & 0 & 8.83 &  0.0329  &  2.94   &  0.9675    & $(0.299, -0.135, 0.561)$ & 0.650 &  $(-471.57, 54.81, -896.04)$   & 1014.04\\
 & \texttt{q3.\_0.\_0.\_T\_80\_400} & \texttt{BAM} & 3.& $(0., 0., 0.)$& $(0., 0., 0.)$& 0.& 10.& 0.0177& 2.12  &  0.9721 & $(0., 0., 0.541)$ &0.541 &  $(-64.48, -124.91, 0.)$  &140.57 \\
 \hline
\multirow{2}{*}{29} & \texttt{SXS:BBH:0035}& \texttt{SpEC}  &  3.& $(0.5, 0.03, 0.)$& $(0., 0., 0.)$& 0.& 17.& 0.0132& 0.44  &  0.9704 & $(0.167,0.023,0.581)$ & 0.605 & $(144.12,216.96,186.26)$ & 320.21  \\
& \texttt{SXS:BBH:0030} & \texttt{SpEC} & 3.& $(0., 0., 0.)$& $(0., 0., 0.)$& 0.& 14.& 0.0177& 2.12  & 0.9710 & $(0.,0.,0.541)$ &  0.541 & $(149.78,90.38,0.)$ & 174.94\\
 \hline
 \multirow{2}{*}{30} & \texttt{SXS:BBH:0048} & \texttt{SpEC} & 3.& $(0., 0., 0.5)$& $(0.47, 0.16, 0.)$& 0.375& 14.& 0.0175& 0.20  & 0.9607 &$(0.023,-0.002,0.755)$ & 0.755 & $(2.83,-78.51,-235.65)$ & 248.40 \\
 & \texttt{SXS:BBH:0174} & \texttt{SpEC} & 3.& $(0., 0., 0.5)$& $(0., 0., 0.)$& 0.375& 17.& 0.0132& 0.36  & 0.9607 & $(0.,0.,0.756)$ & 0.756 & $(-100.76,21.66,0.)$ & 103.07 \\
 \hline
\multirow{2}{*}{31} & \texttt{SXS:BBH:0050} & \texttt{SpEC} & 3.& $(0.49, 0.07, 0.)$& $(0.47, 0.18, 0.)$& 0.& 14.& 0.0175& 0.18  & 0.9695 & $(0.243,0.003,0.561)$ & 0.611 & $(-397.86,211.53,-819.3)$ & 935.03 \\
 & \texttt{SXS:BBH:0030} & \texttt{SpEC} & 3.& $(0., 0., 0.)$& $(0., 0., 0.)$& 0.& 14.& 0.0177& 2.12  &  0.971 & $(0.,0.,0.541)$ &  0.541 & $(149.78,90.38,0.)$ & 174.94 \\
 \hline
\multirow{2}{*}{32} &  \texttt{SXS:BBH:0051} & \texttt Spec & 3.& $(0., 0., 0.5)$& $(-0.47, -0.16, 0.)$& 0.375& 14.& 0.0174& 0.16   & 0.9607 & $(-0.023,0.002,0.755)$ & 0.755 & $(6.,-78.14,231.09)$ & 244.02 \\
 & \texttt{SXS:BBH:0174} & \texttt{SpEC} & 3.& $(0., 0., 0.5)$& $(0., 0., 0.)$& 0.375& 17.& 0.0132& 0.36  & 0.9607 & $(0.,0.,0.756)$ & 0.756 & $(-100.76,21.66,0.)$ & 103.07\\
 \hline
\multirow{2}{*}{33} & \texttt{SXS:BBH:0053} & \texttt{SpEC} & 3.& $(0.49, 0.07, 0.)$& $(-0.46,-0.18, 0.)$& 0.& 14.& 0.0176& 0.20  &  0.9707  & $(0.201,-0.006,0.562)$ &  0.597 & $(-201.97,216.74,-405.19)$ & 501.94\\
 & \texttt{SXS:BBH:0030} & \texttt{SpEC} & 3.& $(0., 0., 0.)$& $(0., 0., 0.)$& 0.& 14.& 0.0177& 2.12  & 0.9710 & $(0.,0.,0.541)$ &  0.541 & $(149.78,90.38,0.)$ & 174.94\\
 \hline
\multirow{2}{*}{34} & \texttt{q4.\_\_0.5\_0.\_-0.3\_\_0.\_0.5\_-0.3\_\_pcD12} & \texttt{ET} & 4.& $(0.5, 0.0, -0.3)$& $(0.,0.5, -0.3)$& -0.3& 11.75 & 0.0225 & 2.82  & 0.9807  & $(0.216,0.054,0.380)$ & 0.44 &$(-171.36,46.93,-242.82)$  & 300.88\\
 &   \texttt{q4.\_\_-0.3\_-0.3\_\_pcD12} & \texttt{ET} & 4.& $(0., 0., -0.3)$& $(0., 0., -0.3)$& -0.3& 11.85& 0.0225 & 0.95  & 0.9814 &$(0.,0.,0.305)$ & 0.305 &  $(50.33,-175.04,0.)$   & 182.13\\
 \hline
\multirow{2}{*}{35} & \texttt{q4.\_\_0.\_0.5\_-0.3\_\_0.\_0.5\_-0.3\_\_pcD12} & \texttt{ET} & 4.& $(0., 0.5, -0.3)$& $(0.,0.5, -0.3)$& -0.3& 11.81 & 0.0224 & 1.41  &  0.9804 & $(-0.033,0.232,0.386)$ & 0.452 &  $(28.33,-215.29,-310.75)$  & 379.10\\
 &   \texttt{q4.\_\_-0.3\_-0.3\_\_pcD12} & \texttt{ET} & 4.& $(0., 0., -0.3)$& $(0., 0., -0.3)$& -0.3& 11.85& 0.0225 & 0.95  & 0.9814 &$(0.,0.,0.305)$ & 0.305  & $(50.33,-175.04,0.)$   &  182.13 \\
 \hline
\multirow{2}{*}{36} &\texttt{SXS:BBH:0058} & \texttt{SpEC} & 5.& $(0.5, 0.03, 0.)$& $(0., 0., 0.)$& 0.& 15.& 0.0158& 2.12  & 0.9815 & $(0.287,0.014,0.455)$ &  0.538 & $(274.63,-163.17,210.99)$ & 382.84 \\
 & \texttt{SXS:BBH:0056} & \texttt{SpEC} & 5.& $(0., 0., 0.)$& $(0., 0., 0.)$ & 0.& 15.& 0.0159& 0.50  & 0.9824 & $(0.,0.,0.417)$ &  0.417 & $(-68.11,-122.09,0.)$ & 139.81 \\
\hline
\hline
\end{tabular}
}
\end{table*}
\end{turnpage}

\section{Mismatches of higher order modes}
\label{sec:AppendixB}
Complementary to Sec.~\ref{sec:systematic} here we present the results of single mode mismatches for the remaining higher order modes. 

\begin{figure}[!]
\centering
\includegraphics[scale=0.36]{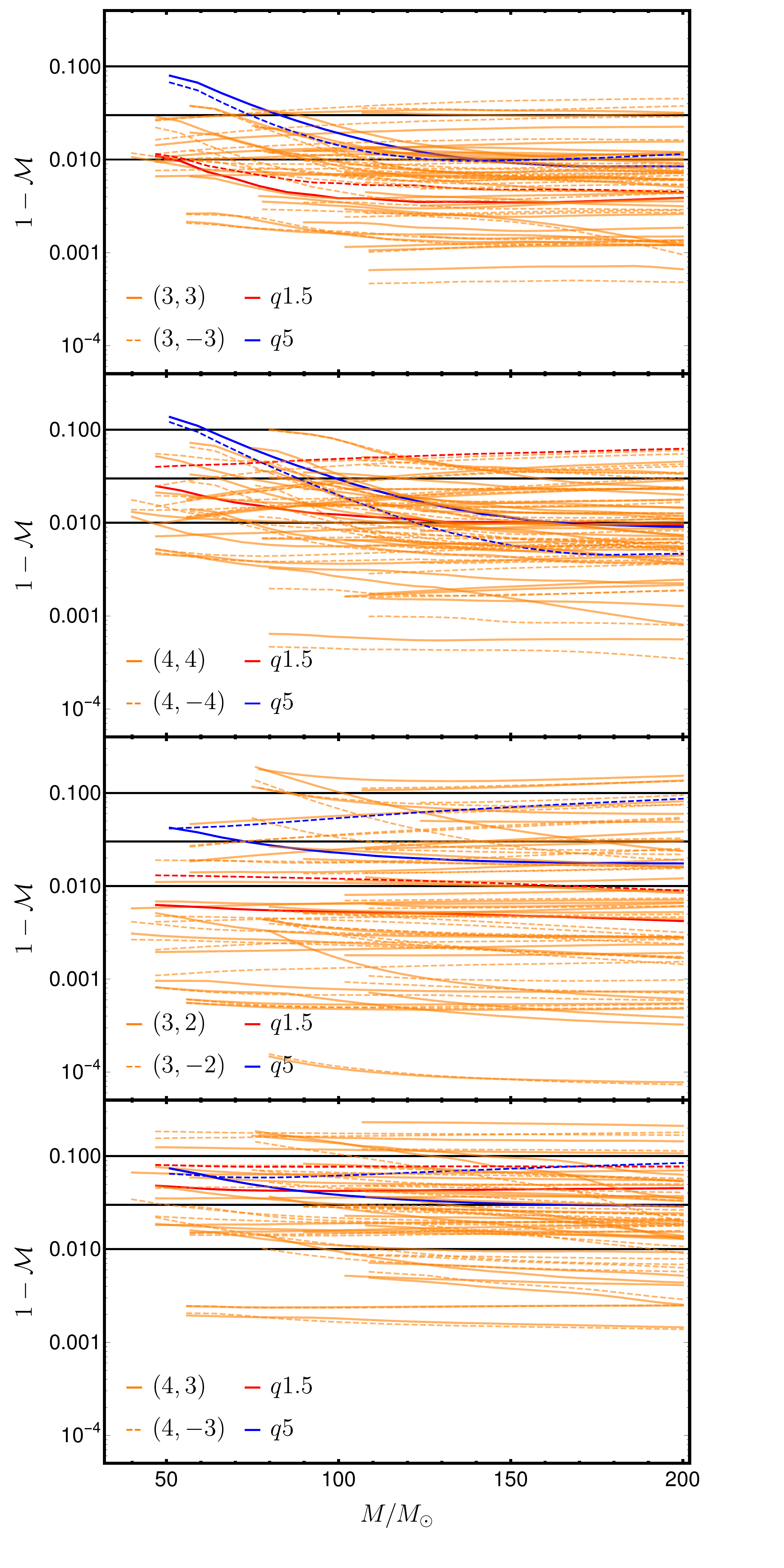}
\caption{Mode by mode mismatches between QA and AS modes for all NR configurations as a function of the total mass of the system. From the top to the bottom we show the results for the $(3,\pm 3)$, $(4,\pm 4)$,  $(3,\pm 2)$,$(4,\pm 3)$ modes, respectively. The configurations with IDs \SXSqonepointfiveID{} and \SXSqfiveID{} are highlighted with red and blue colors, respectively. The thick and dashed curves correspond to positive and negative m modes, respectively. The horizontal lines mark the $1\%$, $3 \%$ and $10\%$ value of the mismatch.}
\label{fig:mmQAHMs}
\end{figure}

Figure \ref{fig:mmQAHMs} shows the results for the mismatches between QA and AS modes following Sec.~\ref{sec:QAvsSA}. From top to bottom, the plots refer to the $(3,|3|)$, $(4,|4|)$, $(3,|2|)$, and $(4,|3|)$ modes. Additionally, mismatches for the configurations with IDs \SXSqonepointfiveID{} and \SXSqfiveID{} are highlighted with red and blue colors, respectively. The horizontal lines mark the $1\%$, $3 \%$ and $10 \%$ value of the mismatch. We find overall increase in the mismatch values in the two lowest panels, corresponding to $(3,|2|)$ and $(4,|3|)$ modes compared to the $(3,|3|)$, $(4,|4|)$ modes (top two panels). As discussed in the main text, this increase is caused by the strong mode-mixing effect in the $(3,|2|)$ and $(4,|3|)$ modes which is not captured properly by (APX1). 

\begin{figure*}[hbt!]
\centering
\includegraphics[scale=0.4]{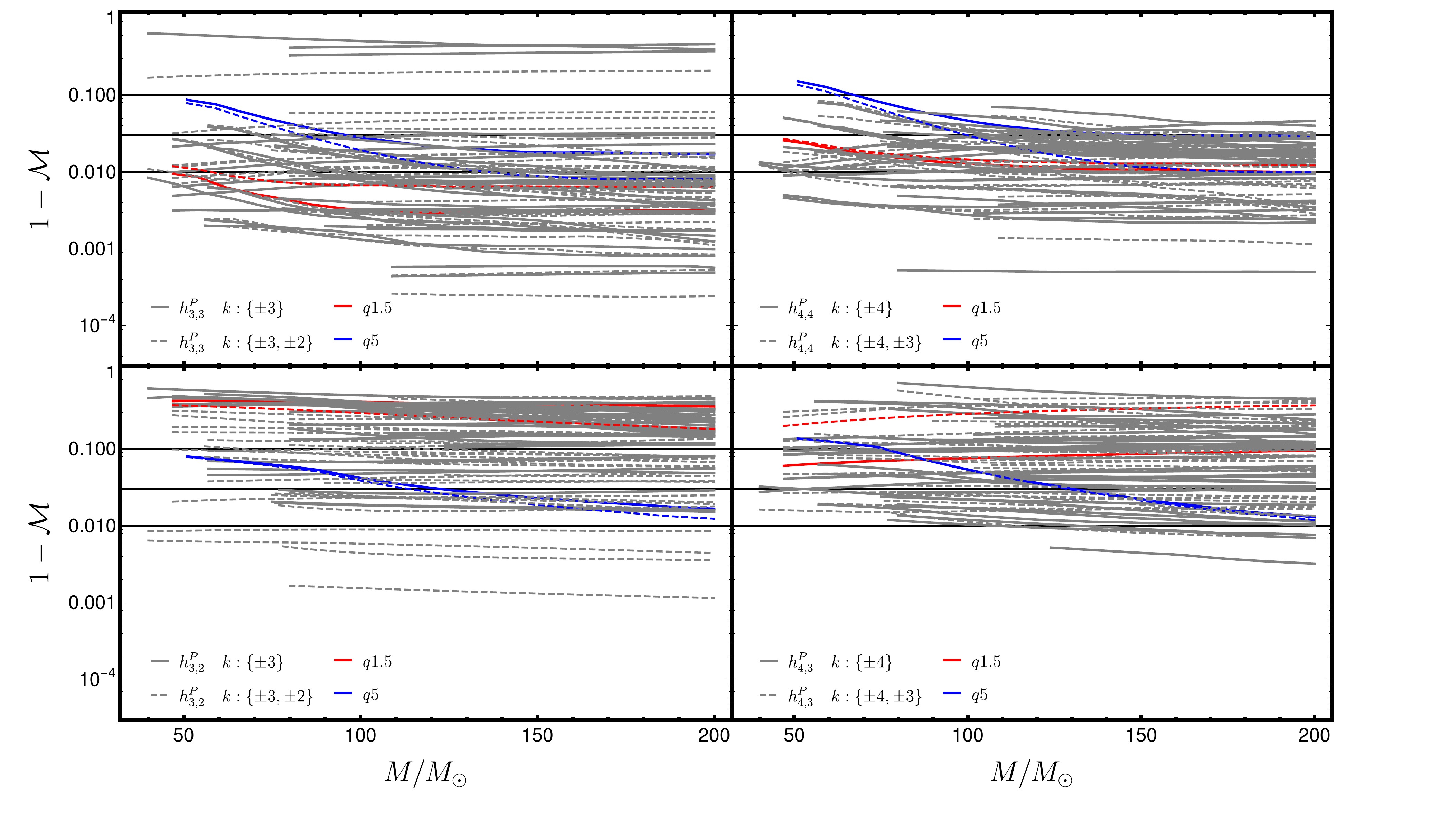}
\vspace*{-0.75cm}
\caption{Mode by mode mismatches between precessing and approximate precessing modes for all configurations in Tab. \ref{tab:tabNR3} as a function of the total mass of the system. The top left and right panels show the results for the $(3,3)$ and $(4,4)$ modes; the bottom left and right plots for the $(3,2)$ and $(4,3)$ modes. The thick and dashed lines correspond to taking two or four aligned-spin waveforms to generate the approximate precessing waveforms, respectively. The letter $k$ represents the index of the rotation operator given in Eq. \eqref{eq01}. In addition, configurations  with  IDs \SXSqonepointfiveID{} and \SXSqfiveID{} are highlighted with red and blue colors, respectively. The horizontal lines mark the $1\%$, $3 \%$ and $10 \%$ value of the mismatch.}
\label{fig:mmPrecHMs}
\end{figure*}

Single mode mismatches between approximate precessing and precessing waveforms for the higher order modes $\{\ell,m\}=\{(3,2),(3,3),(4,3),(4,4)\}$ are shown in Fig. \ref{fig:mmPrecHMs}. The top left and right panels correspond to the $(3,3)$ and $(4,4)$ modes; the bottom left and right panels show the results for the $(3,2)$ and $(4,3)$ modes. The configurations with IDs \SXSqonepointfiveID{} and \SXSqfiveID{} are highlighted with red and blue colors, respectively. The thick and dashed lines correspond to taking 2 and 4 AS waveforms to generate the approximate precessing waveforms, respectively. For instance, in the case of the $(3,2)$ mode the thick lines correspond to taking the AS $(3,\pm 3)$ modes, while the dashed lines to taking the AS $(3,\pm 3)$ and $(3,\pm 2)$ modes into account in the construction (see Sec.~\ref{sec:PrecvsP} for details). 
For the higher order modes, we find that the modes affected by mode-mixing, $(3,|2|)$ and $(4,|3|)$, have high mismatches with less than $30\%$ of cases below $3\%$ (see Tab. \ref{tab:tabmmPrec}). The other subdominant modes, $(3,3)$ and $(4,4)$, have mismatches below $3\%$ for more than $80 \%$ of cases. Furthermore, the inclusion of more AS modes, although it has a moderate impact, tends to improve the mismatches.

\begin{figure*}[hbt!]
\centering
\includegraphics[scale=0.4]{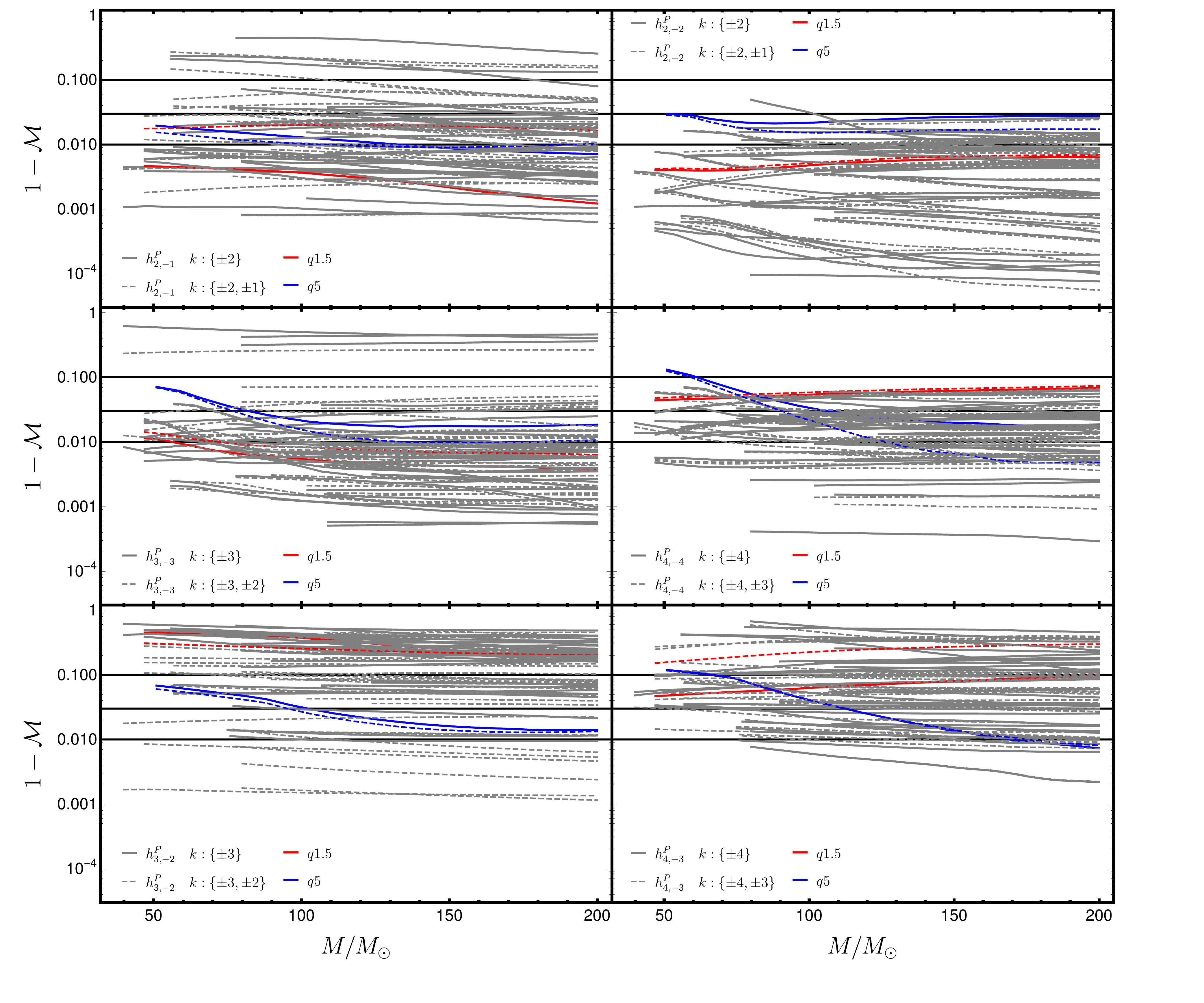}
\vspace*{-0.5cm}
\caption{Mode by mode mismatches between negative $m$ precessing and approximate precessing modes for all NR configurations as a function of the total mass of the system. We show results for the following modes: $(2,-1)$ (top left), $(2,-2)$ (top right), $(3,-3)$ (middle left), $(4,-4)$ (middle right), $(3,-2)$ (bottom left) and $(4,-3)$ (bottom right). The thick and dashed lines correspond to taking two and four aligned-spin waveforms to generate the approximate precessing waveforms, respectively. The letter $k$ represents the index of the rotation operator given in Eq. \eqref{eq01}. Configurations  with  IDs \SXSqonepointfiveID{} and \SXSqfiveID{} are highlighted in red and blue, respectively. The horizontal lines mark mismatches of $1\%$, $3 \%$ and $10 \%$.}
\label{fig:mmNegPrec}
\end{figure*}

Figure \ref{fig:mmNegPrec} shows the results for single mode mismatches between the approximate precessing and precessing negative $m$-modes $\{\ell,m\}=\{(2,-2),(2,-1),(3,-2),(3,-3),(4,-3),(4,-4)\}$ for all NR pairs as a function of the total mass of the system. Comparing Fig. \ref{fig:mmPrec} and \ref{fig:mmNegPrec} we identify some asymmetries between the positive and negative $m$-modes. For instance, focusing on the highlighted configurations, IDs ~\SXSqfiveID ~and~\SXSqonepointfiveID, we find slightly smaller mismatches for the negative $m$-modes than for the positive ones.

\begin{figure*}[hbt!]
\centering
\includegraphics[scale=0.4]{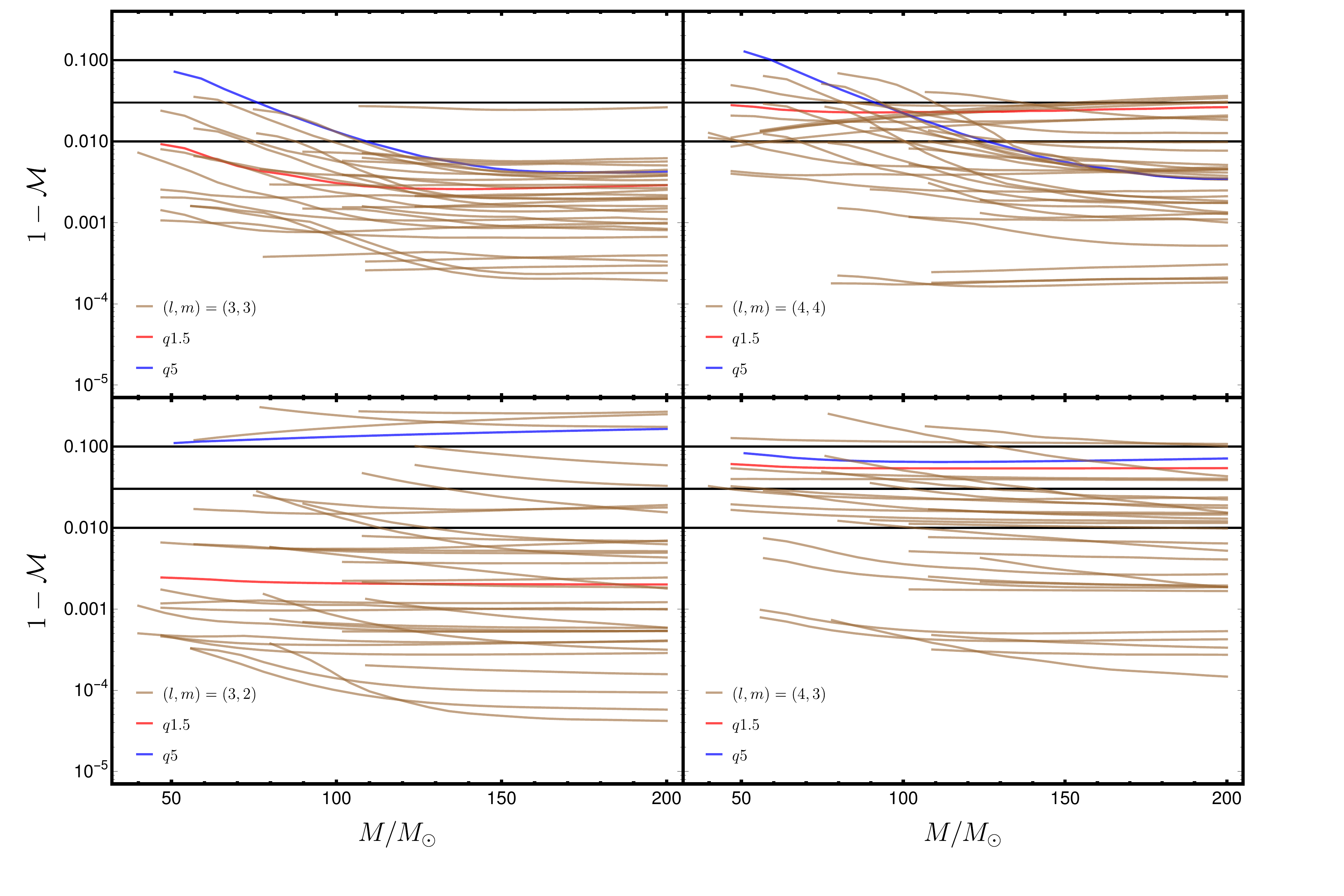}
\vspace*{-0.75cm}
\caption{Single mode mismatches between the AS modes, $h^{AS}_{\ell m}$, and the symmetric QA modes, $h^+_{\ell m}$, as a function of the total mass of the system for all configurations in Tab. \ref{tab:tabNR3}. Top row: Results for the $(3, 3)$ (left) and $(4,4)$ modes (right). Bottom row: Result for the $(3, 2)$ (left) and $(4, 3)$ modes (right). The configurations with IDs \SXSqonepointfiveID{} and \SXSqfiveID{} are highlighted with red and blue colors, respectively. The horizontal lines mark the $1\%$, $3 \%$ and $10 \%$ value of the mismatch.}
\label{fig:mmQASymvsASHMs}
\end{figure*}
 
\begin{figure*}[hbt!]
\centering
\hspace*{-0.75cm}
\includegraphics[scale=0.3]{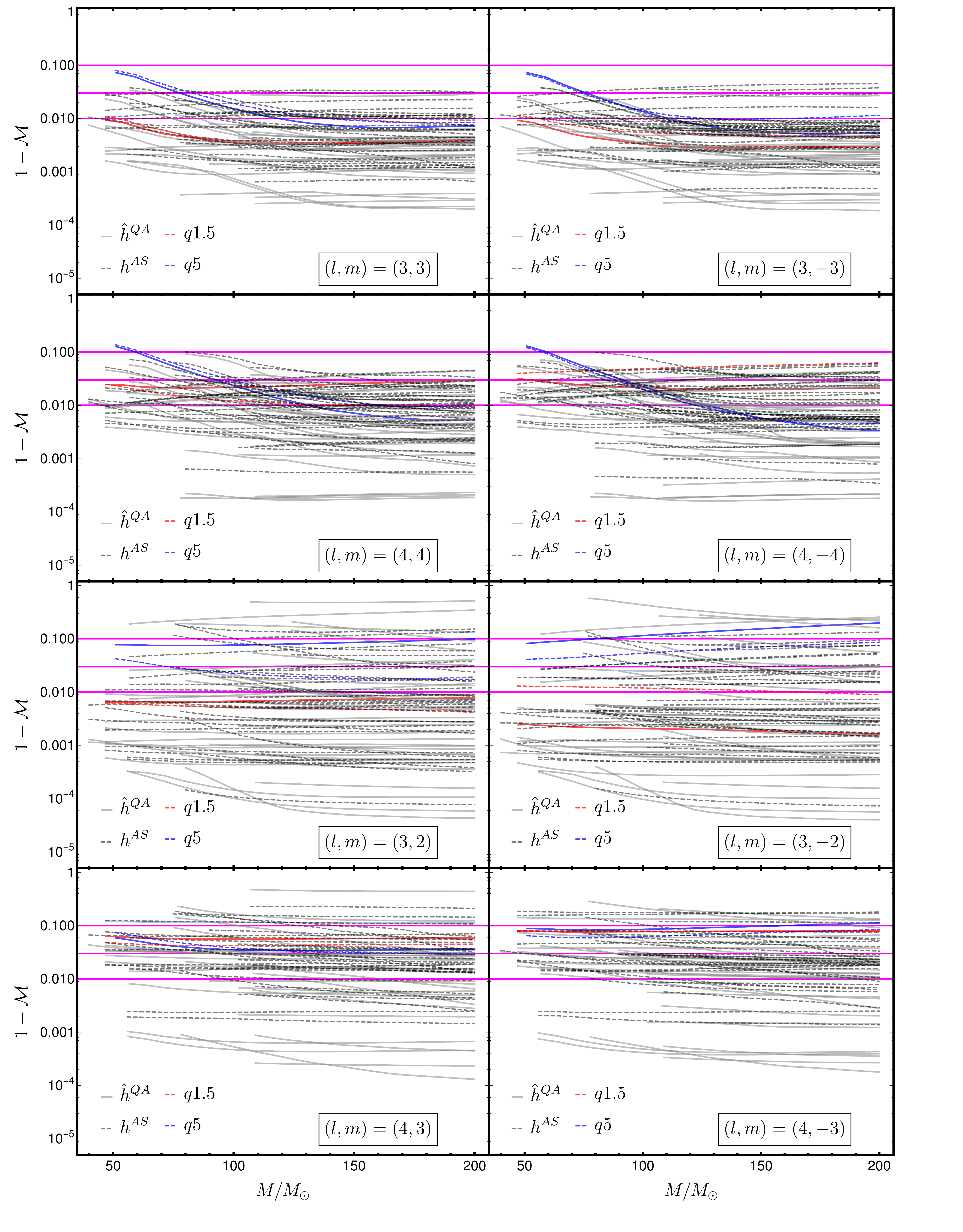}
\vspace*{-0.75cm}
\caption{
Single mode mismatches between the QA modes, $h^{QA}_{\ell m}$, the approximate QA modes, $\hat{h}^{QA}_{\ell m}$, and the AS modes, $h^{AS}_{\ell m}$, as a function of the total mass of the system for all NR configurations. In the left [right] panels from top to bottom we show the results for the $\{(3,3)$, $(4,4)$,$(3,2)$,$(4,3)\}$ [$\{(3,-3)$, $(4,-4)$, $(3,-2)$,$(4,-3)\}$] modes. The thick gray (dashed black) lines correspond to mismatches between $h^{QA}_{\ell m}$ and $\hat{h}^{QA}_{\ell m}$ ($h^{AS}_{\ell m}$). In addition, configurations  with  IDs \SXSqonepointfiveID{} and \SXSqfiveID{} are highlighted with red and blue colors, respectively. The horizontal lines mark the $1\%$, $3 \%$ and $10 \%$ value of the mismatch. In the odd-$m$ panels the cases with PI symmetry have been removed}
\label{fig:mmQAvsQAASHMs}
\end{figure*}
 
In Sec. \ref{sec:WaveformDecomp} we have further investigated the time domain decomposition of co-precessing waveforms used by precessing surrogate models~\cite{Blackman:2017dfb, PhysRevD.96.024058}. We show the results of this analysis for higher order modes in Figs. \ref{fig:mmQAvsQAASHMs} and \ref{fig:mmQASymvsASHMs}. The identification between AS and the slowly varying part of the QA modes, referred to as symmetric QA modes defined as $h^{+}_{\ell m}=A^+_{\ell m} e^{i \phi^-_{\ell m}}$, is quantified through mismatches displayed in Fig. \ref{fig:mmQAvsQAASHMs}. Overall, we find that this approximation gets worse for higher order modes, especially for the modes affected significantly by mode mixing. Given this first approximation, we then constructed approximate QA modes (see Sec. \ref{sec:WaveformDecomp} for details), $\hat{h}^{QA}_{\ell m}$, replacing the slowly-varying part of the QA modes by the AS amplitude and phase. The mismatches between the approximate QA and the QA modes for higher order modes are shown in Fig.~\ref{fig:mmQASymvsASHMs}. Similarly to the first approximation, we find an increase in mismatch, in particular for the modes affected by mode-mixing.

\section{PI symmmetry and waveform systematics}   
\label{sec:AppendixC}
\begin{figure*}[!h]
       \includegraphics[scale=0.38]{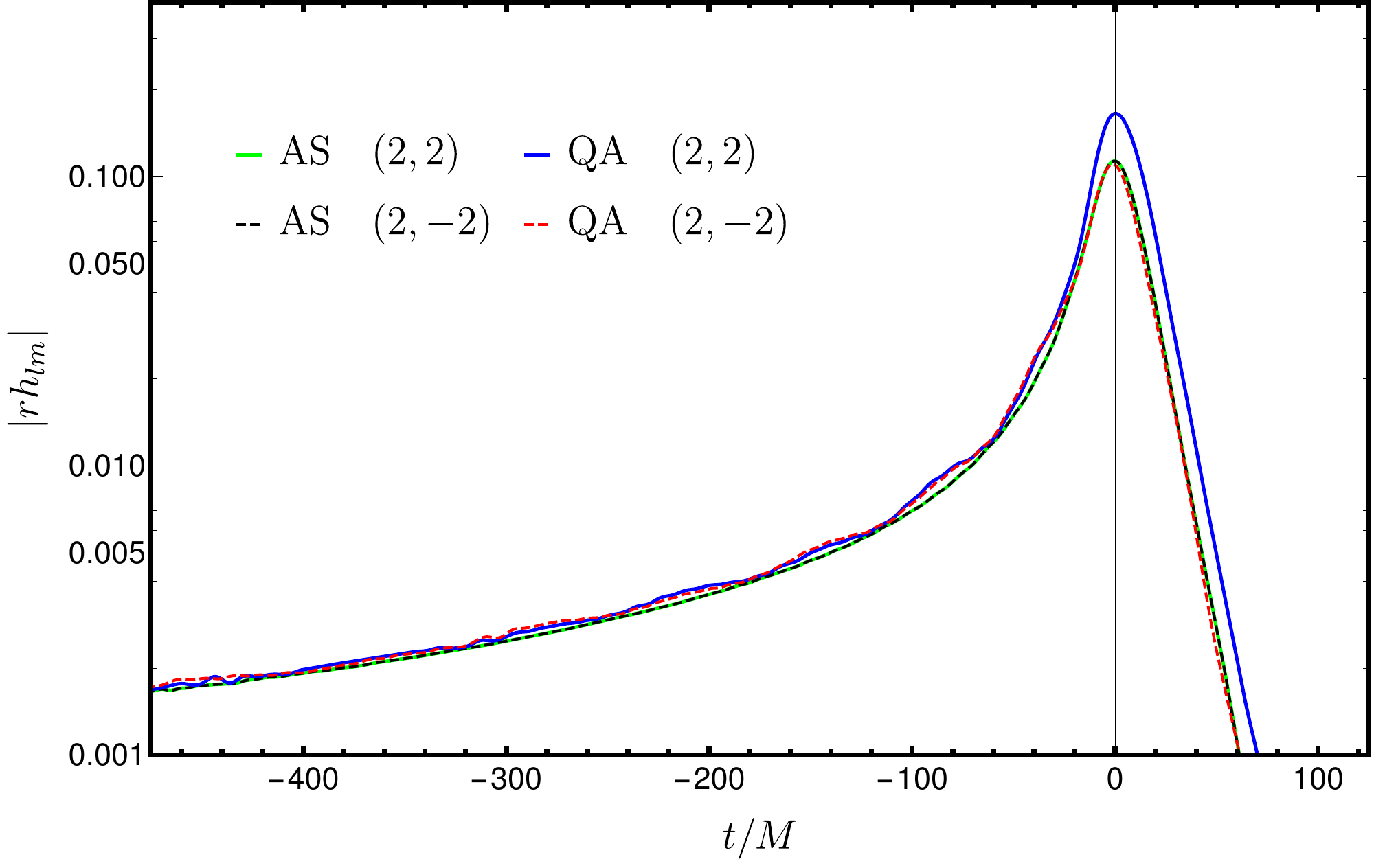} \quad
       \includegraphics[scale=0.38]{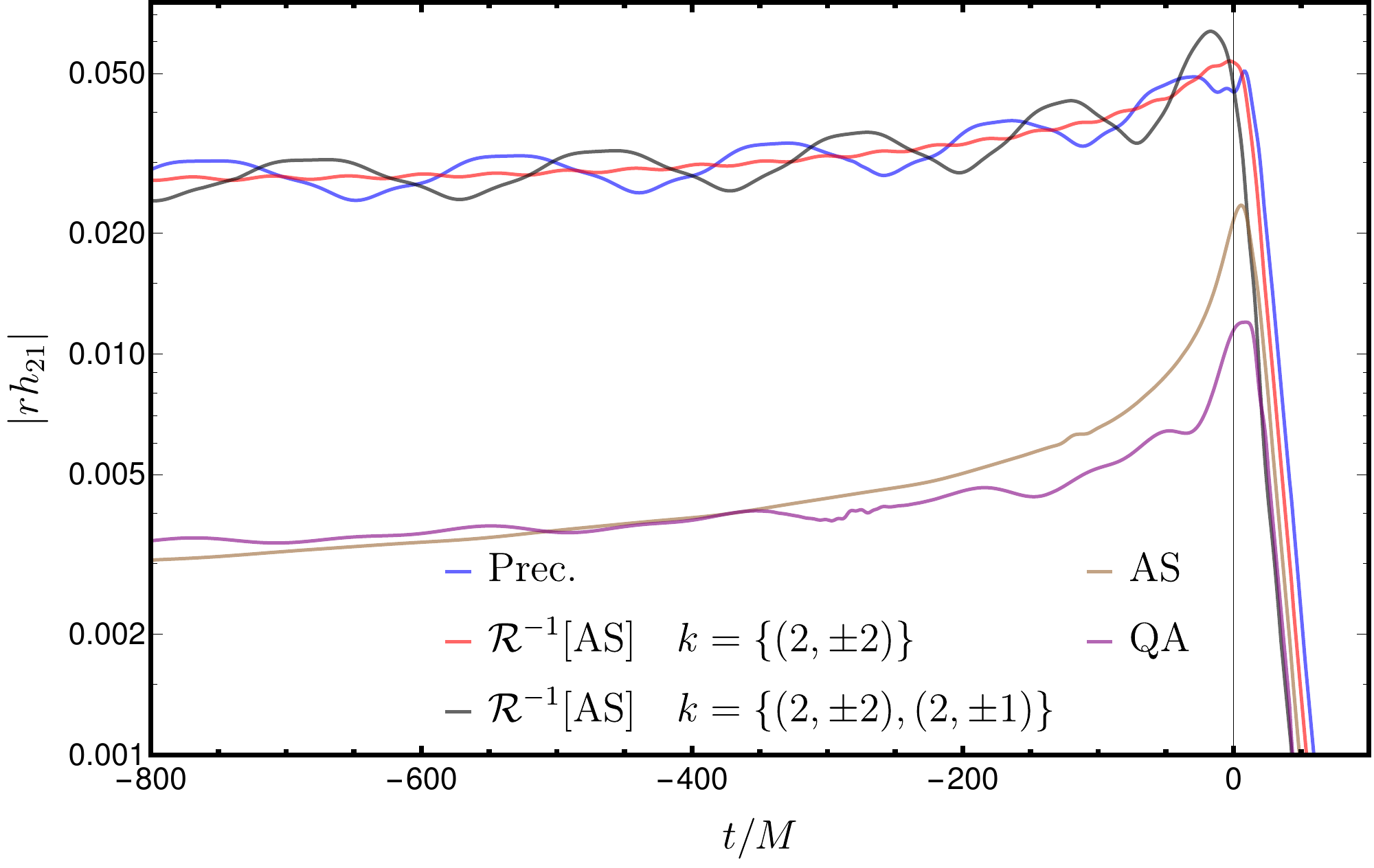}
       \caption{Left: Time domain amplitudes of the $\{\ell, m\}=\{(2,2),(2,-2)\}$-modes for the configuration with ID 28 in Tab. \ref{tab:tabNR3}. 
       Right: Time domain amplitude of the aligned-spin (AS, brown), quadrupole-aligned (QA, purple), precessing (Prec., blue) and approximate precessing ($\mathcal{R}^{-1}\left[AS \right]$, red and black) $(2,1)$-mode for the configuration with ID 10.
       In both panels Tthe vertical line indicates the peak of the AS $(2,2)$-mode.
       }
       \label{fig:plotq3}
\end{figure*}

In Sec.~\ref{sec:systematic} we found (APX1) to be particularly poor for certain binary configurations. 
Once such case is the configuration with ID 28 in Tab. \ref{tab:tabNR3}. The time domain amplitude of $r h_{\ell m}$ for the AS and QA $\{\ell, m\}=\{(2,2),(2,-2)\}$ are shown in the left panel Fig. \ref{fig:plotq3}. The solid and dashed lines represent the positive and negative m modes, respectively. 
In this particular case, the QA $(2,2)$-mode has more power at merger than the corresponding AS mode, which causes the mismatch to rise above the $3\%$. However, the QA $(2,-2)$ mode accurately reproduces the AS mode through merger and ringdown. The mode asymmetry is inherent to precession and is exacerbated by the high $\chi_p$ value of this particular precessing configuration.

In contrast to Fig. \ref{fig:plotASvsQA}, we do not observe time shifts between the QA and AS modes as the QA modes shown in Fig. \ref{fig:plotq3} have been constructed from $\psi_{4,\ell m}$ via fixed frequency integration \cite{Reisswig:2010di}, therefore reducing the amount of time-shift. Note also that these time shifts do not affect the result of the mismatch calculations as they are computed taking into account time shifts between waveforms by performing an inverse Fourier transform.

The right panel of Fig.~\ref{fig:plotq3} shows the $(2,1)$-modes for the configuration with ID 10, a case with a mass averaged mismatch above $10\%$ (see In Sec. \ref{sec:QAvsSA}). We observe a clear difference between the QA (purple) and AS (brown) amplitudes, demonstrating that (APX1) is unable to capture the strong interaction at merger for this configuration. The approximate precessing waveform generated with the either two or four AS modes, resembles the precessing $(2,1)$-mode (blue) better but still not accurately throughout the late inspiral but in particular during the merger. These large differences are the source of the high mismatch.

We have found in Sec.~\ref{sec:QAvsSA} that the case with ID 4 has a very high mismatch for the odd $m$-modes due to the PI symmetry exhibited by equal mass equal spin black holes. For configurations with PI symmetry the odd $m$-modes vanish identically, however, in NR simulations these modes are not zero due to numerical error, although they are extremely small compared to the even $m$-modes. For precessing configuration, however, this symmetry is broken and the odd $m$ QA modes will not vanish. 
As a consequence, the mismatches between the QA and AS odd $m$-modes for such configurations are high. From the four configurations with PI symmetry, IDs 1,2,3 show lower mismatches than ID 4 due to fact that the negative aligned spin component diminishes the difference in the amplitude between the modes resulting in a much lower mismatch when compared to the one of ID 4 ($\chi_{\text{eff}}=0$). This also poses a clear limitation when rotating the $(2,1)$ precessing mode to form the QA $(2,1)$ because the mode mixing in the rotation leaves the QA with more power than the corresponding AS mode. Moreover, it is also a tight constraint in the inverse transformation because the approximate precessing modes can only be generated with the information of the even m modes. This is a clear limitation of (APX1).

Finally, in Sec. \ref{sec:PrecvsP} when analyzing the single mode mismatches of the $(2,2)$-mode (top panel of Fig. \ref{fig:mmPrec}) we found a case, ID 28, with the mismatch curve above the $3\%$ threshold. The configuration with ID 28 is the same as in the co-precessing frame has a mismatch slightly above $3\%$. In the inertial frame it occurs the same situation as in Fig. \ref{fig:plotq3}. The asymmetries between positive and negative $m$ precessing modes are not accurately reproduced by the approximate precessing waveforms. As a consequence, the mismatch of the $(2,2)$ mode is much higher than the mismatch of the $(2,-2)$ mode, which is below the $3\%$ horizontal line as seen in the top right panel of Fig. \ref{fig:mmNegPrec}. 

\section{Contour Plots matches including higher order modes}  
\label{sec:AppendixE}
\begin{figure*}[!]

\noindent\begin{minipage}{\textwidth}

\noindent\begin{minipage}[H]{.45\textwidth}
\includegraphics[scale=0.35]{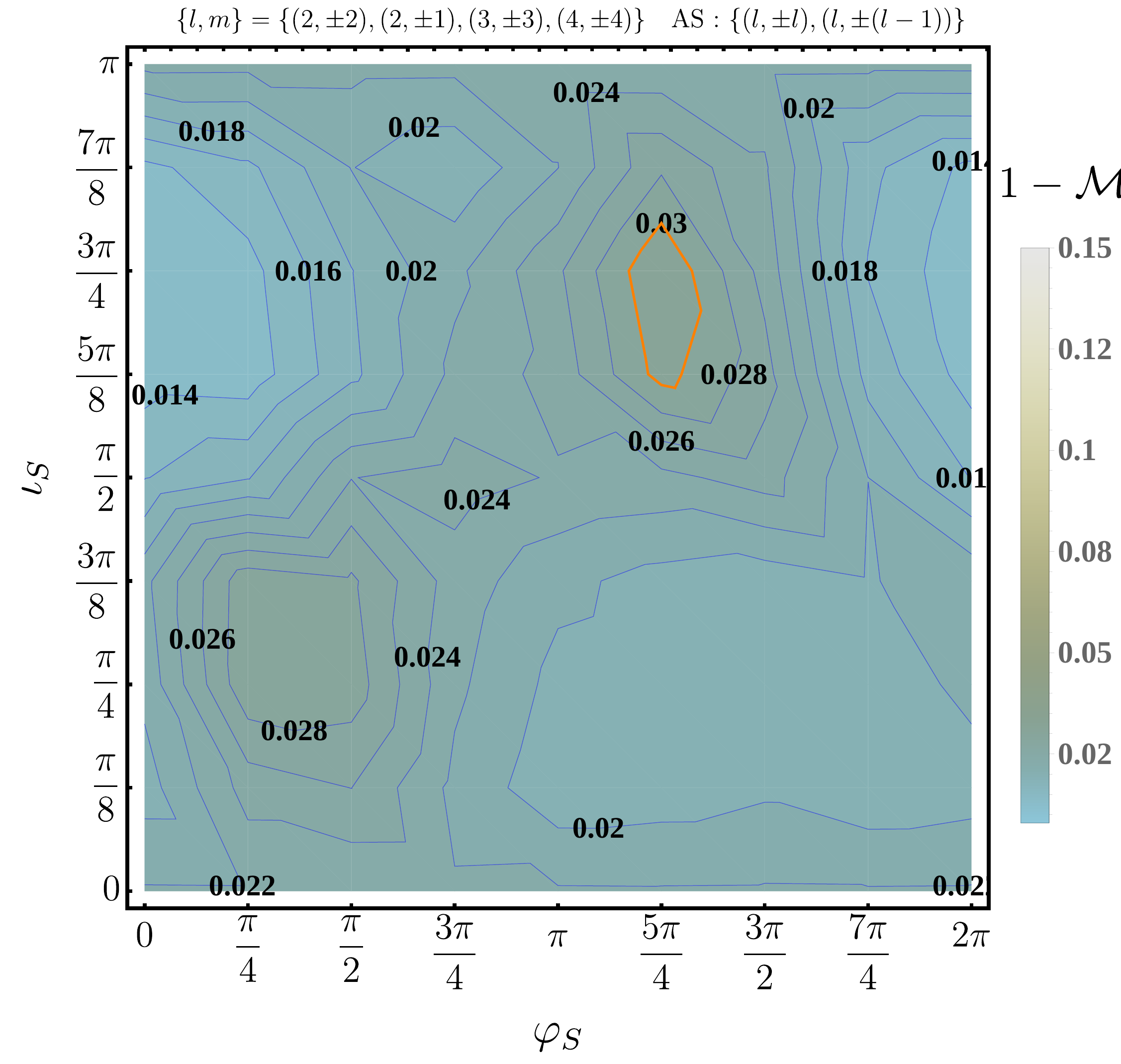}
\end{minipage} 
\hspace{1cm}
\begin{minipage}[H]{.45\textwidth}
\includegraphics[scale=0.35]{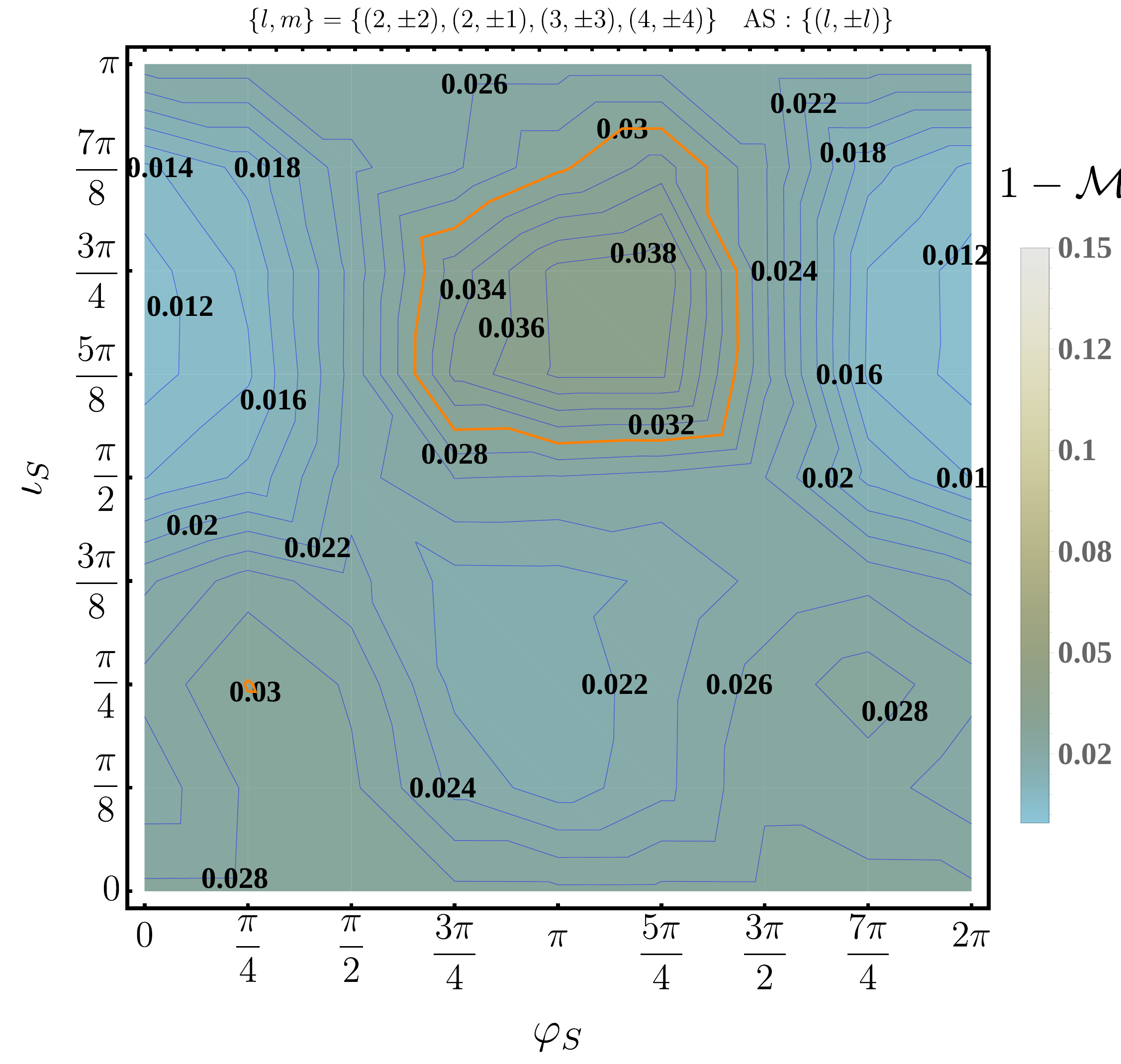}
\end{minipage}

\end{minipage}


\noindent\begin{minipage}{\textwidth}

\noindent\begin{minipage}[H]{.45\textwidth}
\includegraphics[scale=0.35]{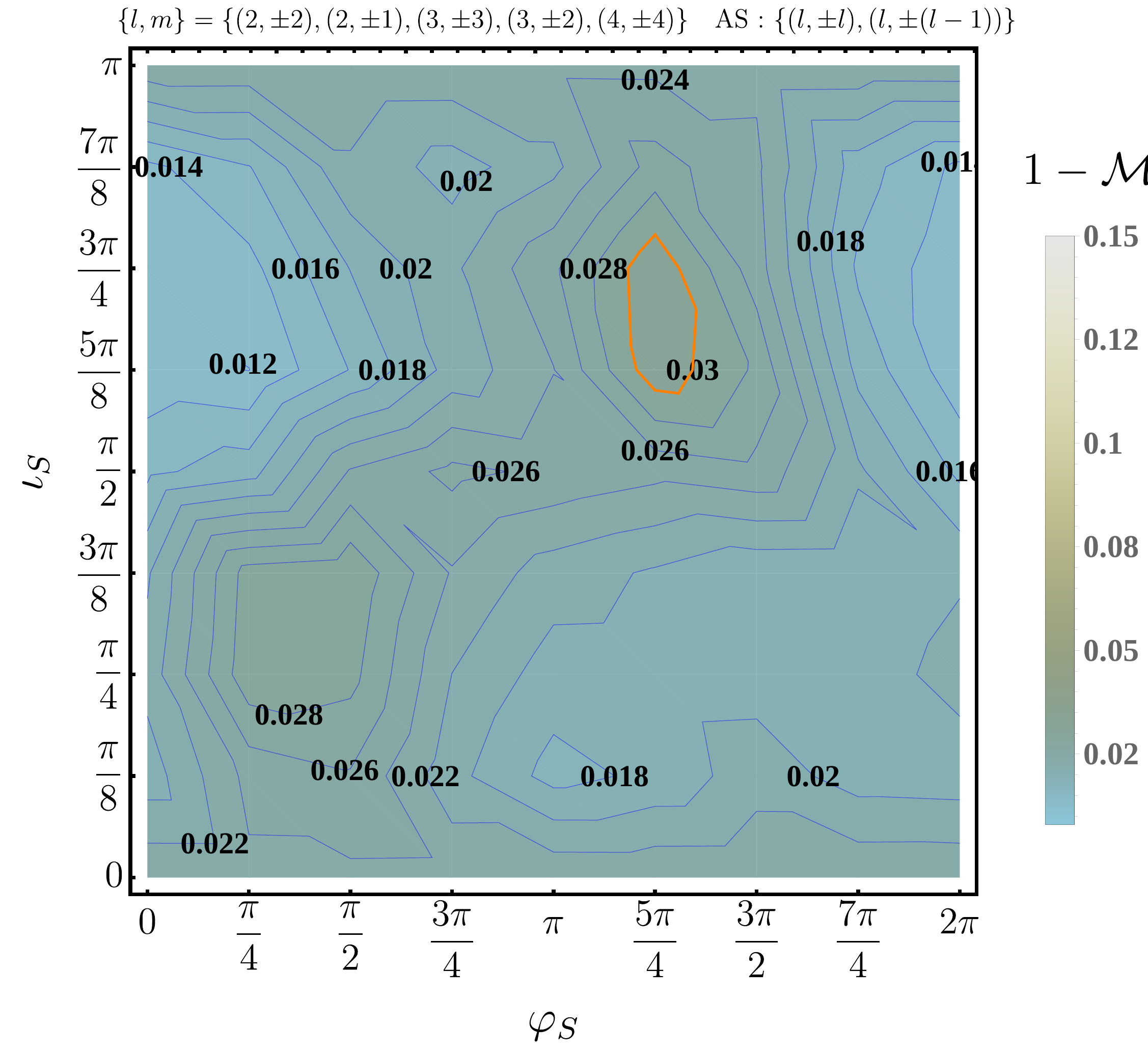}
\end{minipage} 
\hspace{1cm}
\begin{minipage}[H]{.45\textwidth}
\includegraphics[scale=0.35]{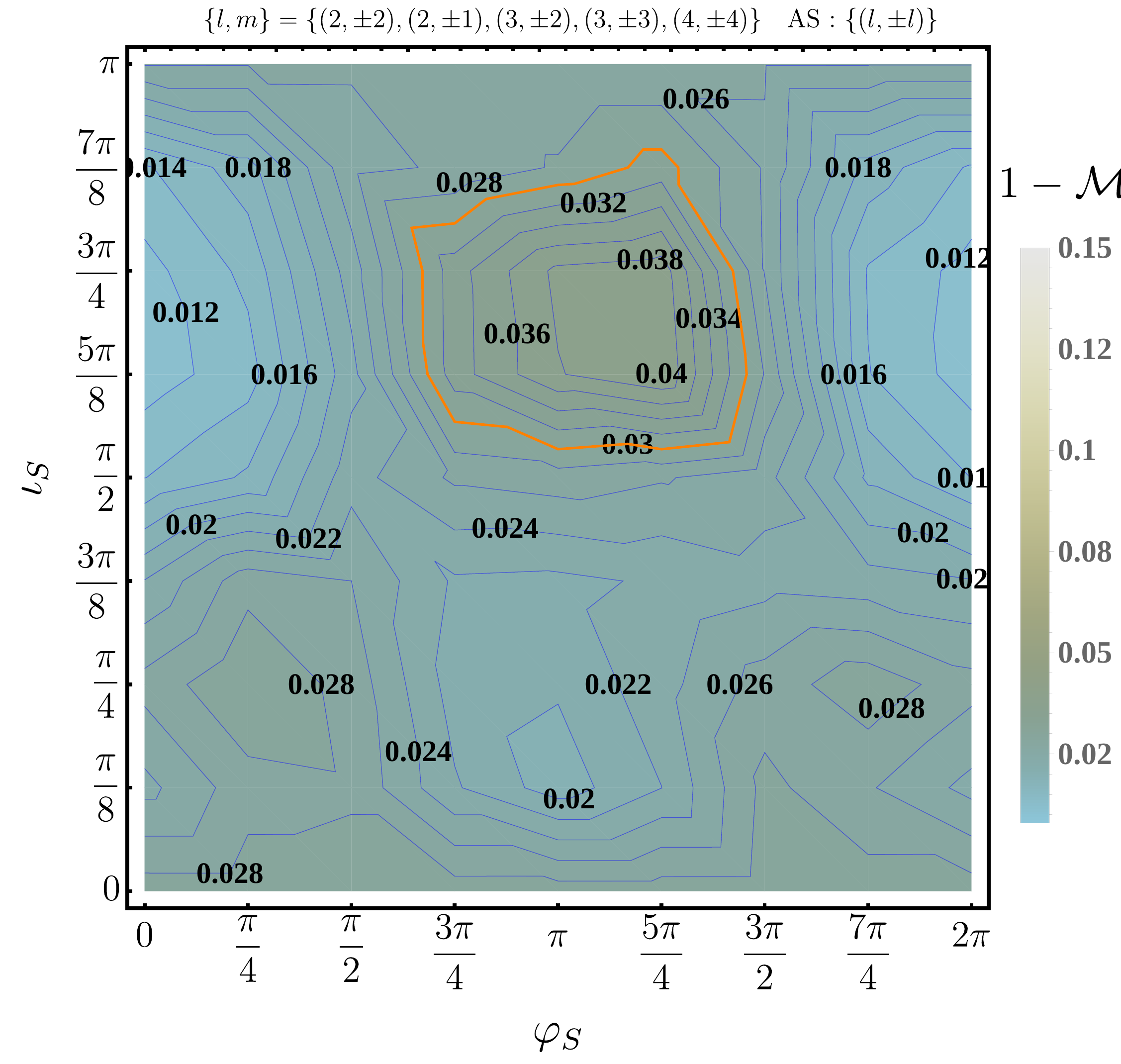}
\end{minipage}

\end{minipage}

\noindent\begin{minipage}{\textwidth}

\noindent\begin{minipage}[H]{.45\textwidth}
\includegraphics[scale=0.38]{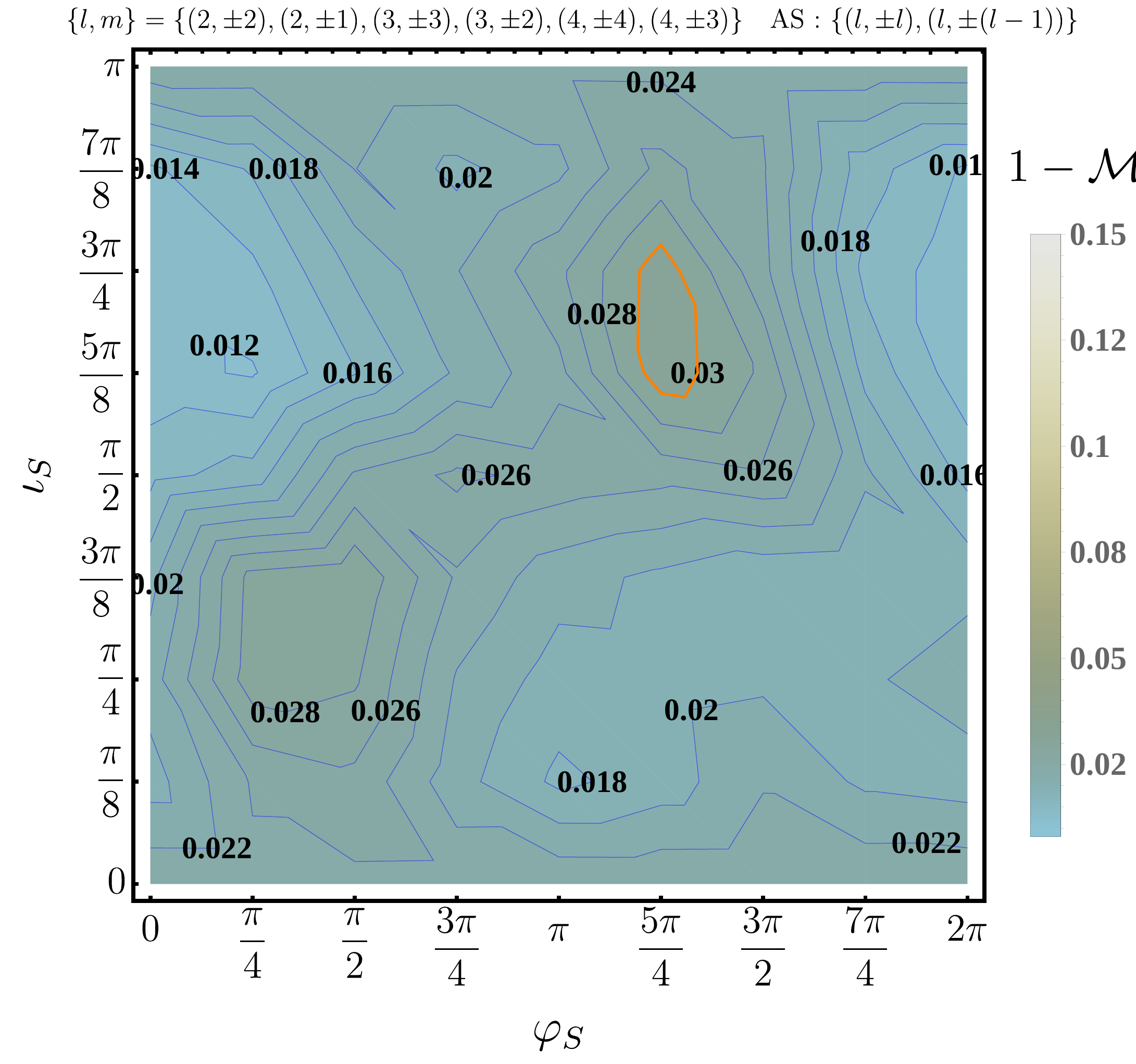}
\end{minipage} 
\hspace{1cm}
\begin{minipage}[H]{.45\textwidth}
\includegraphics[scale=0.35]{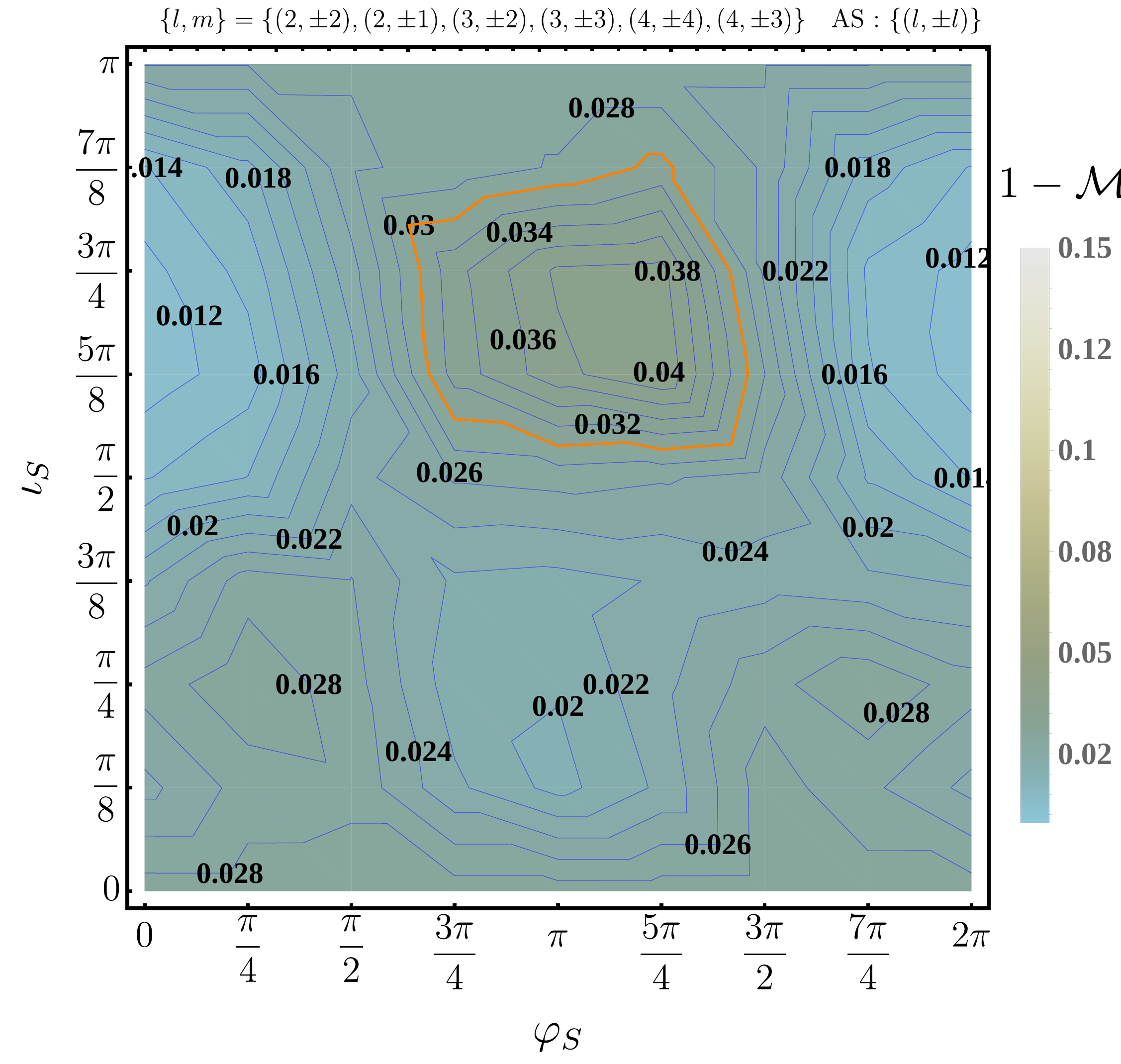}
\end{minipage}
\end{minipage}
\caption{Strain mismatch between precessing and approximate precessing waveforms in the inertial frame averaged over the angle $\kappa_S$ for a total mass of 65 $M_\odot$ for the configuration  with  ID \SXSqfiveID{} as a function of the inclination and the azimuthal angle of the signal (precessing waveform). In the plot labels $\{\ell, m\}$ denotes the modes used in the sum of the complex strain given in Eq. \eqref{eq2}, while AS represent the aligned-spin modes taken into account in Eq. \eqref{eq01}. In addition, the  $3 \%$ and $10 \%$  mismatch values are highlighted with orange and red curvess, respectively.}

\label{fig:strainPrecHMs}
\end{figure*}

Figure \ref{fig:strainPrecHMs} contour plots of the strain mismatches between precessing and approximate precessing waveforms, averaged over the angle $\kappa_S$ for a total mass of 65 $M_\odot$ for the configuration  with  ID \SXSqfiveID{}. In the figure, the label $\{\ell, m\}$ refers to the modes used in the sum of the complex strain of Eq. \eqref{eq2}, while $AS$ represents the aligned-spin modes taken into account in Eq. \eqref{eq01}. In addition, the  $3 \%$ and $10 \%$  mismatch values are highlighted with orange and red curves, respectively. 
In the top, middle and bottom panels the $(2, \pm 2),(2, \pm 1),(3, \pm 3),(4, \pm 4)$; $(2, \pm 2),(2, \pm 1),(3, \pm 3),(3, \pm 2),(4, \pm 4)$ and $(2, \pm 2),(2, \pm 1)$, $(3, \pm 3)$, $(3, \pm 2)$, $(4, \pm 4)$, $(4, \pm 3)$  modes are taken into account in the sum of the complex strain, respectively. In the left and right panels the $(\ell,|\ell|),(\ell,|\ell-1|)$ AS modes are taken into account, respectively. 
The results are similar to the bottom panels of Fig. \ref{fig:strainPrec}. The addition of higher order modes in the complex strain increases the mismatch overall for all inclinations, while the inclusion of more AS higher order modes tends to lower the mismatches.

\clearpage

\bibliography{biblio}

\end{document}